\documentclass[11pt,a4paper]{article}

\makeatletter

\@addtoreset{equation}{section}
\makeatother

\usepackage[margin=3cm]{geometry}
\usepackage[english]{babel}
\usepackage[utf8]{inputenc}
\usepackage[T1]{fontenc}
\usepackage{latexsym} 
\usepackage{amssymb}
\usepackage{amsmath}
\usepackage{graphicx,wrapfig}
\usepackage{color}
\usepackage{transparent}
\usepackage{subcaption}
\usepackage[authoryear]{natbib}

\begin{document}
\phantom{ }
\vskip35mm
\begin{center}
\Huge{Stability Analysis of Ecomorphodynamic Equations} 
\vskip5mm
\Large{The role of riverbed vegetation in river pattern formation}
\vskip15mm
\Large{
Master thesis in Environmental Engineering at EPF Lausanne\\
\vskip10mm
February -- June 2014
\vskip20mm
Fabian Baerenbold 
\vskip45mm
\begin{table*}[h]
       \centering
       \large{
       \begin{tabular}{l l}
       Supervision:   & Dr. Benoît Crouzy (AHEAD)\\
                              & Prof. Paolo Perona (AHEAD)\\
        \end{tabular}
        }
\end{table*}
}
\end{center}
\addtolength{\textheight}{1cm}
\thispagestyle{empty}
\newpage
\section*{Acknowledgements}
I would like to thank Benoît and Paolo for their very competent supervision of this master thesis.
\thispagestyle{empty}

\newpage
\thispagestyle{empty}
\begin{abstract}
Although riparian vegetation is present in or along many water courses of the world, its active role resulting from the interaction with flow and sediment processes has only recently become an active field of research. Especially, the role of vegetation in the process of river pattern formation has been explored and demonstrated mostly experimentally and numerically until now. In the present work, we shed light on this subject by performing a linear stability analysis on a simple model for riverbed vegetation dynamics coupled with the set of classical river morphodynamic equations. The vegetation model only accounts for logistic growth, local positive feedback through seeding and resprouting, and mortality by means of uprooting through flow shear stress. Due to the simplicity of the model, we can transform the set of equations into an eigenvalue problem and assess the stability of the linearized equations when slightly perturbated away from a spatially homogeneous solution. If we couple vegetation dynamics with a 1D morphodynamic framework, we observe that instability towards long sediment waves is possible due to competitive interaction between vegetation growth and mortality. Moreover, the domain in the parameter space where perturbations are amplified was found to be simply connected. Subsequently, we proceed to the analysis of vegetation dynamics coupled with a 2D morphodynamic framework, which can be used to evaluate instability towards alternate and multiple bars. It is found that two kinds of instabilities, which are discriminated mainly by the Froude number, occur in a connected domain in the parameter space. At lower Froude number, instability is mainly governed by sediment dynamics and leads to the formation of alternate and multiple bars while at higher Froude number instability is driven by vegetation dynamics, which only allows for alternate bars.
\end{abstract}
\newpage
\thispagestyle{empty}
\phantom{ }

\newpage
\addtolength{\textheight}{-1cm}
\tableofcontents
\thispagestyle{empty}

\newpage
\thispagestyle{empty}
\phantom{ }

\newpage
\clearpage
\pagenumbering{arabic} 

\section{Introduction}
\begin{figure}
\centering
\includegraphics[width=90mm]{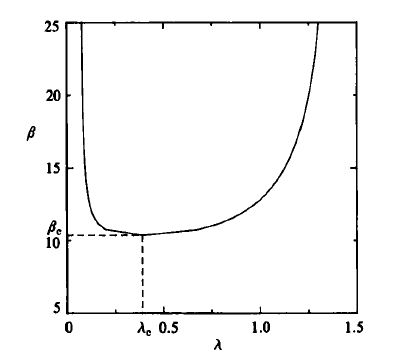}
\caption{Neutral curve for alternate bar formation (instability towards alternate bars above the line, no instability below) taken from \cite{CST}; $\beta$ is the river's aspect ratio and $\lambda$ is the dimensionless longitudinal wavenumber which characterizes the spatial periodicity of the bars}
\label{fig:CST}
\end{figure}

\begin{figure}
\centering
\includegraphics[width=140mm]{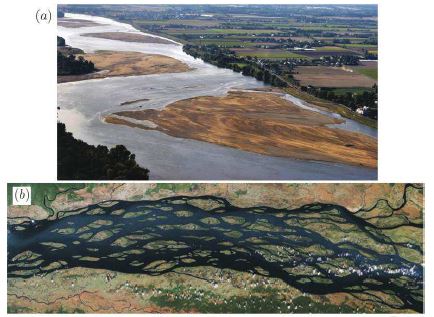}
\caption{Alternate bars in the Loire river (a) and multiple bars in the Congo river (b); Pictures taken from \cite{Chiodi}}
\label{fig:bars}
\end{figure}

\begin{figure}
\centering
\includegraphics[width=160mm]{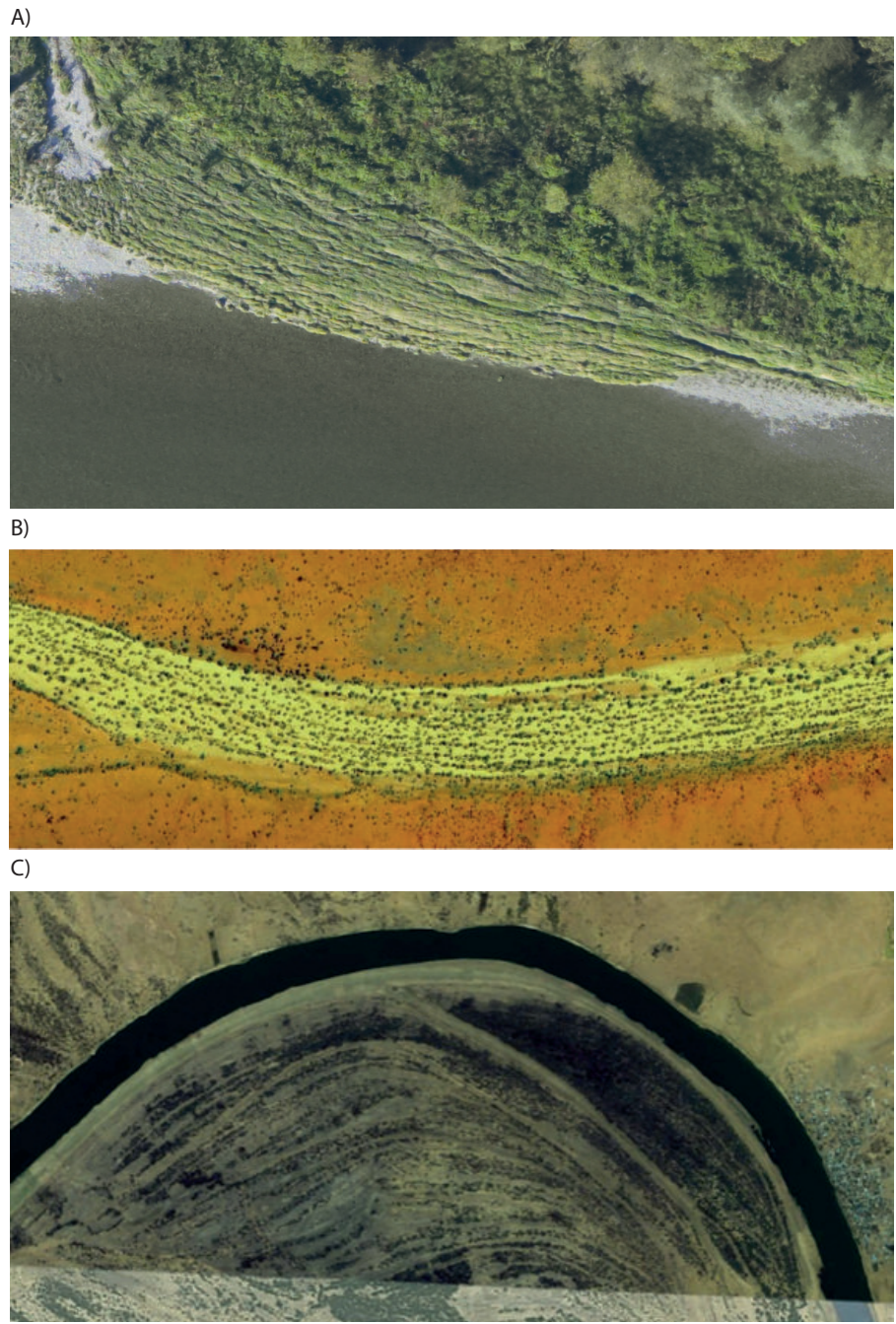}
\caption{Examples of river bed and flood plain vegetation patterns emerging in different fluvial environments: A) rills observed on a bar of the Thur River, Switzerland; B) anabranching of the entire river bed (Marshall River, Australia); C) scroll bars on the inside of a meandering bend (Senegal river, near Bakel), Map data: Google, Digitalglobe} \label{fig:anabranch}
\end{figure}

River landscapes exhibit many different forms in all climatic regions of the world. Nevertheless, one is also able to observe common features which often can be found to be the product of specific water and sediment interaction (see for example \cite{Se} and references therein). The science studying the coupled water and sediment dynamics to explain formation and alteration of river courses is called river morphodynamics and includes research on mainly longitudinal structures like long sediment waves and more complex 2-dimensional structures, among them are alternate and multiple bars (see Figure \ref{fig:bars}). In the past, these research areas have been investigated using experimental setups (see \cite{FePa} for bar experiments for example), numerical simulations (see \cite{Fe} for example) and theoretical approaches based on linear stability analysis (\cite{Callander}, \cite{EnSk} and \cite{Parker}). Linear stability analysis is a concept that allows to study the asymptotic fate ($t\rightarrow \infty$) of a linear or linearized system which is slightly perturbated away from a spatially homogeneous solution. The method was originally developed by \cite{Turing} and has been applied frequently to hydrodynamic topics since. For example, \cite{CST} used linear stability analysis to show that the capacity of a river to develop alternate or multiple bars is closely linked to its aspect ratio (river width divided by depth). This benchmark result is depicted in Figure \ref{fig:CST} and we will reproduce it later in the present work.\\
It is well-known though that in addition to water and sediment transport, riparian vegetation can play a crucial role in river pattern development (see \cite{Gu}). In particular, it is recognized that riparian vegetation affects river morphology through modification of the flow field, bank strength and erosion/sedimentation processes in the riverbed/floodplain (\cite{CaPe}). However, due to the very complex nature of the dynamic interactions between vegetation and sediment transport and flow, riparian vegetation evolution was often not taken into account explicitly. Instead, it was added as a correcting factor for bed roughness and bank stability (\cite{Ja}). While the treatment of vegetation as a correction factor may be justified when looking at short timescales where riparian vegetation density does not change much, this is not the case for river pattern formation which occurs over much longer timescales and where vegetation takes an active role in the process. For instance, dynamic interaction between riparian vegetation and flow and sediment is thought to be crucial in the formation of anabranching river patterns on vegetated bars and in ephemeral rivers in dry regions (Figure \ref{fig:anabranch} A and B). Additionally, we can find similar patterns on the inside of a meandering bend in large streams (scroll bars, Figure \ref{fig:anabranch} C).\\
Recently, researchers have added riparian vegetation dynamics to numerical morphodynamic models and included some of the feedback mechanisms that are thought to occur in nature. Namely, \cite{MuPa} took into account vegetation induced impedance to sediment transport and increase in bank stability and \cite{Perucca} modeled the interaction of bank stabilizing vegetation and a meandering riverbed. Furthermore, \cite{Perona} proposed an analytical morphodynamic model coupled with an equation for riverbed vegetation dynamics. But, until today vegetation dynamics was never included in a stability analysis of morphodynamic equations. In fact, several difficulties arise when trying to formulate a physical vegetation model suitable for stability analysis. For example, in modeling the sediment stabilizing effect of plant root systems is often taken into account as a threshold in the sediment transport function below which no erosion occurs. However, such a threshold possesses mathematical properties that are not suitable for a stability analysis .\\
In the present work, we propose a minimal model for the evolution of riverbed vegetation density which takes into account only very basic mechanisms and is thus suitable for a stability analysis. Using this vegetation model and a standard morphodynamic framework, we would like to explore the possibility of such a coupled morphodynamic-vegetation system (ecomorphodynamic equations) to explain the formation of anabranching patterns. We would like to know which are the determining variables and to what extent the ecomorphodynamic analysis differs from the state of the art river morphodynamics. Hence, we perform an analytical linear stability analysis on the linearized set of ecomorphodynamic equations which describe a model river whose riverbed is colonized by plants. This river is assumed to be of constant width with inerodible banks, the riverbed consists of cohesionless, erodible material (sand/gravel) of uniform size and the river's sediment transport capacity is thought to always exceed the threshold above which sediment transport occurs. Additionally we assume sediment transport to be mainly bedload.\\
We begin by formulating an equation which describes the evolution of riverbed vegetation density (section \ref{sec:veg}) and discuss the different terms and its validity. Important mechanisms to be considered are vegetation growth, distribution by means of seeding and resprouting, and death through flow impact induced uprooting. This equation is then coupled with a standard 1-dimensional and 2-dimensional river morphodynamic framework (sections \ref{sec:1D_gov} and \ref{sec:2D_gov} respectively) which consists of depth-averaged fluid and sediment continuity as well as a formulation for momentum balance in the fluid. These systems are subsequently linearized and perturbated around a spatially homogeneous solution and the conditions for which the wavelike perturbations amplify are investigated using linear stability analysis. This is done for a 1D-framework to study instability towards long sediment waves in section \ref{sec:1D} and for a 2D-framework to study the formation of alternate and multiple bars. The main focus in this work is on highlighting the fundamental role that vegetation dynamics can have in this process together with known mechanisms of sediment dynamics.
\newpage
\phantom{}
\newpage
\section{Modeling Dynamics of Riverbed Vegetation in a Stream}\label{sec:veg}
Stability analysis of morphodynamic equations generally does not include the active role of vegetation explicitly due to the complex nature of the interaction mechanisms. Hereafter, we develop an analytic model for riverbed vegetation dynamics and discuss its validity for different conditions. For simplicity, we model vegetation as rigid, non-submerged cylinders with constant radius and submerged height equal to water depth.
\begin{figure}
\centering
\includegraphics[width=120mm]{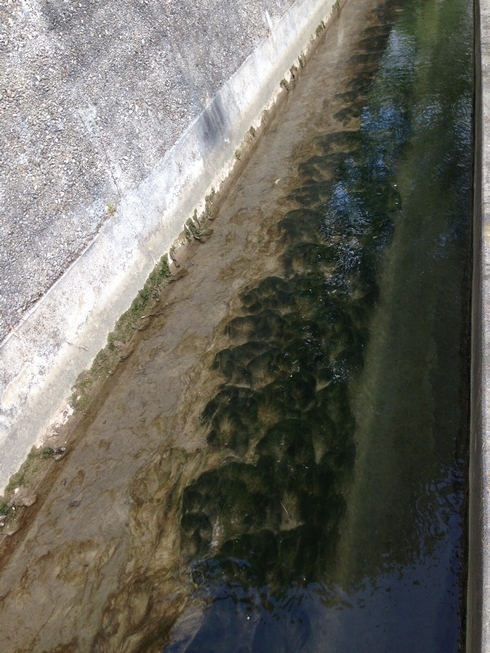}
\caption{Concrete channel with some sediment transport and presence of longitudinal vegetation patterns; Picture credits: Paolo Perona}
\label{fig:veg_paolo}
\end{figure}
We then call $\tilde{\phi}$ the vegetation density defined as number of plants per unit area of riverbed and we model its growth by the logistic term 
\begin{equation}\label{eq:veg1}
\tilde{\alpha}_g\tilde{\phi}(\tilde{\phi}_m-\tilde{\phi})
\end{equation}
with carrying capacity $\tilde{\phi}_m$ and specific vegetation growth rate $\tilde{\alpha}_g$. Furthermore, we assume that vegetation growth is stimulated by nearby existing vegetation by means of seeding and resprouting (i.e. positive local feedback) and we model it using the diffusion term
\begin{equation}\label{eq:veg2}
\tilde{D}\frac{\partial{^2\tilde{\phi}}}{\partial{\tilde{s}^2}},
\end{equation}
with $\tilde{D}$ the streamwise vegetation diffusion constant and $\tilde{s}$ the streamwise coordinate.
We finally want to quantify vegetation death caused by flow drag for which we only consider the direct uprooting effect of flow drag on non-submerged and rigid vegetation (Type I mechanism after \cite{Ed}). In this case, a fluid parcel which impacts on the vegetation is decelerated from mean stream velocity to zero. Furthermore, the rate of fluid that impacts on the vegetation is also proportional to stream velocity while the vegetation cross-section per cubic meter of river is proportional to water depth and vegetation density. We therefore propose the vegetation uprooting term (see also \cite{Wu})
\begin{equation}\label{eq:veg3}
-\tilde{\alpha}_d\tilde{Y}\tilde{U}^2\tilde{\phi},
\end{equation}
where $\tilde{\alpha}_d$ is a proportionality constant, $\tilde{Y}$ the water depth and $\tilde{U}$ the streamwise velocity. Putting together equations (\ref{eq:veg1}) to (\ref{eq:veg3}) we get the rate of change of vegetation density as
\begin{equation} \label{eq:veg_const}
\frac{\partial{\tilde{\phi}}}{\partial{\tilde{t}}}=\tilde{\alpha}_g\tilde{\phi}(\tilde{\phi}_m-\tilde{\phi})+\tilde{D}\frac{\partial{^2\tilde{\phi}}}{\partial{\tilde{s}^2}}-\tilde{\alpha}_d\tilde{Y}\tilde{U}^2\tilde{\phi}.
\end{equation}
\begin{figure}
  \centering
  \def\svgwidth{400pt}
  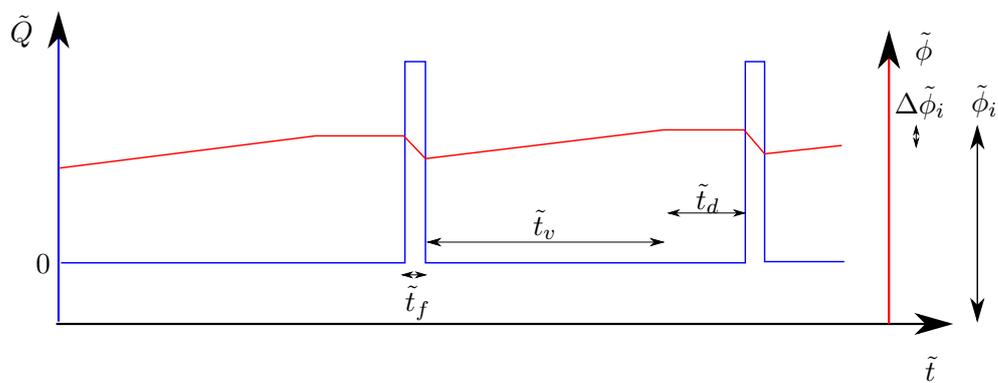
  \caption{Generalization of the constant flow approach to non-constant flows}
  \label{fig:timescales}
\end{figure}
However, in real rivers flow is not constant throughout the year. Typically, large parts of a river's cross-section are only flooded during a limited amount of time per year which allows vegetation to colonize these surfaces during non-flooded periods. Therefore, equation (\ref{eq:veg_const}), except for certain special cases (see Figure \ref{fig:veg_paolo}, where vegetation seems to grow while being completely submerged most of the time), is not really applicable for vegetation growth in natural streams since it considers all processes to happen simultaneously. In reality however, vegetation grows and seeds during the vegetation period (which is part of the non-flooded period) and is uprooted during the flooding period. To simplify our analysis, we assume constant and continuous flow and thus we have to integrate growth and seeding into the flooding period. In the following, we call $\tilde{t}_d$ the drought period without vegetation growth, $\tilde{t}_v$ the vegetation period and $\tilde{t}_f$ the duration of the flooding (see Figure \ref{fig:timescales} for illustration where $\tilde{t}_v$ and $\tilde{t}_d$ have been separated for simplicity). The duration of a complete cycle (for example a year or half a year depending on the specific conditions) is then given by $\tilde{t}_d+\tilde{t}_v+\tilde{t}_f$. If we assume that vegetation density does not vary much during a complete cycle i.e.
\begin{equation}
(\tilde{\phi}_i-\tilde{\phi}_{i-1})<<\tilde{\phi}_i,
\end{equation}
where $\tilde{\phi}_i$ is the value of $\tilde{\phi}$ at the end of cycle i, then we can approximate the difference $\frac{\tilde{\phi}_i-\tilde{\phi}_{i-1}}{\tilde{t}_d+\tilde{t}_v+\tilde{t}_f}$ by the continuous time derivative $\frac{\partial{\tilde{\phi}}}{\partial{\tilde{t}}}$. We write the change of $\tilde{\phi}$ after one cycle as
\begin{equation}
\frac{\tilde{\phi}_i-\tilde{\phi}_{i-1}}{\tilde{t}_d+\tilde{t}_v+\tilde{t}_f}=\left[\tilde{\alpha}_g\tilde{\phi}_i(\tilde{\phi}_m-\tilde{\phi}_i)+\tilde{D}\frac{\partial{^2\tilde{\phi}_i}}{\partial{\tilde{s}^2}}\right]\frac{\tilde{t}_v}{\tilde{t}_d+\tilde{t}_v+\tilde{t}_f}-\tilde{\alpha}_d\tilde{Y}\tilde{U}^2\tilde{\phi}_i\frac{\tilde{t}_f}{\tilde{t}_d+\tilde{t}_v+\tilde{t}_f}.
\end{equation}
By approximating the finite differences by derivatives we get
\begin{equation}
\frac{\partial{\tilde{\phi}}}{\partial{\tilde{t}}}=\left[\tilde{\alpha}_g\tilde{\phi}(\tilde{\phi}_m-\tilde{\phi})+\tilde{D}\frac{\partial{^2\tilde{\phi}}}{\partial{\tilde{s}^2}}\right]\frac{\tilde{t}_v}{\tilde{t}_d+\tilde{t}_v+\tilde{t}_f}-\tilde{\alpha}_d\tilde{Y}\tilde{U}^2\tilde{\phi} \frac{\tilde{t}_f}{\tilde{t}_d+\tilde{t}_v+\tilde{t}_f}
\end{equation}
and since we assumed $\tilde{t}_d$,$\tilde{t}_v$ and $\tilde{t}_f$ to be constant, we can integrate them into the proportionality constants to end up with
\begin{equation}\label{eq:veg_fin}
\frac{\partial{\tilde{\phi}}}{\partial{\tilde{t}}}=\alpha_g\tilde{\phi}(\tilde{\phi}-\phi_m)+D\frac{\partial{^2\tilde{\phi}}}{\partial{\tilde{s}^2}}-\alpha_d\tilde{Y}\tilde{U}^2\tilde{\phi}
\end{equation}
where $\alpha_g=\tilde{\alpha}_g\frac{\tilde{t}_v}{\tilde{t}_d+\tilde{t}_v+\tilde{t}_f}$, $D=\tilde{D}\frac{\tilde{t}_v}{\tilde{t}_d+\tilde{t}_v+\tilde{t}_f}$ and $\alpha_d=\tilde{\alpha}_d\frac{\tilde{t}_f}{\tilde{t}_d+\tilde{t}_v+\tilde{t}_f}$.
\begin{figure}
\centering
\includegraphics[width=120mm]{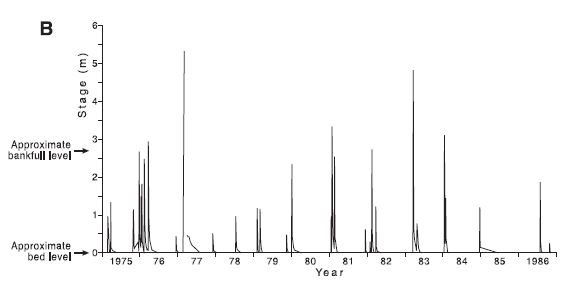}
\caption{Hydrograph of the Marshall River taken from \cite{To}}
\label{fig:hydrograph}
\end{figure}
We can see that merging together the different mechanisms results in a relative increase of the growth and diffusion constant with respect to the uprooting constant if $\tilde{t}_v\gg \tilde{t}_f$. So even if in general the vegetation uprooting coefficient is much higher than the growth coefficient, this can be compensated by the small timescale ratio $\frac{\tilde{t}_f}{\tilde{t}_v}$ to get a regime where mutual feedback is possible. This is the case for example in the Marshall River (see hydrograph in Figure \ref{fig:hydrograph}) and also for bar flooding in the Thur River (see for example \cite{Thur}). Thus, the differential equation (\ref{eq:veg_fin}) may also be valid in the case of non-constant flow if the modeling assumptions are met.\\
We quickly want to discuss two of the most important modeling assumptions adopted above, namely:
\begin{itemize}
\item Vegetation density change during a cycle is small compared to its actual value
\item The only uprooting effect is due to direct flow drag on non-submerged rigid vegetation
\end{itemize}
The first assumption can be assumed to be valid if one considers the case of well developed vegetation. The vegetation coverage is dense enough to not allow much more biomass to be produced and at the same time a large part of the vegetation is robust enough to outlive the flooding period. The second point refers to the fact that we only consider direct flow drag (thus neglecting erosion which exposes the root system). Additionally, we need rigid vegetation like small trees or bushes with mean vegetation height $\tilde{h}_v$ greater than water depth $\tilde{Y}$ in order for our assumption to be valid. For non-rigid vegetation, the exponent of $\tilde{U}$ in the uprooting term (equation \ref{eq:veg3}) should be somewhere between 1 and 2 while in the case of completely submerged vegetation the surface impacted by flow drag would be reduced by a factor of $\frac{\tilde{h}_v}{\tilde{Y}}$ and thus, $\tilde{Y}$ would be replaced by $\tilde{h}_v$ in equation (\ref{eq:veg3}).
\newpage
\section{Stability Analysis of 1D Ecomorphodynamic Equations}\label{sec:1D}
In this section we perform a linear stability analysis of the 1-dimensional ecomorphodynamic equations. The 1D-framework is valid in case flow, bed and vegetation can be assumed to be homogeneous in the direction transverse to the flow. After the derivation of the dimensionless governing equations, linear stability is assessed. We first reproduce some well-known results (\cite{La1} and \cite{Ca}) and then we go on to evaluate the effect that riverbed vegetation dynamics has on these results.
\begin{figure}
  \centering
  \def\svgwidth{400pt}
  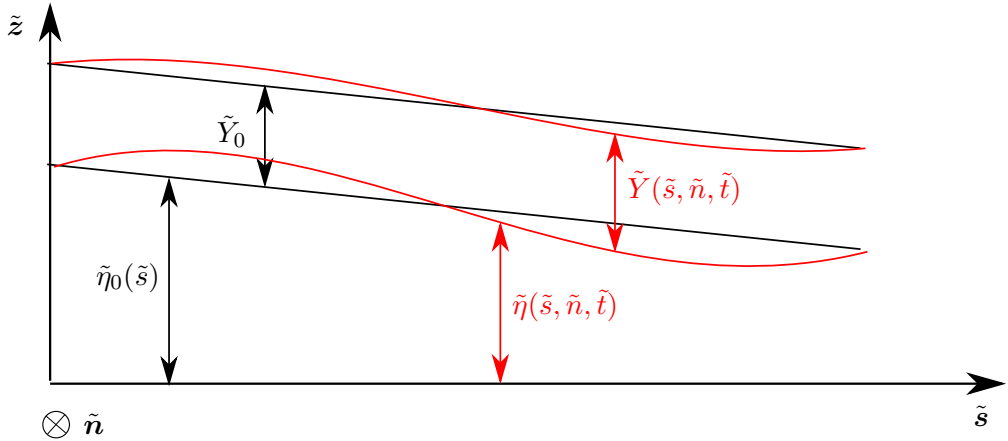
  \caption{Scheme with uniform flow (in black) and perturbated flow (in red)}
  \label{fig:saintvenant}
\end{figure}
\subsection{Governing Equations}\label{sec:1D_gov}
Figure \ref{fig:saintvenant} depicts the model scheme adopted with the streamwise coordinate $\tilde{s}$, the lateral (normal) coordinate $\tilde{n}$ (not used in the 1D-analysis) and the vertical coordinate $\tilde{z}$. The riverbed of constant width is assumed to consist of sandy, non-cohesive material which causes friction and may be transported by the fully turbulent flow. Furthermore, we consider the case of a straight channel (see for example \cite{BS} for curved channels) with non-erodible banks. Additionally, vegetation as described in section \ref{sec:veg} is able to colonize the whole riverbed. Then, assuming hydrostatic pressure distribution and the river width to be considerably larger than the flow depth, flow velocity may be depth-averaged and thus we get the well-known 1-dimensional de Saint-Venant momentum conservation law
\begin{equation}\label{hans11}
\frac{\partial{\tilde{U}}}{\partial{\tilde{t}}}+U\frac{\partial{\tilde{U}}}{\partial{\tilde{s}}}+g\left[\frac{\partial{\tilde{Y}}}{\partial{\tilde{s}}}+\frac{\partial{\tilde{\eta}}}{\partial{\tilde{s}}}\right]+\frac{\tilde{\tau}}{\tilde{Y}}=0.
\end{equation}
where the first and second term are the local and convective acceleration respectively, the third term represents hydrostatic pressure distribution, term 4 is the streamwise slope of the river and term 5 is the bed friction term. Recall that $\tilde{U}$ is the streamwise velocity and $\tilde{Y}$ is the water depth, while $\tilde{\eta}$ is the bed elevation. We have to keep in mind that we are talking about long waves throughout our analysis (pattern wavelength is larger than channel width) in order for the depth-average as well as the 1D-formulation to make sense.
As a closure relationship for bed friction $\tilde{\tau}$, we choose the simple Chezy equation and write
\begin{equation}
\tilde{\tau}_s=\frac{g}{\chi^2}\tilde{U}^2,
\end{equation}
where $\chi$ is the overall Chezy coefficient. The overall Chezy coefficient depends on both, the bed roughness and the roughness induced by vegetation. According to \cite{Baptist}, it can be expressed for non-submerged and rigid vegetation as
\begin{equation}\label{hans12}
\chi=\sqrt{\frac{1}{\frac{1}{\chi_b^2}+\frac{c_D\,d\,\tilde{\phi} \tilde{Y}}{2g}}},
\end{equation}
where $\chi_b=\frac{1}{n}\tilde{Y}^{1/6}$ is the bed roughness which can be calculated by fixing Manning coefficient $n$, $c_D$ is the Stokes drag coefficient and $d$ is the vegetation diameter.\\
Subsequently, flow continuity is formulated as
\begin{equation}\label{hans13}
\frac{\partial{\tilde{Y}}}{\partial{\tilde{t}}}+\frac{\partial{(\tilde{Y}\tilde{U})}}{\partial{\tilde{s}}}=0
\end{equation}
thus neglecting flow diversion by vegetation and assuming that sediment density in the water is low, therefore omitting the sediment term. Note that the flow diversion effect could easily be added but is left out here to keep the analysis simple. In order to account for sediment continuity, we then write the well-known 1D-Exner equation, valid for non-cohesive sediment with uniform grain size as
\begin{equation}\label{hans14}
(1-p)\frac{\partial{\tilde{\eta}}}{\partial{\tilde{t}}}+\frac{\partial{\tilde{Q}_s}}{\partial{\tilde{s}}}=0,
\end{equation}
where $p$ is bed porosity and $\tilde{Q}_s$ is sediment flux per unit width. We assume well-developed sediment transport (always above the critical threshold), mainly in the form of bed load transport and therefore, as was done by \cite{La1}, we adopt $\tilde{Q}_s=a\tilde{U}^3$ where $a$ is a parameter. This is an approximation of the original Meyer-Peter/M$\ddot{\mathrm{u}}$ller formula which states $\tilde{Q}_s=8(\theta-\theta_{cr})^{3/2}$ with $\theta$ the dimensionless shear stress and $\theta_{cr}$ the critical dimensionless shear stress. Omitting the threshold $\theta_{cr}$ (assuming sediment transport to be always above the threshold) and knowing that $\theta$ is proportional to $\tilde{U}^2$ we get back our simplified power law.\\
Finally, we model riverbed vegetation dynamics using
\begin{equation}\label{hans15}
\frac{\partial{\tilde{\phi}}}{\partial{\tilde{t}}}=\alpha_g\tilde{\phi}(\tilde{\phi}_m-\tilde{\phi})+D\frac{\partial{^2\tilde{\phi}}}{\partial{\tilde{s}^2}}-\alpha_d\tilde{Y}\tilde{U}^2\tilde{\phi}
\end{equation}
as explained in section \ref{sec:veg}. Equations (\ref{hans11}) and equation (\ref{hans13}) are conventionally called the de Saint-Venant's (SV) equations. If sediment dynamics is added, we speak of de Saint-Venant-Exner equations (SVE) or morphodynamic equations. Since we added vegetation dynamics to SVE, we name it the de Saint-Venant-Exner-Vegetation equations (SVEV) or the ecomorphodynamic equations.
\subsection{Governing Equations in Dimensionless Variables}\label{sec:1D_dim}
To perform a linear stability analysis, it is convenient to work with dimensionless quantities. Therefore, to write equations (\ref{hans11}) and (\ref{hans13}) to (\ref{hans15}) in dimensionless form, we introduce the change of variables (motivated by the approach of \cite{Ca})
\begin{subequations}
\label{eq:1D_cha}
\begin{align}
U&=\frac{\tilde{U}}{\tilde{U}_0}\\
Y&=\frac{\tilde{Y}}{\tilde{Y}_0}\\
\eta&=\frac{\tilde{\eta}}{\tilde{Y}_0}\\
\phi&=\frac{\tilde{\phi}}{\tilde{\phi}_m}\\
t&=\tilde{t}\frac{\tilde{Y}_0}{\tilde{U}_0}\\
s&=\frac{\tilde{s}}{\tilde{Y}_0},
\end{align}
\end{subequations}
where $\tilde{Y}_0$ is the normal water depth and $\tilde{U}_0$ is the velocity at normal water depth. Using change of variables (\ref{eq:1D_cha}), we obtain (arranged in a way to have the time derivative on the left-hand side)

\begin{subequations}
\label{eq:1D_dim}
\begin{align}
\frac{\partial{U}}{\partial{t}}&=-U\frac{\partial{U}}{\partial{s}}-\frac{1}{F_0^2}\left[\frac{\partial{Y}}{\partial{s}}+\frac{\partial{\eta}}{\partial{s}}\right]-c_b\frac{U^2}{Y}-c_v\phi U^2\\
\frac{\partial{Y}}{\partial{t}}&=-Y\frac{\partial{U}}{\partial{s}}-U\frac{\partial{Y}}{\partial{s}}\\
\frac{\partial{\eta}}{\partial{t}}&=-\gamma U^2\frac{\partial{U}}{\partial{s}}\\
\frac{\partial{\phi}}{\partial{t}}&=\nu_g \phi(1-\phi)+\nu_D\frac{\partial{^2\phi}}{\partial{s^2}}-\nu_d\phi
YU^2,
\end{align}
\end{subequations}
where $F_0=\frac{\tilde{U}_0}{\sqrt{g\tilde{Y}_0}}$, $c_b=\frac{g}{\chi_b^2}$, $c_v=\frac{c_Dd\tilde{\phi}_m \tilde{Y}_0}{2}$, $\gamma=\frac{3\tilde{Q}_{s0}}{(1-p)\tilde{U}_0\tilde{Y}_0}$, $\nu_g=\frac{\alpha_g \tilde{\phi}_m \tilde{Y}_0}{\tilde{U}_0}$, $\nu_{D}=\frac{D}{\tilde{Y}_0 \tilde{U}_0}$ and  $\nu_d=\alpha_d\tilde{Y}_0^2\tilde{U}_0$.

\subsection{Linear Stability Analysis}\label{sec:1D_lin}
A linear stability analysis consists of studying the behavior of a linearized system when slightly perturbated away from a spatially homogeneous solution (see \cite{Turing}). In the case of river morphology, a common choice for a homogeneous solution consists of a river with flat bed and constant slope under constant, uniform flow conditions (see section \ref{sec:veg} for the generalization to non-constant flow). Then, the reaction of the linearized system to small perturbations on every state variable is investigated whose physical meaning may be a variation in sediment supply or channel width for example (\cite{La1}. Regardless of its shape, such a local perturbation can readily be interpreted as a wave packet and thus a velocity perturbation wave packet $U_1$ can be written as a Fourier series with continuous wavenumber k
\begin{equation} \label{eq:wavepacket}
U_1(s,t)=\frac{\epsilon}{2\pi}\int_{-\infty}^\infty u(t,k)\exp(iks)\,dk.
\end{equation}
where $\epsilon$ is the perturbation amplitude and $u$ the velocity perturbation.
In a linear system, each sinusoidal component of the perturbation wave packet can then be treated separately to evaluate if there is growth towards periodic spatial patterns of the riverbed. In the following, we first derive the homogeneous solution of (\ref{eq:1D_dim}) and then linearize and perturbate the equations around the homogeneous solution.

\subsubsection{Homogeneous Solutions}\label{sec:1D_lin_hom}
We begin with looking for spatially homogeneous solutions $\{U_0,Y_0,\eta_0,\phi_0\}$ using normal flow conditions. So, $U_0=1$, $Y_0=1$ and $\eta_0=-J_0\cdot s$ (where $J_0$ is the slope at normal flow conditions). Using the dimensionless governing equations, we can find $J_0$ and $\phi_0$ as
\begin{subequations}\label{eq:1D_hom}
\begin{align}
J_0=&F_0^2\left[c_b+c_v\left(\frac{\nu_g-\nu_d}{\nu_g}\right)\right]\\
\phi_0=&\frac{\nu_g-\nu_d}{\nu_g}.
\end{align}
\end{subequations}
Note that the equations also allow a trivial solution with $\phi_0=0$ which corresponds to a riverbed without vegetation. This solution becomes the only physically relevant solution in case $(\nu_g-\nu_d)<0$. Since the aim of this work is to evaluate the influence of vegetation on river patterns, the solution with $\phi_0=0$ is not interesting. The non-zero dimensionless homogeneous solution can finally be summarized as
\begin{equation}\label{hans33}
\{U_0,Y_0,\eta_0(s),\phi_0\}=\{1,1,-J_0s,\phi_0\}
\end{equation}
with $J_0$ and $\phi_0$ as defined in (\ref{eq:1D_hom}).

\subsubsection{Linearization of Perturbated Equations}\label{sec:1D_lin_lin}
\begin{figure}
  \centering
  \def\svgwidth{300pt}
  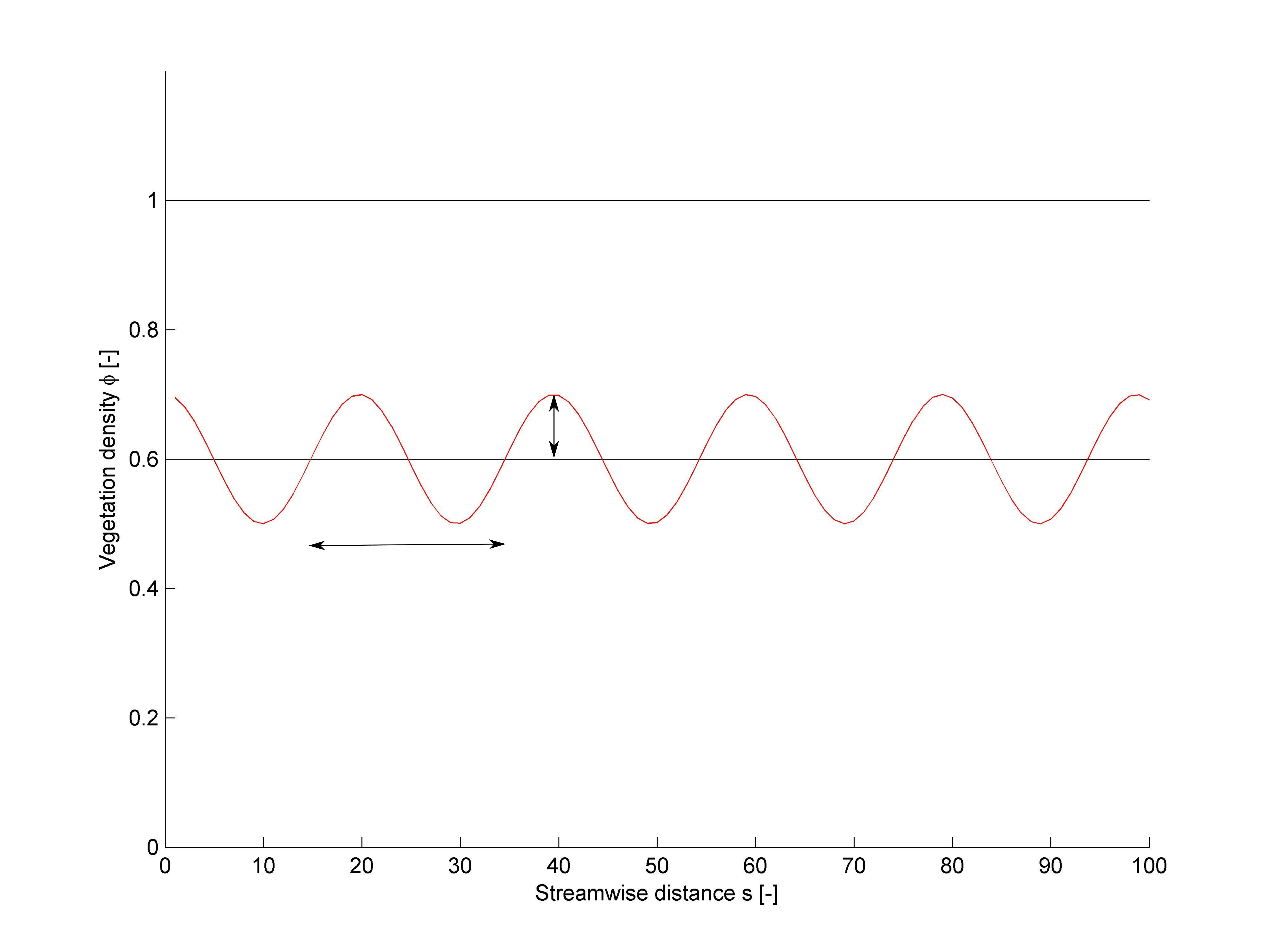
  \caption{Illustration of vegetation waves around the homogeneous solution $\phi_0$}
  \label{fig:phi0}
\end{figure}

The linearization is done by introducing into the dimensionless equations (\ref{eq:1D_dim}) the perturbated homogeneous solution
\begin{align} \label{hans34}
 \{U_0,Y_0,\eta_0,\phi_0\}+\epsilon\{U_1,Y_1,\eta_1,\phi_1\}
\end{align} 
with $\epsilon$ the perturbation parameter and $\{U_1,Y_1,\eta_1,\phi_1\}$ the perturbation Ansatz. As we want to look for regular spatial patterns, we choose the perturbation ansatz as
\begin{equation}\label{eq:1D_ansatz}
\{U_1,Y_1,\eta_1,\phi_1\}=\{u(t),y(t),h(t),f(t)\}\cos(ks)
\end{equation}
with $k$ the real dimensionless perturbation wavenumber and $\{u(t),y(t),h(t),f(t)\}$ the perturbation vector. While we are used to deal with sinusoidal patterns of velocity, water depth and bed elevation this is less common for vegetation density. Figure \ref{fig:phi0} depicts sinusoidal vegetation density patterns around a mean vegetation density of $\phi_0$. We can see that this formulation only is valid if $\phi_0$ is larger than the vegetation perturbation amplitude. In fact, if this is not the case we get locally negative values for vegetation density which does not make sense physically. So we have to bear in mind that vegetation needs to be well-developed in order for our analysis to be valid.\\
The cosine of equation \ref{eq:1D_ansatz} can then be written as
\begin{equation}
\cos(ks)=\frac{\exp(iks)+\exp(-iks)}{2}.
\end{equation}
It can easily be seen that we get a complex conjugated system of equations when inserting the perturbation Ansatz into (\ref{eq:1D_dim}). Thus, one can write the perturbated homogeneous solution as
\begin{equation} \label{eq:1D_perhom}
\{1,1,-J_0s,\phi_0\}+\epsilon\{u(t),y(t),h(t),f(t)\}\exp(iks)+c.c.
\end{equation}
where c.c. denotes the complex conjugate. Note that the perturbation term of (\ref{eq:1D_perhom}) for a given wavenumber $k$ is nothing else than one component of the wave packet introduced in (\ref{eq:wavepacket}).
Then by only keeping the $\mathcal{O}(\epsilon)$ terms we get
\begin{subequations}\label{eq:1D_perteq}
\begin{align}
\frac{du}{dt}&=(-ik-2c_b-2c_v\phi_0)u+\left(\frac{-ik}{F_0^2}+c_b\right)y+\left(\frac{-ik}{F_0^2}\right)h+(-c_v)f\\
\label{hans36}
\frac{dy}{dt}&=(-ik)u+(-ik)y\\
\label{hans37}
\frac{dh}{dt}&=(-i\gamma k)u\\
\label{hans38}
\frac{df}{dt}&=(-2\nu_d\phi_0)u+(-\nu_g\phi_0)y+(-(\nu_g-\nu_d)-\nu_Dk^2)f.
\end{align}
\end{subequations}
The system of equations (\ref{eq:1D_perteq}) can then be written as
\begin{equation}\label{eq:1D_system}
\begin{pmatrix} \frac{du}{dt} \\ \frac{dy}{dt} \\ \frac{dh}{dt} \\ \frac{df}{dt} \end{pmatrix}
=A
\begin{pmatrix} u \\ y \\ h \\ f \end{pmatrix},
\end{equation}
where A is the following 4 x 4 matrix:
\begin{equation}\label{eq:1D_matrix}
A=
\begin{pmatrix} -ik-2c_b-2c_v\phi_0 & \frac{-ik}{F_0^2}+c_b & \frac{-ik}{F_0^2} & -c_v\\ -ik & -ik & 0 & 0 \\ -i\gamma k & 0 & 0 & 0 \\ -2\phi_0\nu_d &  -\phi_0\nu_d & 0 & -\phi_0\nu_g-\nu_Dk^2 \end{pmatrix}.
\end{equation}
Equations (\ref{eq:1D_system}) and operator (\ref{eq:1D_matrix}) define a system of ordinary, homogeneous differential equations with constant coefficients which describes the initial (linear) temporal evolution of the initially perturbated system. To find general solutions of this system, we have to introduce the concept of a normal operator: an operator is normal if $AA^*=A^*A$, where $A^*$ is the complex conjugate transpose of $A$. If $A$ was a normal operator, the matrix' eigenfunctions would form an orthogonal basis and we could write the general solution as
\begin{equation}\label{eq:1D_solution}
\sum_i c_i\exp(\omega_it)
\end{equation}
where i is the rank of the matrix (4 in this case), $c_i$ are coefficients and $\omega_i$ are the complex eigenvalues of $A$. In the limit of large t, this solution is dominated by the exponential with the largest temporal growth rate (maximum of the real parts of $\omega_i$) and thus the solution decays to zero if the maximum growth rate is below zero and it diverges for a positive maximum growth rate. However, in the context of river morphology A is not a normal operator and therefore its eigenfunctions do not form an orthogonal basis. That is, transient growth occurs (\cite{Ca}) and (\ref{eq:1D_solution}) is not generally valid anymore. However, asymptotically the exponential with the largest real part of the eigenvalues is still going to dominate and thus describes the behavior of the system. As in this work we are only interested in the long-term behavior of perturbations, we thus can still state that the initially small perturbations will be amplified in the long-term linear regime if the real part of any $\omega_i$ is positive. And if the largest growth rate occurs for a finite wavenumber $k$, this mode would be amplified stronger than all other modes contained in the wave packet and thus would dominate after some time due to the exponential character of the growth rate. Thus, we can retain the following important points:
\begin{itemize}
\item The system is stable (perturbation is not amplified) with respect to a perturbation mode with wavenumber $k$ if $\mathrm{Max}(\mathrm{Re}(\omega(k)))<0$
\item The system is unstable (perturbation is amplified) with respect to a perturbation mode with wavenumber $k$ if $\mathrm{Max}(\mathrm{Re}(\omega(k)))>0$
\item The system is unstable towards regular spatial patterns if the highest growth rate $\mathrm{Max}(\mathrm{Re}(\omega(k)))$ occurs at finite wavenumber $k$
\end{itemize}
Additionally, the phase velocity of a perturbation can be computed using the imaginary part of the eigenvalues as 
\begin{equation}
v_p(k)=\frac{-\mathrm{Im}(\omega(k))}{k}
\end{equation} which gives information about the propagation of the perturbation: if $v_p>0$ then the perturbation propagates downstream and conversely if $v_p<0$ the perturbation propagates upstream.

\newpage
\subsection{Linear Stability Analysis: Results}\label{sec_1D_Sta}
In this section, the results of the stability analysis of matrix A, which was derived in section \ref{sec:1D_lin}, are presented and interpreted. The eigenvalues are calculated numerically and plotted by Mathematica for different parameter values while the pattern images are computed using Matlab. Additionally, in the simplest case of no vegetation and no sediment transport, the instability condition can be calculated analytically. The aim of the analysis is to find parameter regions where the fastest growing initial perturbation has a finite wavenumber and thus the system can evolve to a regular pattern upon perturbation. In the following two subsections, we first repeat the calculations done by Lanzoni et al., 2006 for the cases of hydrodynamic equations (SV) and hydrodynamic equations coupled with sediment dynamics (SVE). Then in \ref{sec:1D_sta_SVV} and \ref{sec:1D_sta_SVEV}, we analyze the hydrodynamic equations coupled with vegetation dynamics (SVV) and finally the hydrodynamic equations coupled with sediment and vegetation dynamics (SVEV).\\
\subsubsection{SV Equations (fixed bed without vegetation)}\label{sec:1D_sta_SV}
\begin{figure}
\centering
\begin{subfigure}{.5\textwidth}
    \centering
    \includegraphics[width=75mm]{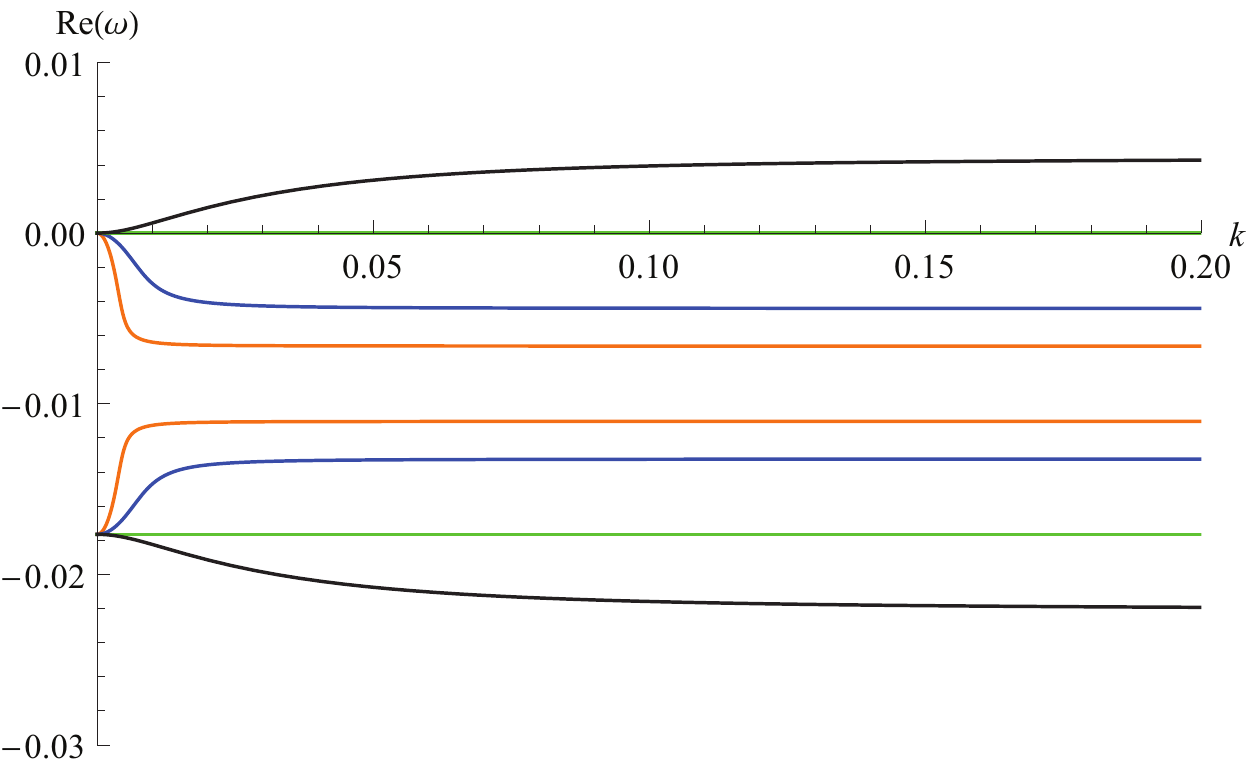} \\
    \caption{Temporal growth rate}
    \label{fig_sv1}
\end{subfigure}%
\begin{subfigure}{.5\textwidth}
    \centering
    \includegraphics[width=75mm]{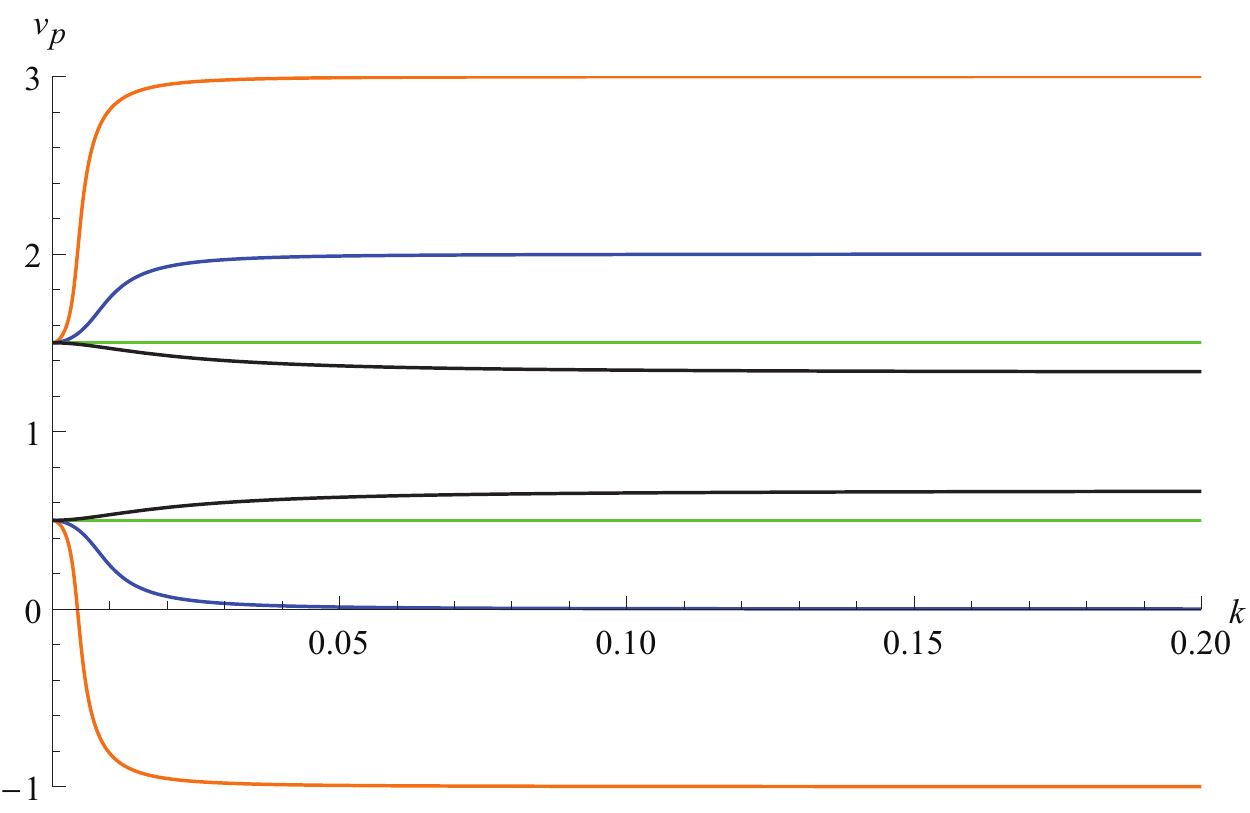} \\
    \caption{Phase velocity}
    \label{fig_sv2}
\end{subfigure}
\caption{Temporal growth rate $\mathrm{Re}(\omega)$ and phase velocity $v_p$ of SV equations as a function of wavenumber $k$ for $F_0=0.5$ (orange), $F_0=1$ (blue), $F_0=2$ (green), $F_0=3$ (black); parameter values are $\tilde{Y}_0=1\,\mathrm{m}$ and $n=0.03$}
\label{fig_sv}
\end{figure}
The stability analysis of the de Saint-Venant equations only consists of analyzing a 2 x 2 matrix (taking the upper left part of matrix A with $\phi_m=0$) which gives the following characteristic equation for the eigenvalues $\omega$:
\begin{equation}\label{hans41}
\omega^2+(2c_b+2ik)\omega+k^2\left(\frac{1-F_0^2}{F_0^2}\right)+3ic_bk=0.
\end{equation}
Solving this equation for $Re(\omega)=0$, we get the condition $F_0=2$, independently from $c_b$ and $k$. Figure \ref{fig_sv1} shows the temporal growth rate for different values of $F_0$. For $F_0=2$ the growth rate is zero. If $F_0=3$ however, the growth rate increases asymptotically with increasing $k$ while for $F_0<2$ it is always negative.This means that perturbations are amplified only if $F_0>2$ and that the most unstable mode is the one where the wavenumber $k$ tends to infinity and so the wavelength is equal to zero. According to \cite{La1} this instability is linked to the formation of roll waves. But, in the linear regime no instability towards regular patterns with finite wavelength is possible in a river with fixed bed and without riverbed vegetation.\\
Additionally, Figure \ref{fig_sv2} shows that perturbations propagate downstream only if the flow is subcritical and in both directions if flow is supercritical.

\subsubsection{SVE Equations (movable bed without vegetation)}\label{sec:1D_sta_SVE}
\begin{figure}
\centering
\begin{subfigure}{.5\textwidth}
   \centering
    \includegraphics[width=75mm]{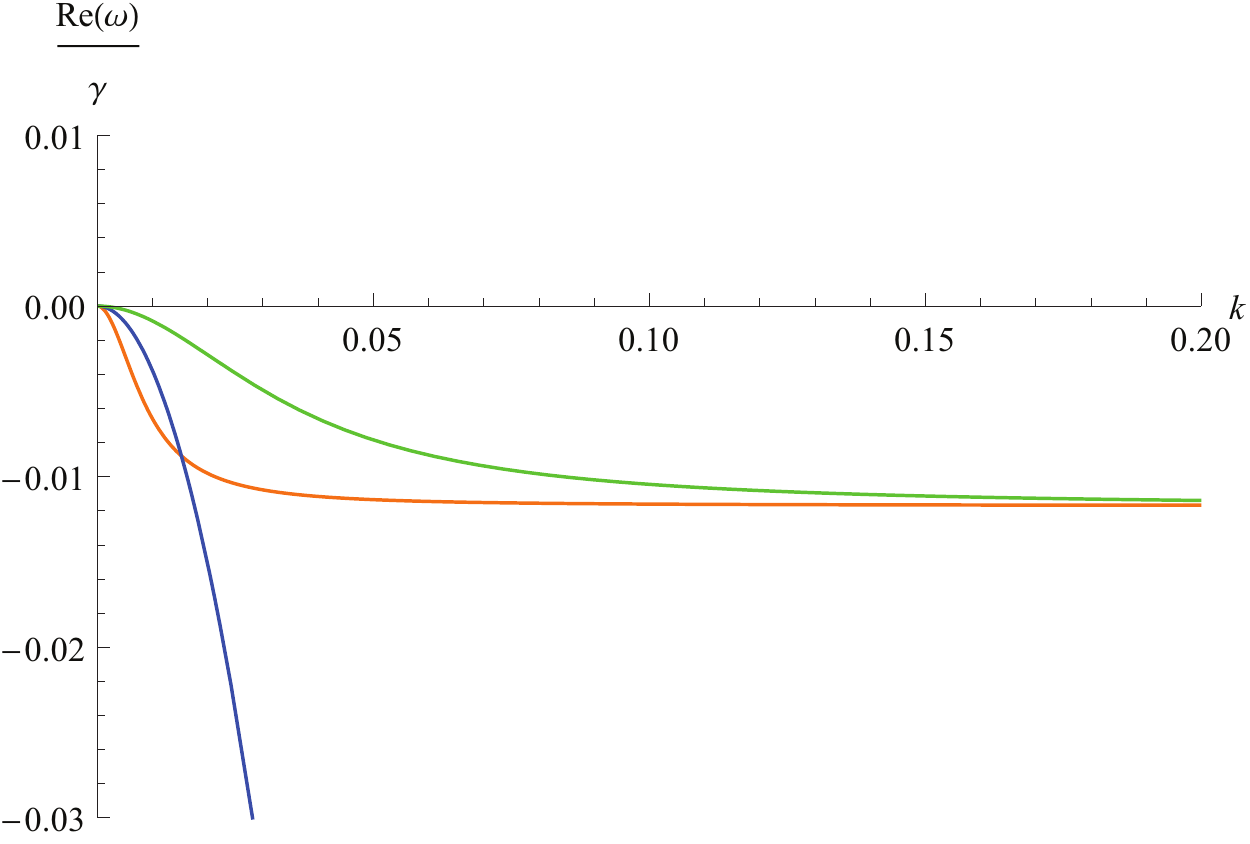}
    \caption{Temporal growth rate}
    \label{fig_sve2}
\end{subfigure}%
\begin{subfigure}{.5\textwidth}
    \centering
    \includegraphics[width=75mm]{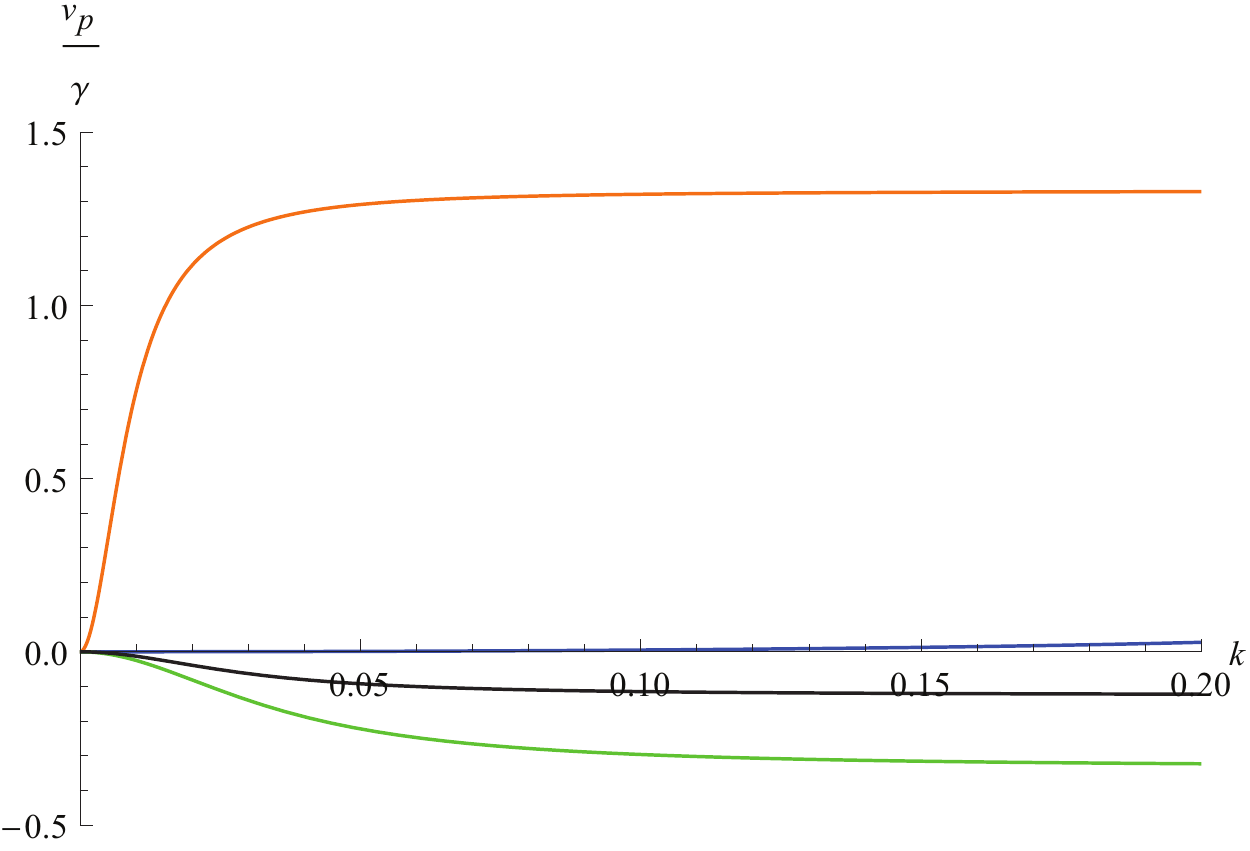}
    \caption{Phase velocity}
    \label{fig_sve4}
\end{subfigure}
\caption{Temporal growth rate $\mathrm{Re}(\omega)$ and phase velocity $v_p$ of the morphodynamic mode of SVE equations as a function of wavenumber $k$ for $F_0=0.5$ (orange), $F_0=1$ (blue), $F_0=2$ (green), $F_0=3$ (black); parameter values are $\tilde{Y}_0=1\,\mathrm{m}$, $n=0.03$ and $\gamma=10^{-3}$}
\label{fig_sve}
\end{figure}

If sediment dynamics is added to the de Saint-Venant equations (which means that the bed material may be transported by the flow), the eigenvalues of a 3 x 3 matrix have to be computed (upper left part of A). It turns out that if the morphodynamic timescale is small compared to the hydrodynamic timescale which is normally the case ($\gamma\sim\mathcal{O}(10^{-3}-10^{-4})$, see \cite{Parker} for some values), the first two (hydrodynamic) modes are essentially the same as in the previous paragraph. There is however a third mode (called morphodynamic mode) that appears because of sediment dynamics. The temporal growth rate and the phase velocity are depicted in Figure \ref{fig_sve} for the morphodynamic mode only (scaled by $\gamma$).\\
The growth rate of the morphodynamic mode is below zero (equal to zero at k=0) for all values of $F_0$ and $k$ which means that the morphodynamic mode is (as were the hydrodynamic modes) not able to produce instability towards finite patterns. Finally, as we can see from Figure \ref{fig_sve4} the migration of the perturbations is downstream if the flow is subcritical and upstream if the flow is supercritical. All results of paragraphs \ref{sec:1D_sta_SV} and \ref{sec:1D_sta_SVE} are in agreement with the findings of \cite{La1} which confirms the correctness of the stability analysis performed.

\subsubsection{SVV Equations (fixed bed with vegetation}\label{sec:1D_sta_SVV}
\begin{figure}
\begin{center}
    \includegraphics[width=130mm]{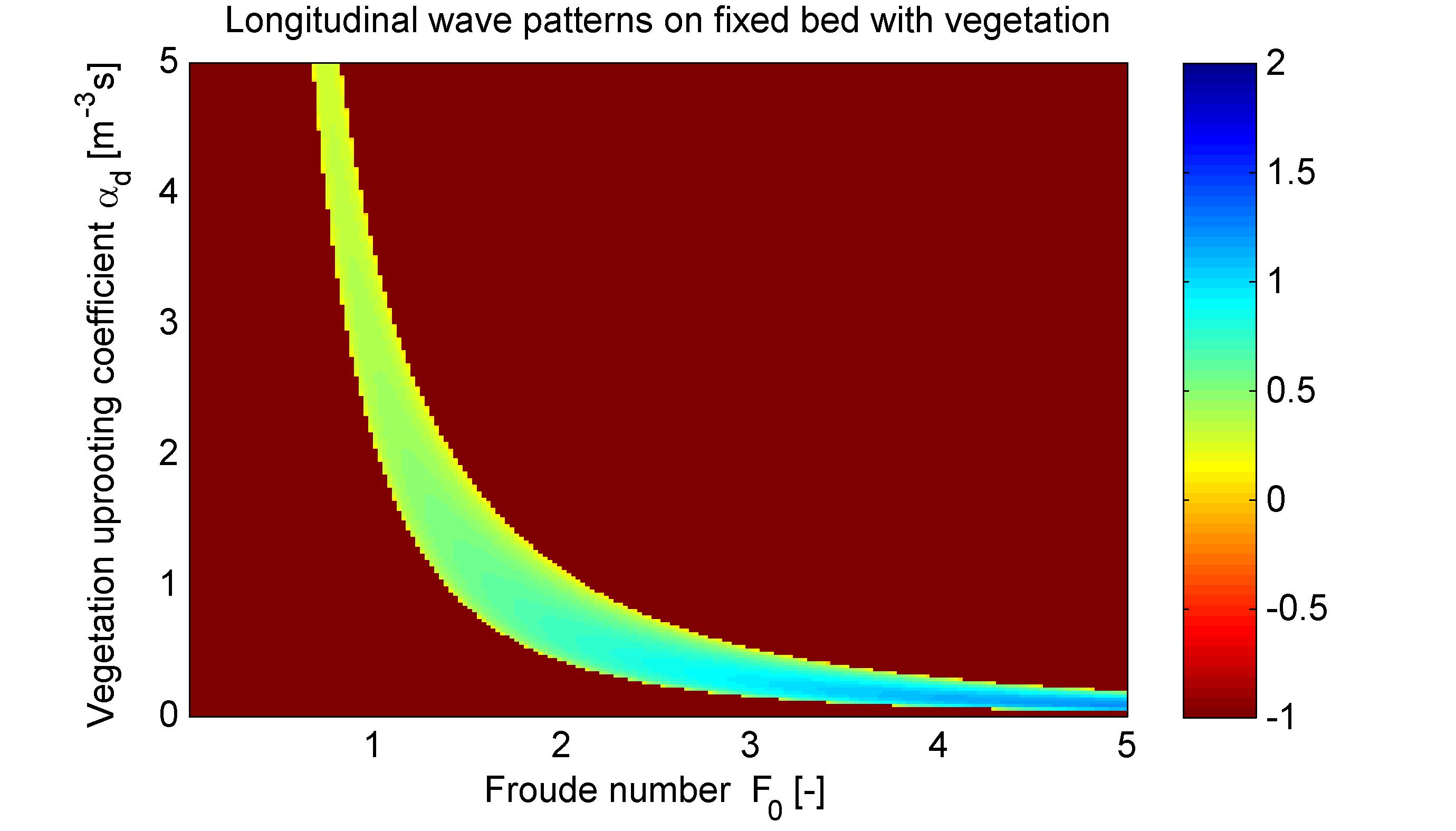} \\[1\baselineskip]
    \caption{Pattern domain as a function of $F_0$ and $\alpha_d$: the value of the wavenumber with maximum growth rate is indicated by the color code, -1 means that there are no patterns or patterns with non-physical meaning (negative $\phi_0$); parameter values are $\tilde{\phi}_m=50\,\mathrm{m^{-2}}$, $\alpha_g=1\,\mathrm{m^2s{-1}}$, $D=0\,\mathrm{m^2s{-1}}$ and Table \ref{table_param}}
    \label{fig_pattern_aD}
\end{center}
\end{figure}

\begin{figure}
\begin{center}
    \includegraphics[width=130mm]{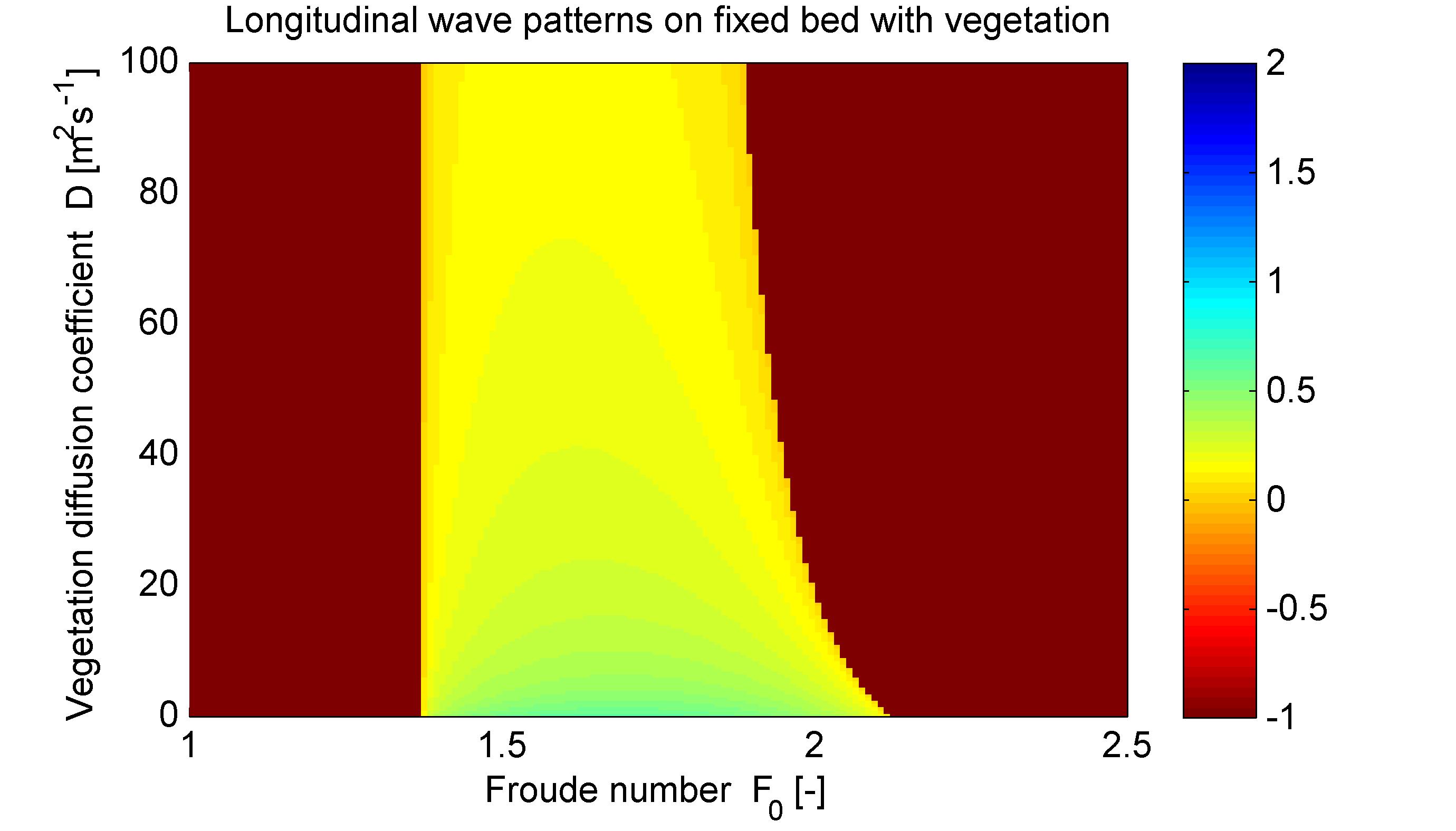} \\[1\baselineskip]
    \caption{Pattern domain as a function of $F_0$ and $D$: the value of the wavenumber with maximum growth rate is indicated by the color code, -1 means that there are no patterns or patterns with non-physical meaning (negative $\phi_0$); parameter values are $\tilde{\phi}_m=50\,\mathrm{m^{-2}}$, $\alpha_g=1\,\mathrm{m^2s{-1}}$, $\alpha_d=1\,\mathrm{m^{-3}s}$ and Table \ref{table_param}}
    \label{fig_pattern_D}
\end{center}
\end{figure}

The effect of taking sediment dynamics into account was shown in the previous section. Now, we analyze the de Saint-Venant equations coupled with vegetation dynamics (but with a fixed bed geometry). Again, no analytical solution is available and thus we analyze
\begin{equation}\label{hans42}
A_{svv}=
\begin{pmatrix} -ik-2c_b-2c_v\phi_0 & \frac{-ik}{F_0^2}+c_b & -c_v\\ -ik & -ik & 0 \\ -2\phi_0\nu_d &  -\phi_0\nu_d  & -\phi_0\nu_g-\nu_Dk^2 \end{pmatrix}
\end{equation}
numerically for different parameters.\\
\begin{table}
\caption{Values for constant parameters of the analysis}
\centering
\begin{tabular}{c c c c}
\hline\hline
Parameter name & Variable & Value & Units\\[0.5ex]
\hline
Normal water depth & $\tilde{Y}_0$ & 1 & m \\
Stokes drag coefficient & $c_D$ & 1.5 & - \\
Vegetation diameter & $d$ & 0.01 & m \\
Manning coefficient & $n$ & 0.03 & $\mathrm{m^{-1/3}s}$ \\
\hline
\end{tabular}
\label{table_param}
\end{table}
Aside from wavenumber $k$, this matrix contains 6 dimensionless parameters: $F_0$ which characterizes the flow, the bed friction coefficient  $c_b$, the vegetation friction coefficient $c_v$ and the vegetation dynamics parameters $\nu_g$, $\nu_{D}$ and $\nu_d$ which determine the influence of growth, diffusion and uprooting by flow on the vegetation respectively. These dimensionless parameters are composed of physical parameters of which some have been assigned typical values (see Table \ref{table_param}), while the physical vegetation coefficients $\{\tilde{\phi}_m, \alpha_d, \alpha_d, D\}$  have been varied for different Froude numbers. The parameters which were given typical values do not give any additional insight into the problem when varied and therefore are held constant for the entire analysis.\\[1\baselineskip]
We could identify the set of physical parameters
\begin{equation}\label{hans43}
\{\tilde{\phi}_m,\alpha_g,\alpha_d,D\}=\{50,1,1,0\}
\end{equation}
to give local (finite) maxima in the $k$-$Re(\omega)$ plot for a certain range of Froude numbers.
\begin{figure}
\begin{center}
    \includegraphics[width=150mm]{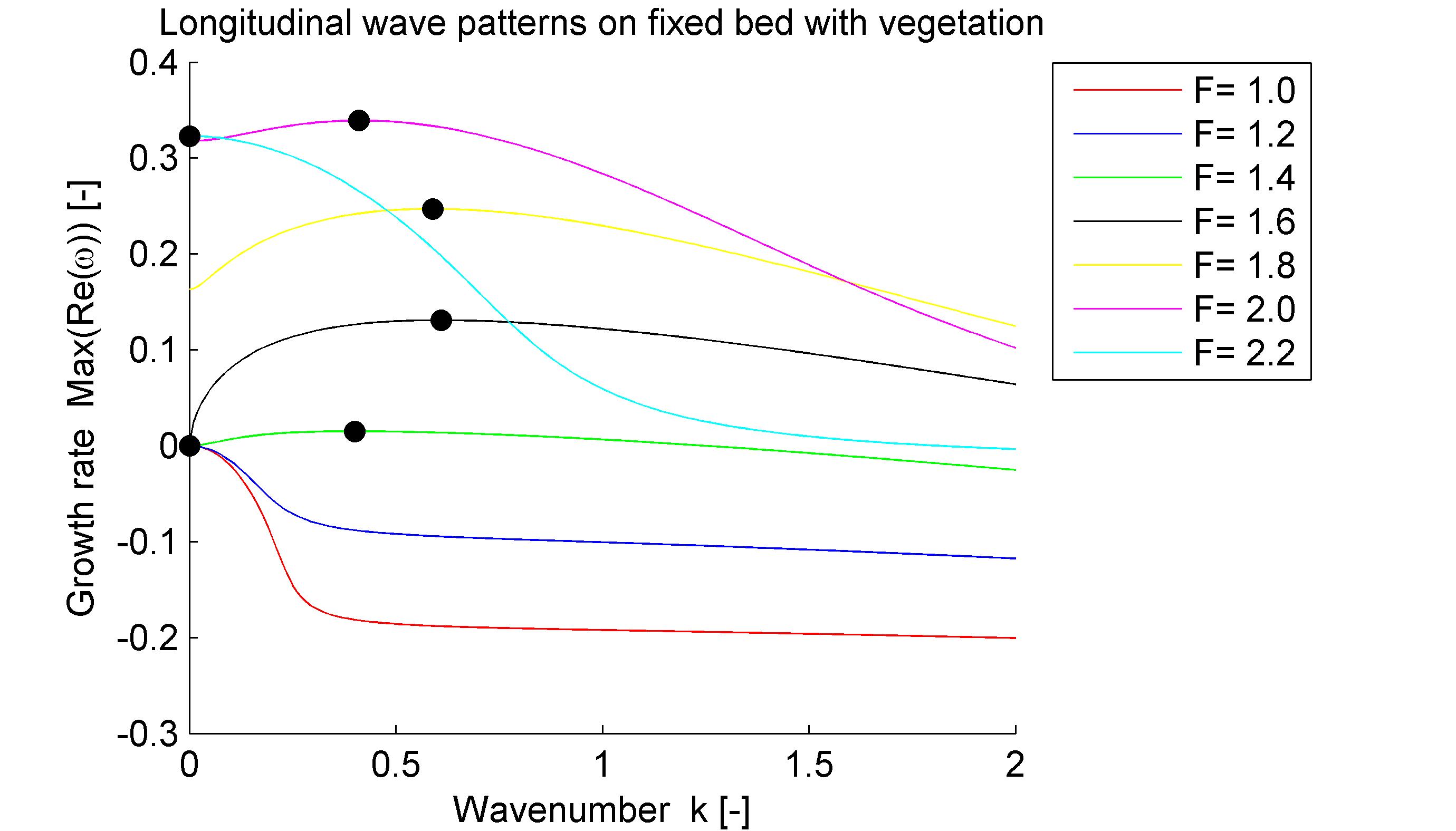}
    \caption{Maximum temporal growth rate (maximum eigenvalue) of SVV equations as a function of wavenumber $k$ for different Froude numbers, the black dots mark the maximum of each curve; parameter values are $\tilde{\phi}_m=50\,\mathrm{m^{-2}}$, $\alpha_g=1\,\mathrm{m^2s^{-1}}$, $\alpha_d=1\,\mathrm{m^{-3}s}$, $D=0\,\mathrm{m^2s^{-1}}$ and Table \ref{table_param}}
    \label{fig_svv_F}
\end{center}
\end{figure}
\begin{figure}
\centering
\begin{subfigure}{.5\textwidth}
\centering
    \includegraphics[width=75mm]{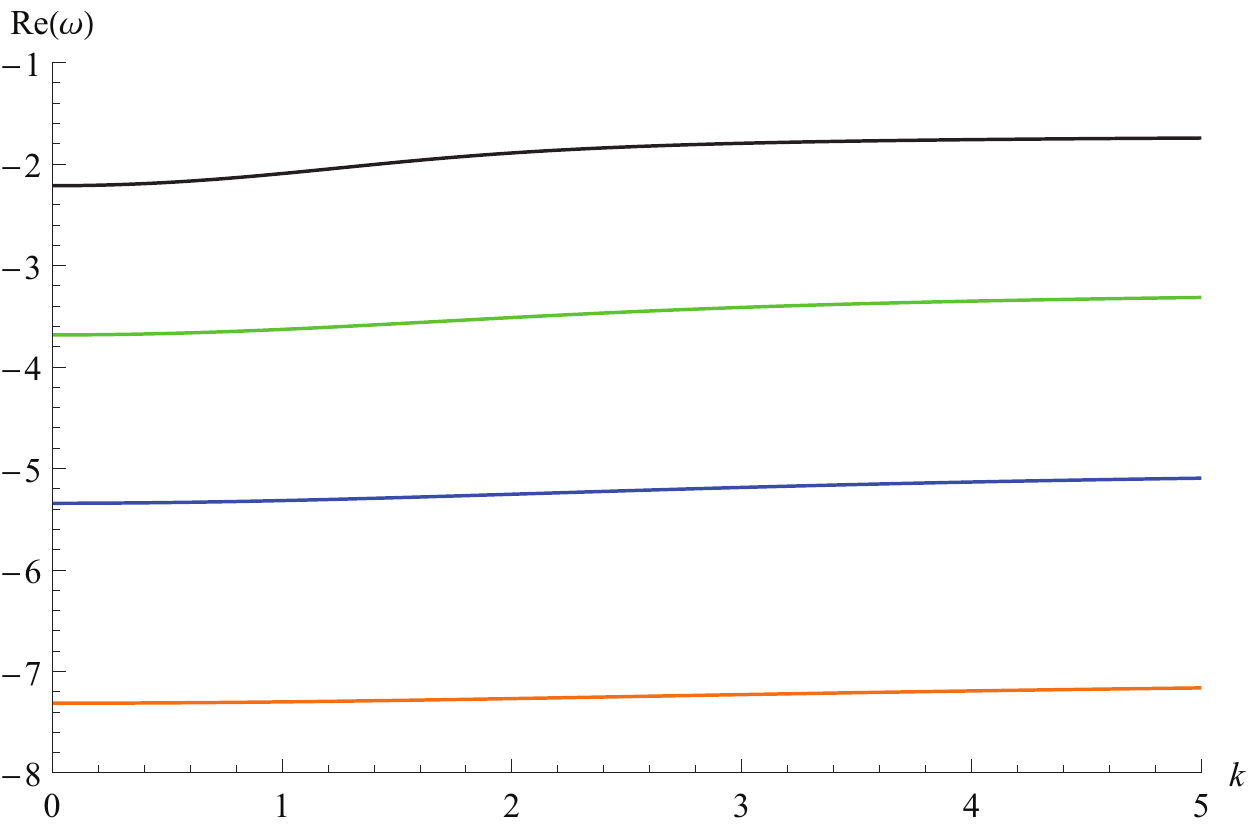}
    \caption{Temporal growth rate}
    \label{fig_svv1}
\end{subfigure}%
\begin{subfigure}{.5\textwidth}
\centering
    \includegraphics[width=75mm]{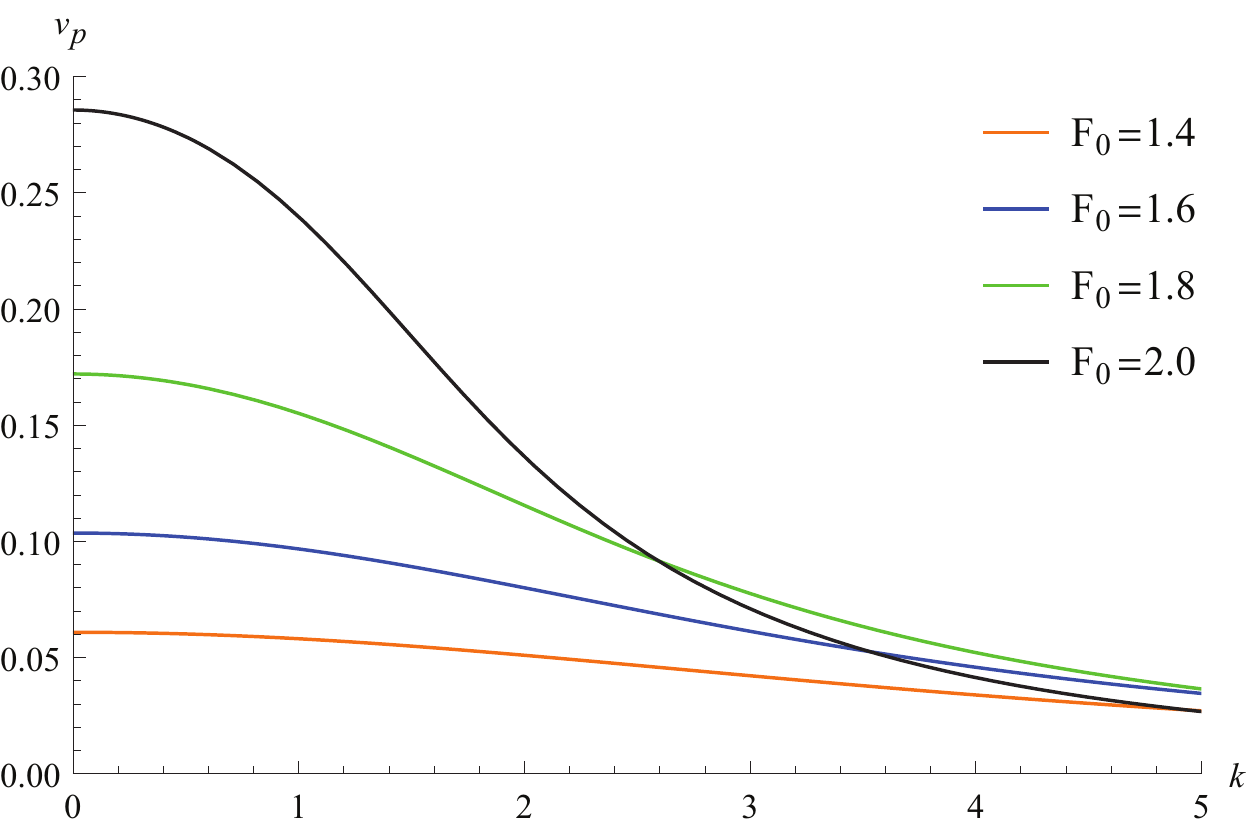}
    \caption{Phase velocity}
    \label{fig_svv3}
\end{subfigure}
\caption{Temporal growth rate $\mathrm{Re}(\omega)$ and phase velocity $v_p$ of the vegetation mode of SVV equations as a function of wavenumber $k$ for different Froude numbers; parameter values are $\tilde{\phi}_m=50\,\mathrm{m^{-2}}$, $\alpha_g=1\,\mathrm{m^2s^{-1}}$, $\alpha_d=1\,\mathrm{m^{-3}s}$, $D=0\,\mathrm{m^2s^{-1}}$ and Table \ref{table_param}}
\label{fig_svv}
\end{figure}
Figure \ref{fig_svv_F} then shows $\mathrm{Max(Re(\omega))}$ as a function of $k$ for different Froude numbers and with the parameters as described above. Note that only the largest eigenvalue is plotted which corresponds to one of the hydrodynamic modes. Comparing these results to the one in Figure \ref{fig_sv1}, one can see that the growth rate is strongly influenced by vegetation. Meanwhile, the growth rate of the vegetation mode itself is strongly negative as is depicted in Figure \ref{fig_svv1} and it reaches positive values only for non-physical parameter configurations. Additionally, the phase velocity of the vegetation mode (Figure \ref{fig_svv3}) can be seen to be positive for the whole range of Froude numbers from Figure \ref{fig_svv_F}. \\
It is clear from Figure \ref{fig_svv_F} that in order to have the maximum growth rate not at $k=0$, some minimum value is required for the Froude number for a given set of parameters. Furthermore, for increasing Froude numbers the wavenumber with maximum growth rate first increases to reach some maximum value between $F_0=1.6$ and $F_0=1.8$ and then decreases to reach zero again. This clearly indicates that, for this parameter configuration, there exists some domain of Froude numbers for which the most unstable perturbation wavenumber is not equal to zero, meaning that a pattern with a finite wavelength should be established upon perturbation. This part of the parameter space can be understood as the domain where plant growth and uprooting are in competition to give rise to persisting vegetation patterns. A lower Froude number leads to overwhelming plant growth (thus not allowing regular vegetation patterns), while a too high Froude number increases uprooting to the point where no vegetation can survive the flow drag anymore.\\
Having found a domain in the parameter space which shows instability towards pattern, we now discuss the shape of this domain by varying $\tilde{\phi}_m$, $\alpha_g$, $\alpha_d$ and $D$ for different Froude numbers (keeping the other parameters as indicated in Table \ref{table_param}). The conditions for a pattern region are the following:
\begin{itemize}
\item The wavenumber $k$ with maximum instability (highest $\mathrm{Max(Re(\omega))}$) is finite
\item In order to have solutions with physical meaning, $\phi_0$ has to be positive which gives the condition $(\nu_g-\nu_d)>0$
\end{itemize}
\begin{figure}
  \centering
  \def\svgwidth{400pt}
  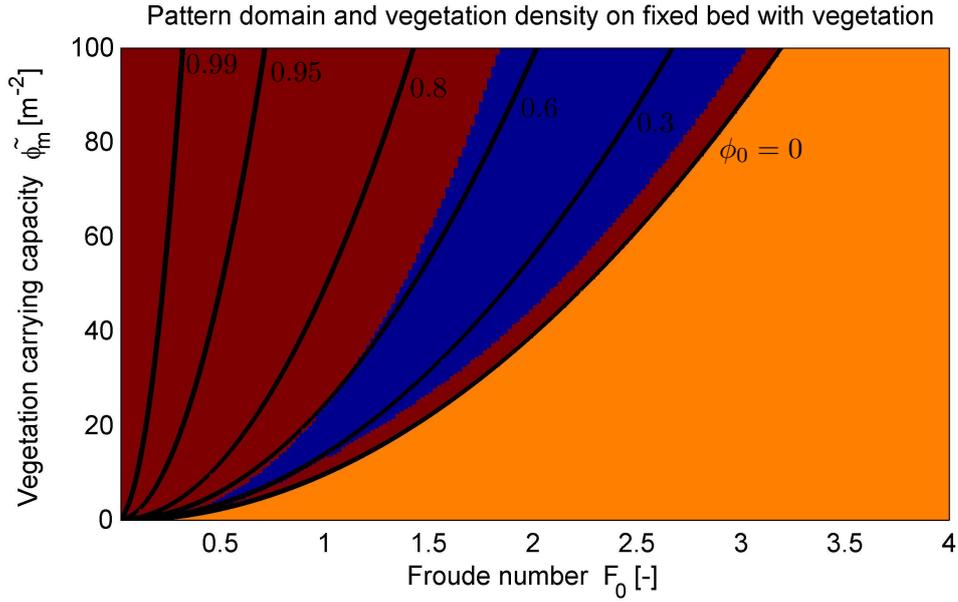
  \caption{Pattern domain as a function of $F_0$ and $\tilde{\phi}_m$ and contour lines of homogeneous vegetation density $\phi_0$ in black: the domain with patterns is depicted in blue, the domain without patterns in red and the domain where $\phi_0$ is negative in orange; parameter values are $\alpha_g=1\,\mathrm{m^2s^{-1}}$, $\alpha_d=1\,\mathrm{m^{-3}s}$, $D=0\,\mathrm{m^2s^{-1}}$ and Table \ref{table_param}}
  \label{fig_pattern_old}
\end{figure}
\begin{figure}
\begin{center}
    \includegraphics[width=130mm]{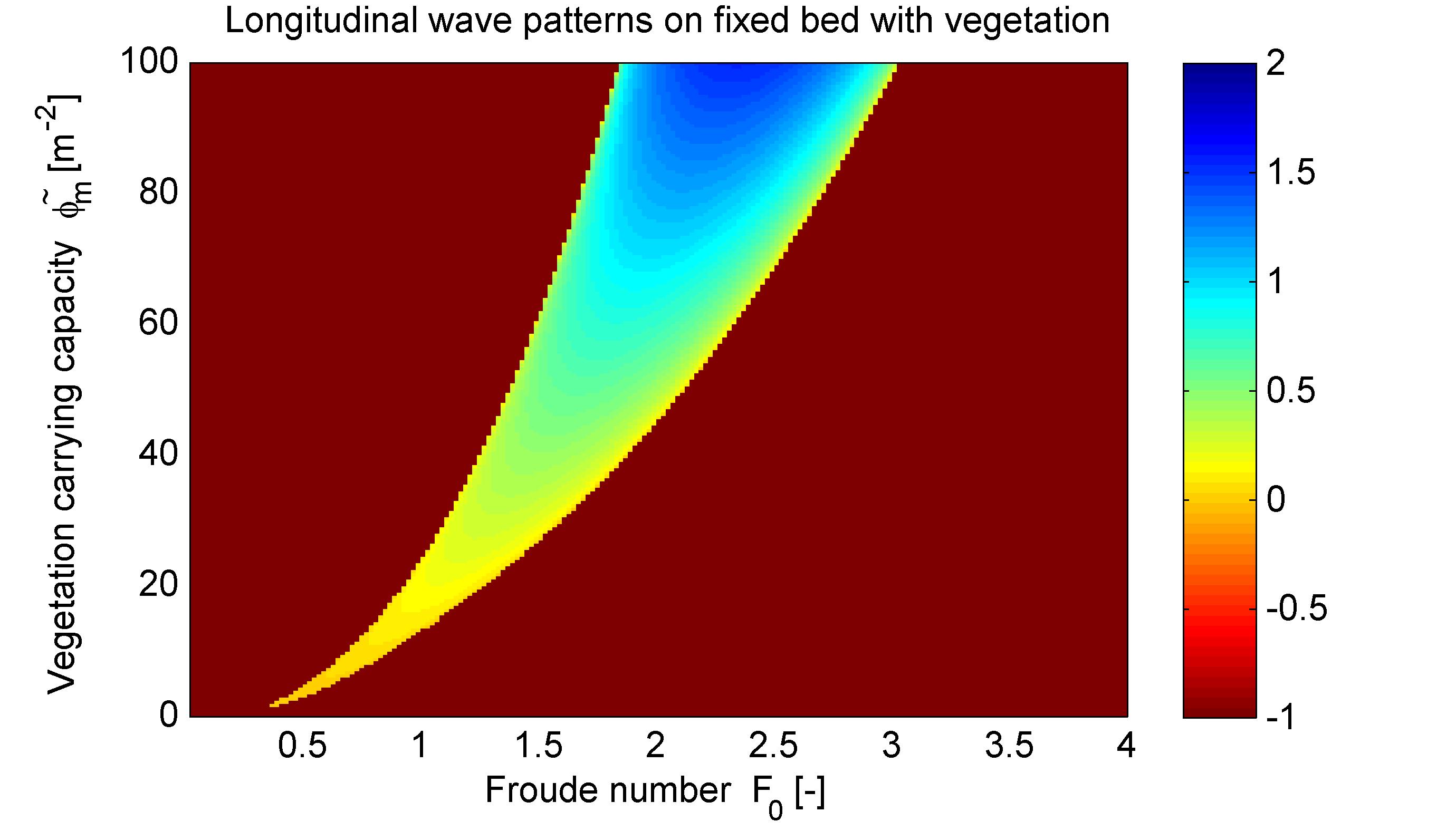} \\[1\baselineskip]
    \caption{Pattern domain as a function of $F_0$ and $\tilde{\phi}_m$: the value of the wavenumber with maximum growth rate is indicated by the color code, -1 means that there are no patterns or patterns with non-physical meaning (negative $\phi_0$); parameter values are $\alpha_g=1\,\mathrm{m^2s^{-1}}$, $\alpha_d=1\,\mathrm{m^{-3}s}$, $D=0\,\mathrm{m^2s{-1}}$ and Table \ref{table_param}}
    \label{fig_pattern_phi}
\end{center}
\end{figure}
\begin{figure}
\begin{center}
    \includegraphics[width=130mm]{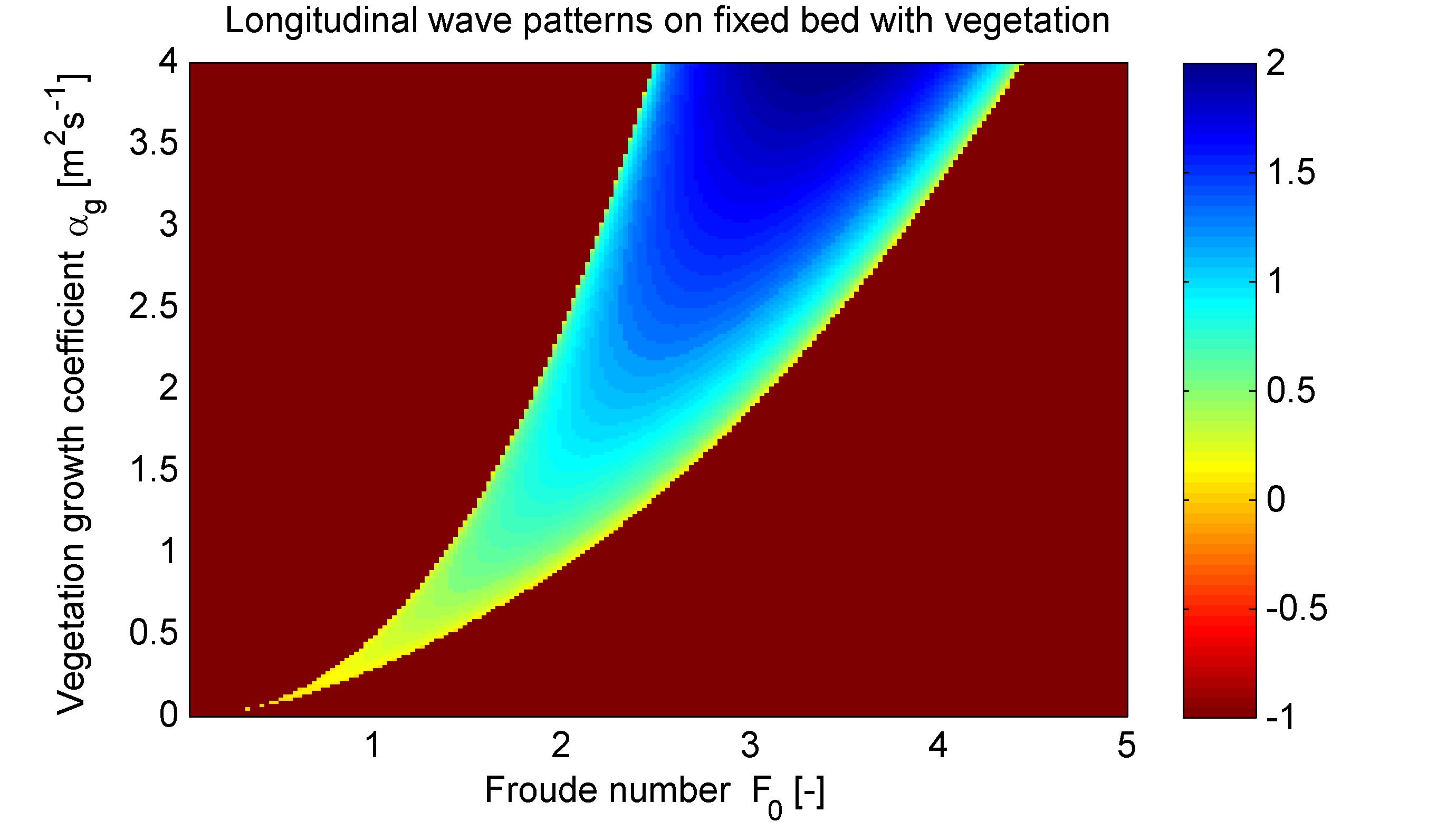} \\[1\baselineskip]
    \caption{Pattern domain as a function of $F_0$ and $\alpha_g$: the value of the wavenumber with maximum growth rate is indicated by the color code, -1 means that there are no patterns or patterns with non-physical meaning (negative $\phi_0$); parameter values are $\tilde{\phi}_m=50\,\mathrm{m^{-2}}$, $\alpha_d=1\,\mathrm{m^{-3}s}$, $D=0\,\mathrm{m^2s{-1}}$ and Table \ref{table_param}}
    \label{fig_pattern_aG}
\end{center}
\end{figure}
Using Matlab to implement the pattern search algorithm based on the two conditions named above, Figure \ref{fig_pattern_old} is produced for varying $\tilde{\phi}_m$. It confirms that there exists a domain with pattern in the parameter space while part of the parameter space does not allow patterns and another part has no real physical meaning ($\phi_0<0$, therefore the trivial solution $\phi_0=0$ being the only physical solution). While not visible due to finite resolution of the algorithm, further research confirmed that the parameter domain is continuous down to the origin. Also, the domain continues to open up as we increase $\tilde{\phi}_m$ and $F_0$ beyond what is shown in Figure \ref{fig_pattern_old}. Furthermore, the contour lines of $\phi_0$ indicate that patterns occur in the middle range (0.2-0.7) of homogeneous vegetation density $\phi_0$.\\
Figures \ref{fig_pattern_phi}, \ref{fig_pattern_aG} and \ref{fig_pattern_aD} then show the pattern domains in the parameter space for the vegetation parameters  $\{\tilde{\phi}_m, \alpha_g, \alpha_d\}$ as a function of Froude number. The value found for the wavenumber with maximum growth rate is shown by the color code. Quantitatively, the influence of both $\tilde{\phi}_m$ and $\alpha_g$ is similar: If the parameter is increased, more vegetation develops and therefore a higher uprooting capacity (higher Froude number/velocity) is needed to keep the balance. It is the other way around for the parameter $\alpha_d$ whose influence on patterns is depicted in Figure \ref{fig_pattern_aD}: patterns can only exist at high Froude numbers if the uprooting coefficient is low. Furthermore, we can see that higher wavenumbers are reached as the domain continues upwards in Figures \ref{fig_pattern_phi} and \ref{fig_pattern_aG} which corresponds to shorter wavelengths. The same is true for the Froude number midway through the domain for fixed $\tilde{\phi}_m$ or $\alpha_g$: the wavenumber exhibits a local maximum. So generally, higher parameter numbers (more dynamic vegetation equation) and competitiveness between parameters lead to higher wavenumbers (lower wavelengths) and thus to more accentuated patterns. The opposite occurs for the uprooting coefficient (Figure \ref{fig_pattern_aD}) where lower uprooting coefficients and higher Froude numbers lead to more dynamic patterns.
For all Figures, further research has confirmed that the shape of the parameter domains illustrated for a limited parameter range is representative for the general behavior of the parameter domains. \\[1\baselineskip]
Finally, Figure \ref{fig_pattern_D}, which shows the pattern domain for varying vegetation diffusion coefficient $D$, is somewhat different from the aforementioned. In fact, this parameter does only change one boundary of the pattern profile while it leaves unchanged the other one. Further research confirmed that the right boundary tends towards a constant Froude number and so the pattern domain continues as $D$ goes to infinity. It can also be seen that increasing the diffusion constant makes the pattern wavelength increase as well. This is because diffusion increases the local positive feedback of vegetation and thus leads to less variation which results in larger wavelengths. In this sense, vegetation diffusion does not seem to be necessary for pattern development but it can make them disappear if diffusion dominates growth and uprooting (patterns disappear if the wavelength goes to infinity). Moreover, the domain becomes smaller for increasing $D$ but stabilizes again to asymptotically reach a constant Froude number.\\
The analysis of a river on a fixed bed but with riverbed vegetation showed that instability towards finite patterns is possible if vegetation growth and mortality (through uprooting) are in competition. Therefore, riverbed vegetation seems to be a key ingredient in the process of longitudinal river pattern formation. Note however, that a vegetated river with fixed bed is not very likely to be found in nature (an exception can be seen in Figure \ref{fig:veg_paolo}). The patterns found would be patterns in terms of vegetation but on a flat bed. Thus, this conclusion cannot be transposed to field situations yet and we should include sediment dynamics in most of the cases which is done in the next section.
\subsubsection{SVEV Equations (movable bed with vegetation)}\label{sec:1D_sta_SVEV}
\begin{figure}
\begin{center}
    \includegraphics[width=130mm]{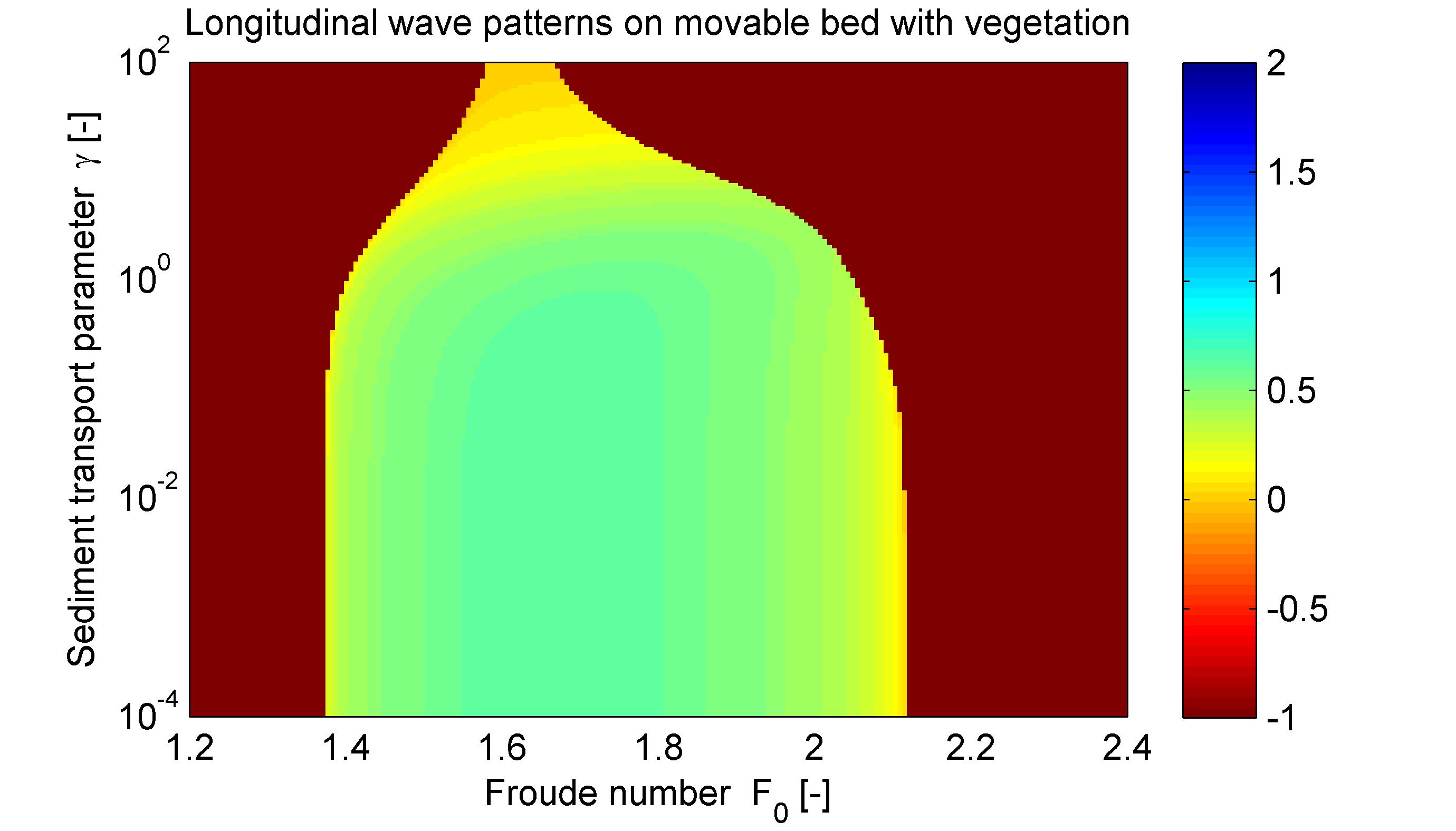} \\
    \caption{Pattern domain as a function of $F_0$ and $\gamma$: The value of the maximum wavenumber is indicated by the color code, -1 means that there are no pattern or pattern without physical meaning; parameters used: $\tilde{\phi}_m=50\,\mathrm{1/m^2}$, $\alpha_g=1\,\mathrm{m^2/s}$, $\alpha_d=1\,\mathrm{s/m^3}$, $D=0\,\mathrm{m^2/s}$ and Table \ref{table_param}}
    \label{fig:patternSVEV}
\end{center}
\end{figure}

Having analyzed the SVE equations where no instability towards patterns exists at the linear level and the SVV equations where patterns could be found depending on parameter values, we now want to combine both sediment dynamics and vegetation growth with the hydrodynamic equations and see whether the pattern regions found before disappear, remain or change. Physically, this consists of analyzing the realistic case of a river with movable bed and vegetation growth.
Figure \ref{fig:patternSVEV} illustrates how the pattern domain found in the previous section changes as a function of the sediment parameter $\gamma$ and Froude number. It is clear that in order to have $\gamma$ influence the domain boundaries and the wavenumber with maximum instability, one needs physically impossibly high values for $\gamma$ (ratio of the sediment timescale to the flow timescale). Recall that typical values for this timescale ratio are $10^{-4}$ to $10^{-3}$. In Figure \ref{fig:patternSVEV}, we can see that major changes occur when $\gamma$ is on the order of 1 (timescales of flow and sediment are comparable) which is not realistic to occur in nature.\\
In conclusion, the proposed vegetation equation is able to produce instability towards longitudinal patterns. Sediment dynamics however does not seem to be crucial for this process in case of a reasonable vegetation density since the roughness induced by vegetation is much higher than the bed roughness. Quantitatively, the most unstable dimensionless wavenumber observed for the basic parameter set (\ref{hans43}) and $F_0=2$ is $k=0.41$. Converting this to a dimensional (physical) wavelength we get: \\[1\baselineskip]
\begin{equation}\label{hans44}
\tilde{\lambda}=\frac{2\pi}{k/\tilde{Y_0}}.
\end{equation}
Equation (\ref{hans44}) yields $\tilde{\lambda}=15.3\,\mathrm{m}$ which seems to be the right order of magnitude for a longitudinal pattern wavelength in a natural river. Nevertheless, a river width of less than 15 meters would be necessary in order for the 1D approximation to be valid in this case.

\subsection{Synthesis and Conclusion of the 1D analysis}
It was found in section \ref{sec_1D_Sta} that, at the level of linear stability analysis, instability towards periodic longitudinal patterns only occurs in vegetated riverbeds. In the case of realistic parameter setups (reasonable vegetation density of around 50 plants/vegetation branches per square meter of riverbed), vegetation roughness completely dominates bed induced roughness and so, the addition of sediment dynamics to a vegetated bed does not significantly change the results. The addition of sediment dynamics then just means that the bed topography follows the vegetation patterns. When looking at Figures \ref{fig_pattern_phi} to \ref{fig_pattern_aD}, we get the impression that regardless which values are assigned to three of the four main parameters ($F_0$,$\phi_m$,$\alpha_g$ and $\alpha_d$) we can always adjust the fourth parameter so as to find river patterns. This was tested and found to be true in the parameter domain indicated in Table \ref{table:test_param}. It is obviously not true for the diffusion coefficient $D$ and for the sediment transport parameter $\gamma$ as these two are not essential ingredients for obtaining patterns and are not able to influence significantly the balance between vegetation growth and death. We can thus end the 1D analysis recalling that riverbed vegetation patterns are the result of dynamic interaction of vegetation growth and mortality through uprooting, whereas bed topography adjusts to vegetation patterns in a passive manner.
\begin{table}
\caption{Parameter domain}
\centering
\begin{tabular}{c c c c c}
\hline\hline
Parameter name & Symbol & Minimum value & Maximum value\\[0.5ex]
\hline
Froude Number & $F_0$ & 0.1 & 20 \\
Carrying capacity & $\tilde{\phi}_m$ & 0 & 1000 \\
Growth coefficient & $\alpha_g$ & 0.1 & 20 \\
Uprooting coefficient & $\alpha_d$ & 0.1 & 20 \\
\hline
\end{tabular}
\label{table:test_param}
\end{table}

\newpage
\phantom{}
\newpage
\section{Stability Analysis of 2D Ecomorphodynamic Equations}\label{sec:2D}
After having shown that instability towards periodic patterns is possible in vegetated rivers that are subject to a 1-dimensional analysis (meaning with laterally homogeneous dynamics), we now move on to the case where lateral flow structure is important. Instability towards 2-dimensional structures (bars) is thus analyzed in this chapter using the 2-dimensional de Saint Venant equation for constant width and movable bed along with appropriately modified equations for flow and sediment continuity and vegetation dynamics.


\subsection{Governing Equations}\label{sec:2D_gov}
The equations for 1-dimensional flow can be extended to describe 2-dimensional flow and vegetation dynamics to yield
\begin{subequations}
\label{eq:2D_gov}
\begin{align}
&\frac{\partial{\tilde{U}}}{\partial{\tilde{t}}}+\tilde{U}\frac{\partial{\tilde{U}}}{\partial{\tilde{s}}}+\tilde{V}\frac{\partial{\tilde{U}}}{\partial{\tilde{n}}}+g\left[\frac{\partial{\tilde{Y}}}{\partial{\tilde{s}}}+\frac{\partial{\tilde{\eta}}}{\partial{\tilde{s}}}\right]+\frac{\tilde{\tau}_s}{Y}=0\\
&\frac{\partial{\tilde{V}}}{\partial{\tilde{t}}}+\tilde{U}\frac{\partial{\tilde{V}}}{\partial{\tilde{s}}}+\tilde{V}\frac{\partial{\tilde{V}}}{\partial{\tilde{n}}}+g\left[\frac{\partial{\tilde{Y}}}{\partial{\tilde{n}}}+\frac{\partial{\tilde{\eta}}}{\partial{\tilde{n}}}\right]+\frac{\tilde{\tau}_n}{Y}=0\\
& \frac{\partial{\tilde{Y}}}{\partial{\tilde{t}}}+\nabla\cdot (\tilde{Y}\mathbf{\tilde{V}})=0\\
&(1-p)\frac{\partial{\tilde{\eta}}}{\partial{\tilde{t}}}+\nabla\cdot \tilde{\mathbf{Q}}_s=0\\
&\frac{\partial{\tilde{\phi}}}{\partial{\tilde{t}}}=\alpha_g\tilde{\phi}(\tilde{\phi}_m-\tilde{\phi})+\nabla^2\cdot (D\tilde{\phi})-\alpha_d\tilde{Y}(\tilde{U}^2+\tilde{V}^2)\tilde{\phi}.
\end{align}
\end{subequations}
where $\tilde{\tau}_s$ and $\tilde{\tau}_n$ are the bed shear stresses and $\tilde{Q}_{ss}$ and $\tilde{Q}_{sn}$ are the sediment transport fluxes in the streamwise and transverse direction respectively. Additionally, $\tilde{\mathbf{V}}$ is the velocity vector $\{\tilde{U},\tilde{V}\}$ with $\tilde{U}$ and $\tilde{V}$ the velocities in the streamwise and transverse (normal) direction respectively (see sections \ref{sec:veg} and \ref{sec:1D_gov} and Figure \ref{fig:saintvenant} to recall the variables/parameters not mentioned here). For the sake of simplicity, the same closure relationships for shear stress and sediment transport are used than in the 1D-formulation. Following \cite{Fe}, we can write total shear stress and total sediment transport as
\begin{subequations}
\label{eq:2D_shear}
\begin{align}
\boldsymbol{\tilde{\tau}}&=\{\tilde{\tau}_s,\tilde{\tau}_n\}=C\{\tilde{U},\tilde{V}\}\sqrt{\tilde{U}^2+\tilde{V}^2}\\
\mathbf{\tilde{Q}_s}&=\{\tilde{Q}_{ss},\tilde{Q}_{sn}\}=\tilde{\Phi}\{\cos(\delta),\sin(\delta)\}
\end{align}
\end{subequations}
where $C=\frac{g}{\chi^2}$ is the total friction coefficient and where we define the sediment transport load $\tilde{\Phi}=a(\sqrt{\tilde{U}^2+\tilde{V}^2})^3$. Furthermore, we can assume particle motion to be the product of shear stress exerted by the fluid and the gravitational effect of a weak lateral slope. The deviation of particle motion from the longitudinal direction is indicated by the angle $\delta$. We may thus write
\begin{subequations}
\label{eq:2D_shear2}
\begin{align}
\cos(\delta)&=\frac{\tilde{U}}{\sqrt{\tilde{U}^2+\tilde{V}^2}}\\
\sin(\delta)&=\frac{\tilde{V}}{\sqrt{\tilde{U}^2+\tilde{V}^2}}-\frac{r}{\beta \sqrt{\tau_\star}}\frac{\partial{\eta}}{\partial{n}}.
\label{eq:weak_slope}
\end{align}
\end{subequations}
where r is an empirical parameter between 0.5 and 0.6 (\cite{Fe}) and $\tau_\star=b\tilde{U}^2$ is the dimensionless Shields stress with $b=\frac{1}{\chi^2_bd_{50}\frac{\rho_s-\rho_w}{\rho_w}}$. Here, $d_{50}$ is the median grain diameter, $\rho_s$ is the sediment density and $\rho_w$ is the water density. Note that this approximation of gravity effects is only valid in the limit of weak transverse slopes where the effect of gravity is small compared to fluid friction effects. Additionally, the deviation angle $\delta$ needs to be small as well in order to perform the simplified Taylor expansion
\begin{equation}
\sin(\delta)=\sin(x-c)\approx \sin(x)-c*\cos(x)\approx \sin(x)-c
\end{equation}
with $\sin(x)=\frac{\tilde{V}}{\sqrt{\tilde{U}^2+\tilde{V}^2}}$, $c=\frac{r}{\beta \sqrt{\tau_\star}}\frac{\partial{\eta}}{\partial{n}}$, $c\ll x$,  $x\ll1$ and thus $\delta \ll1$.

\subsection{Governing Equations in Dimensionless Variables}\label{sec:2D_dim}
To write the system of equations \ref{eq:2D_gov} in dimensionless form, we introduce the change of variables (motivated by the approach of \cite{Fe})
\begin{subequations}
\label{eq:2D_cha}
\begin{align}
U&=\frac{\tilde{U}}{\tilde{U}_0}\\
V&=\frac{\tilde{V}}{\tilde{U}_0}\\
Y&=\frac{\tilde{Y}}{\tilde{Y}_0}\\
\eta&=\frac{\tilde{\eta}}{\tilde{Y}_0}\\
\phi&=\frac{\tilde{\phi}}{\tilde{\phi}_m}\\
t&=\tilde{t}\frac{\tilde{U}_0}{\tilde{B}}\\
s&=\frac{\tilde{s}}{\tilde{B}}\\
n&=\frac{\tilde{n}}{\tilde{B}},
\end{align}
\end{subequations}
where $\tilde{B}$ is half the river width. Recall that $\tilde{Y}_0$ and $\tilde{U}_0$ are the normal depth and velocity respectively. Using change of variables (\ref{eq:2D_cha}) and closure relationships (\ref{hans12}), (\ref{eq:2D_shear}) and (\ref{eq:2D_shear2}), we obtain (arranged in a way to have the time derivative on the left-hand side)
\begin{subequations}
\label{eq:2D_dim}
\begin{align}
\frac{\partial{U}}{\partial{t}}=&-U\frac{\partial{U}}{\partial{s}}-V\frac{\partial{U}}{\partial{n}}-\frac{1}{F_0^2}\left[\frac{\partial{Y}}{\partial{s}}+\frac{\partial{\eta}}{\partial{s}}\right]-\beta\left[\left(c_b\frac{1}{Y}+c_v\phi\right)U(U^2+V^2)^{1/2}\right]\\
\frac{\partial{V}}{\partial{t}}=&-U\frac{\partial{V}}{\partial{s}}-V\frac{\partial{V}}{\partial{n}}-\frac{1}{F_0^2}\left[\frac{\partial{Y}}{\partial{n}}+\frac{\partial{\eta}}{\partial{n}}\right]-\beta\left[\left(c_b\frac{1}{Y}+c_v\phi\right)V(U^2+V^2)^{1/2}\right]\\
\frac{\partial{Y}}{\partial{t}}=&-Y\frac{\partial{U}}{\partial{s}}-U\frac{\partial{Y}}{\partial{s}}-Y\frac{\partial{V}}{\partial{n}}-V\frac{\partial{Y}}{\partial{n}}\\
\begin{split}
\frac{\partial{\eta}}{\partial{t}}=&-\frac{\gamma}{3}\left[(3U^2+V^2)\frac{\partial{U}}{\partial{s}}+2UV\frac{\partial{U}}{\partial{n}}+2UV\frac{\partial{V}}{\partial{s}}+(U^2+3V^2)\frac{\partial{V}}{\partial{n}}\right]\\
& +\frac{\gamma}{3}\frac{r}{\beta \sqrt{\tau_\star}}\left[3(U\frac{\partial{U}}{\partial{n}}+V\frac{\partial{V}}{\partial{n}})(U^2+V^2)^{1/2}\frac{\partial{\eta}}{\partial{n}}+(U^2+V^2)^{3/2}\frac{\partial{^2\eta}}{\partial{n^2}}\right]
\end{split}\\
\frac{\partial{\phi}}{\partial{t}}=&\beta\left[\nu_g\phi(1-\phi)+\nu_{D_s}\frac{\partial{^2\phi}}{\partial{s^2}}+\nu_{D_n}\frac{\partial{^2\phi}}{\partial{n^2}}-\nu_d Y(U^2+V^2)\phi\right].
\end{align}
\end{subequations}
where the aspect ratio\\
where $\beta=\frac{\tilde{B}}{\tilde{Y}_0}$, $F_0=\frac{\tilde{U}_0}{\sqrt{g\tilde{Y}_0}}$, $c_b=\frac{g}{\chi_b^2}$, $c_v=\frac{c_Dd\tilde{\phi}_m \tilde{Y}_0}{2}$, $\gamma=\frac{3\tilde{Q}_{s0}}{(1-p)\tilde{U}_0\tilde{Y}_0}$, $\nu_g=\frac{\alpha_g \tilde{\phi}_m\tilde{Y}_0}{\tilde{U}_0}$,\\ $\nu_d=\alpha_d\tilde{Y}_0^2\tilde{U}_0$, $\nu_{D_s}=\frac{D_s}{\tilde{Y}_0\tilde{U}_0}$ and $\nu_{D_n}=\frac{D_n}{\tilde{Y}_0\tilde{U}_0}$.\\
Note that the dimensionless quantities are essentially the same as in the 1D analysis.

\subsection{Linear Stability Analysis}\label{sec:2D_lin}
To derive the perturbation formulation of the 2D-equations, we essentially proceed in the same manner as was done for the 1D case: spatially homogeneous solutions are computed, perturbated and introduced into the dimensionless governing equations.
\subsubsection{Homogeneous Solutions}\label{sec:2D_lin_hom}
It can easily be found for the non-trivial case ($\phi_0>0$) that 
\begin{subequations}
\label{eq:2D_hom}
\begin{align}
J_0=&\beta F_0^2\left[c_b+c_v\left(\frac{\nu_g-\nu_d}{\nu_g}\right)\right]\\
\phi_0=&\frac{\nu_g-\nu_d}{\nu_g}.
\end{align}
\end{subequations}
The non-zero dimensionless homogeneous solution can therefore be summarized as
\begin{equation}\label{eq:2D_hom2}
\{U_0,V_0,Y_0,\eta_0(s),\phi_0\}=\{1,0,1,-J_0s,\phi_0\}
\end{equation}
with $J_0$ and $\phi_0$ as defined in (\ref{eq:2D_hom}).
\subsubsection{Linearization of Perturbated Equations}\label{sec:2D_lin_lin}
To linearize the 2-dimensional model in the neighborhood of the homogeneous solution, we introduce the perturbated homogeneous solution
\begin{equation} \label{eq:2D_hompert}
\{1,0,1,-J_0s,\phi_0\}+\epsilon\{U_1,V_1,Y_1,\eta_1,\phi_1\}
\end{equation}
into (\ref{eq:2D_dim}). In order to perform a 2-dimensional normal mode analysis, we can write in the most general case (see (\ref{eq:1D_perhom}))
\begin{equation} \label{eq:2D_perturb2}
\{U_1,V_1,Y_1,\eta_1,\phi_1\}=\{u(t),v(t),y(t),h(t),f(t)\}\cos(k_nn+\psi)\exp(ik_ss)+c.c.
\end{equation}
with $k_n$ and $k_s$ the wavenumbers in the normal and streamwise direction respectively and $\psi$ the phase in the normal direction.
However, we can further develop this ansatz by taking into consideration the boundary conditions
\begin{equation} \label{eq:boundary}
\tilde{V}(\pm \tilde{B})=0
\end{equation}
for impermeable lateral boundaries. In dimensionless variables this means
\begin{equation} \label{eq:boundary}
V(\pm 1)=0
\end{equation}
and we thus need the following equality to be true:
\begin{equation}
2=m\frac{\lambda_n}{2},
\end{equation}
where $\lambda_n$ is the dimensionless wavelength of the transverse periodic pattern and $m$ a positive integer and therefore
\begin{equation}
k_n=m\frac{\pi}{2}.
\end{equation}
\begin{figure}
\begin{center}
    \includegraphics[width=130mm]{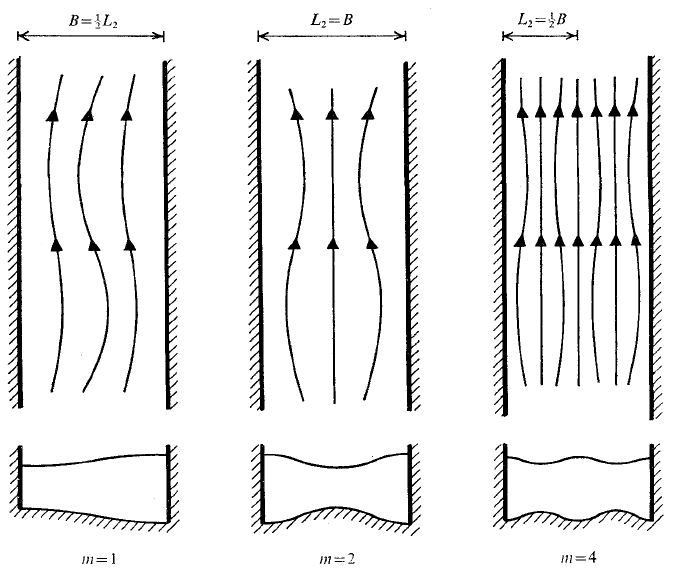} \\
    \caption{Sketch illustrating alternate bars (m=1) and multiple bars (m>1) taken from \cite{EnSk}}
    \label{fig:sketch_EnSk}
\end{center}
\end{figure}
In other words, the river width $2\tilde{B}$ has to be a multiple of half the transverse wavelength for (\ref{eq:boundary}) to be true. Figure \ref{fig:sketch_EnSk} shows streamline patterns and cross-sections for different parameters $m$. $m=1$ is the well-known alternate bar river while m>1 yields multiple bar patterns. Although not shown in Figure \ref{fig:sketch_EnSk}, a sinusoidal pattern still occurs in the longitudinal direction whose fastest growing wavelength depends on the transverse wavelength and thus on $m$. Then replacing the phase angle $\psi$ in equation (\ref{eq:2D_perturb2}), which can only take the values $0$ and $\frac{\pi}{2}$ and by making the distinction between the sine and the cosine we get
\begin{subequations} \label{eq:2D_perturb_v}
\begin{align} 
&V_1=v(t)\sin(m\frac{\pi}{2}n)\exp(ik_s s)	\text{\qquad($m$ odd)}\\
&V_1=v(t)\cos(m\frac{\pi}{2}n)\exp(ik_s s)	\text{\qquad($m$ even)}.
\end{align}
\end{subequations}
Finally, in order to have a perturbation ansatz that is technically convenient, we need the perturbations of the other state variables to be $\pi/2$ out of phase with respect to the perturbation of the transverse velocity $V_1$ (see for example \cite{CST}) and we get
\begin{subequations} \label{eq:2D_perturb}
\begin{align} 
&\{U_1,V_1,Y_1,\eta_1,\phi_1\}=\left\{u(t),v(t)\tan^{-1}(k_n n),y(t),h(t),f(t)\right\}\sin(m\frac{\pi}{2}n)\exp(ik_s s)\\
&\{U_1,V_1,Y_1,\eta_1,\phi_1\}=\left\{u(t),v(t)\tan(k_n n),y(t),h(t),f(t)\right\}\cos(m\frac{\pi}{2}n)\exp(ik_s s)
\end{align}
\end{subequations}
for $m$ odd and even respectively. In fact, this is the only way we can transform our ecomorphodynamic equation system into an eigenvalue problem with the real parts of the eigenvalues determining the asymptotic fate of the system. Substituting (\ref{eq:2D_hompert}) and (\ref{eq:2D_perturb}) into (\ref{eq:2D_dim}) we end up with the following linear system of equations:
\begin{subequations}
\label{eq:2D_lin}
\begin{align}
&\frac{du}{dt}=(-ik_s-2\beta c_b-2\beta c_v\phi_0)u+\left(\frac{-ik_s}{F_0^2}+\beta c_b\right)y+\left(\frac{-ik_s}{F_0^2}\right)h+(-\beta c_v)f\\
&\frac{dv}{dt}=(-ik_s-\beta c_b-\beta c_v\phi_0)v+\left(\frac{-k_n(-1)^{m+1}}{F_0^2}\right)y+\left(\frac{-k_n(-1)^{m+1}}{F_0^2}\right)h\\
&\frac{dy}{dt}=(-ik_s)u+(k_n(-1)^{m+1})v+(-ik_s)y\\
&\frac{dh}{dt}=(-i\gamma k_s)u+(\frac{1}{3}\gamma k_n(-1)^{m+1})v+(-\frac{\gamma}{3}\frac{r}{\beta \sqrt{bU^2_0}}k_n^2)h\\
&\frac{df}{dt}=(-2\beta\nu_d\phi_0)u+(-\beta\nu_g\phi_0)y+(-\beta(\nu_g-\nu_d)-\frac{\nu_{D_s}}{\beta}k_s^2-\frac{\nu_{D_n}}{\beta}k_n^2)f
\end{align}
\end{subequations}
where the lateral wavenumber $k_n=m\frac{\pi}{2}$.
The system of equations (\ref{eq:2D_lin}) can be arranged as
\begin{equation}\label{eq:2D_system}
\begin{pmatrix} \frac{du}{dt} \\ \frac{dv}{dt} \\ \frac{dy}{dt} \\ \frac{dh}{dt} \\ \frac{df}{dt} \end{pmatrix}
=A
\begin{pmatrix} u \\ v \\ y \\ h \\ f \end{pmatrix},
\end{equation}
where A is the following 5 x 5 matrix:
\begin{equation}\label{eq:2D_matrix}
\begin{pmatrix} -ik_s-2\beta c_b-2\beta c_v\phi_0 & 0 & \frac{-ik_s}{F_0^2}+\beta c_b & \frac{-ik_s}{F_0^2} & -\beta c_v\\ 0 & -ik_s-\beta c_b-\beta c_v\phi_0 & \frac{-k_n(-1)^{m+1}}{F_0^2} & \frac{-k_n(-1)^{m+1}}{F_0^2} & 0 \\ -ik_s & k_n(-1)^{m+1} & -ik_s& 0 & 0 \\ -i\gamma k_s & \frac{1}{3}\gamma k_n(-1)^{m+1} & 0 & -\frac{\gamma r}{3\beta \sqrt{bU^2_0}}k_n^2 & 0 \\ -2\beta\nu_d\phi_0 & 0 & -\beta\nu_d\phi_0 & 0 & -\beta\nu_g\phi_0-\frac{\nu_{D_s}}{\beta}k_s^2+\frac{\nu_{D_n}}{\beta}k_n^2 \end{pmatrix}.
\end{equation}
As it was done for an essentially 1-dimensional river, the eigenvalues of this operator can be analyzed to get information about the asymptotic fate of the system. Note that the case where $m=0$ corresponds to the 1-dimensional case (transverse velocity $V$ does not influence the other state variables anymore).

\newpage
\begin{figure}
\begin{center}
    \includegraphics[width=130mm]{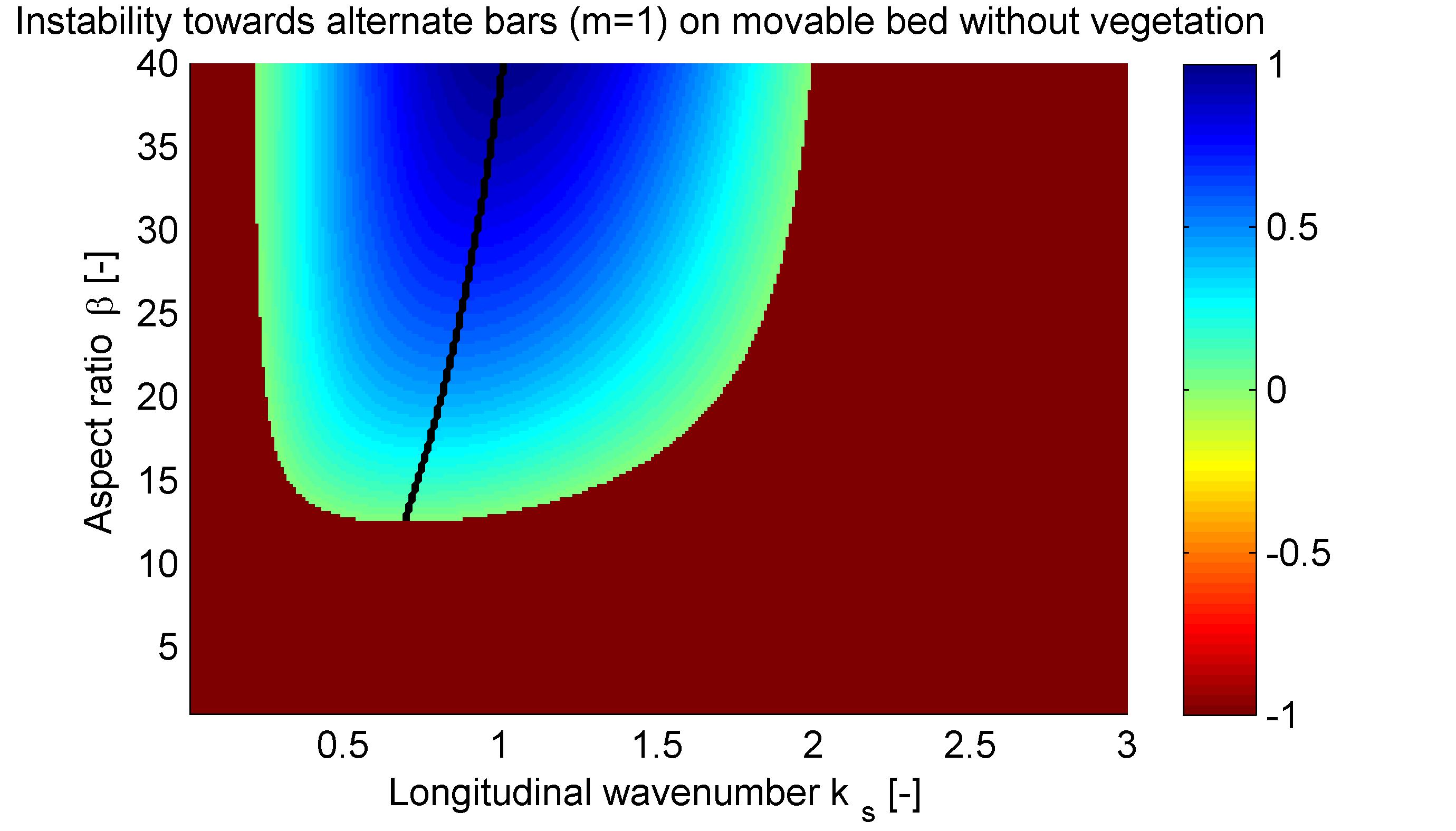} \\
    \caption{Relative growth rate (normalized with respect to the maximum growth rate occurring on the Figure) of alternate bar formation as a function of longitudinal wavenumber $k_s$ and aspect ratio $\beta$: the color code indicates relative growth rate and a value of $-1$ (red) means that no patterns exist; the black line shows the maximum growth rate for given $\beta$ and thus indicates the dominating longitudinal wavenumber; parameter values are $F_0=0.5$ and values indicated in Table \ref{table:2D_sve}}
    \label{fig:2D_sve1}
\end{center}
\end{figure}
\subsection{Linear Stability Analysis: Results}\label{sec:2D_sta}
In this section, we first reproduce the known results for flow dynamics only and flow dynamics coupled with sediment dynamics in order to validate our approach (sections \ref{sec:2D_sta_SV} and \ref{sec:2D_sta_SVE}). Then, we move on to add vegetation dynamics to SV and SVE respectively and analyze the results (sections \ref{sec:2D_sta_SVV} and \ref{sec:2D_sta_SVEV}). Whereas we were only interested in patterns with finite wavenumber in the 1D case, we now also look for patterns where the maximum growth rate occurs for $k_s=0$. This case corresponds to the riverbed divided into longitudinal channels and it can be associated to vegetated ridges in ephemeral rivers for example (Figure \ref{fig:anabranch}B). 
To render the results of the 2D stability analysis comprehensible, we would like to resume here the most important guidelines:
\begin{itemize}
\item The longitudinal wavenumber $k_s$ which dominates in the asymptotic limit is the one with the highest linear growth rate (highest $\mathrm{Max(Re(\omega))}$)
\item Instability towards 2D patterns occurs if for some integer $m\geq1$ we have positive growth rates for finite or zero longitudinal wavenumber $k_s$
\item We have instability towards alternate bars if the highest growth rate occurs for $m=1$ and instability towards multiple bars if the highest growth rate occurs for $m>1$
\item In order to have solutions with physical meaning, $\phi_0$ has to be positive which gives the condition $(\nu_g-\nu_d)>0$
\end{itemize}
\subsubsection{SV Equations (fixed bed without vegetation)}\label{sec:2D_sta_SV}
No instability towards finite patterns is detected when only the hydrodynamic part of the 2-dimensional model is analyzed, which agrees with results from literature (e.g. \cite{Parker}). As a consequence, linear instability does not seem to be inherent to flow dynamics alone (\cite{Parker}).
\subsubsection{SVE Equations (movable bed without vegetation)}\label{sec:2D_sta_SVE}
If the 2D de Saint-Venant equations are coupled with sediment dynamics, we can reproduce the characteristic neutral curve between no instability and instability towards alternate bars (see Figure \ref{fig:CST}). In Figure \ref{fig:2D_sve1}, we added a color code to the instability-region which shows the relative growth rate (scaled by the maximum growth rate that appears in the Figure) and a black line indicating the dominating longitudinal wavenumber $k_s$ for given $\beta$. We can see that the highest growth rate occurs in the middle of the instability domain while lower growth rates are present at the borders. Additionally, a minimum value is required for the aspect ratio in order to produce patterns with finite longitudinal wavenumber. In the case of parameters as indicated in the caption of Figure \ref{fig:2D_sve1}, the aspect ratio has a minimum value of 12 at a dimensionless longitudinal wavenumber of 0.7 which is not that far off from the values found by \cite{CST} (visible on Figure \ref{fig:CST}). The difference in wavenumber at minimum aspect ratio could be due to the different sediment transport closure relationships  or different parameter values that were used in the present work.\\
\begin{table}
\caption{Values for constant parameters of the 2D-SVE-analysis}
\centering
\begin{tabular}{c c c c}
\hline\hline
Parameter name & Variable & Value & Units\\[0.5ex]
\hline
Normal water depth & $\tilde{Y}_0$ & 1 & m \\
Manning coefficient & n & 0.03 & $\mathrm{m^{-1/3}s}$ \\
Transverse slope parameter & r & 0.5 & - \\
Median sediment diameter & $d_{50}$  & 0.005 & m\\
\hline
\end{tabular}
\label{table:2D_sve}
\end{table}
\begin{figure}
\begin{center}
    \includegraphics[width=130mm]{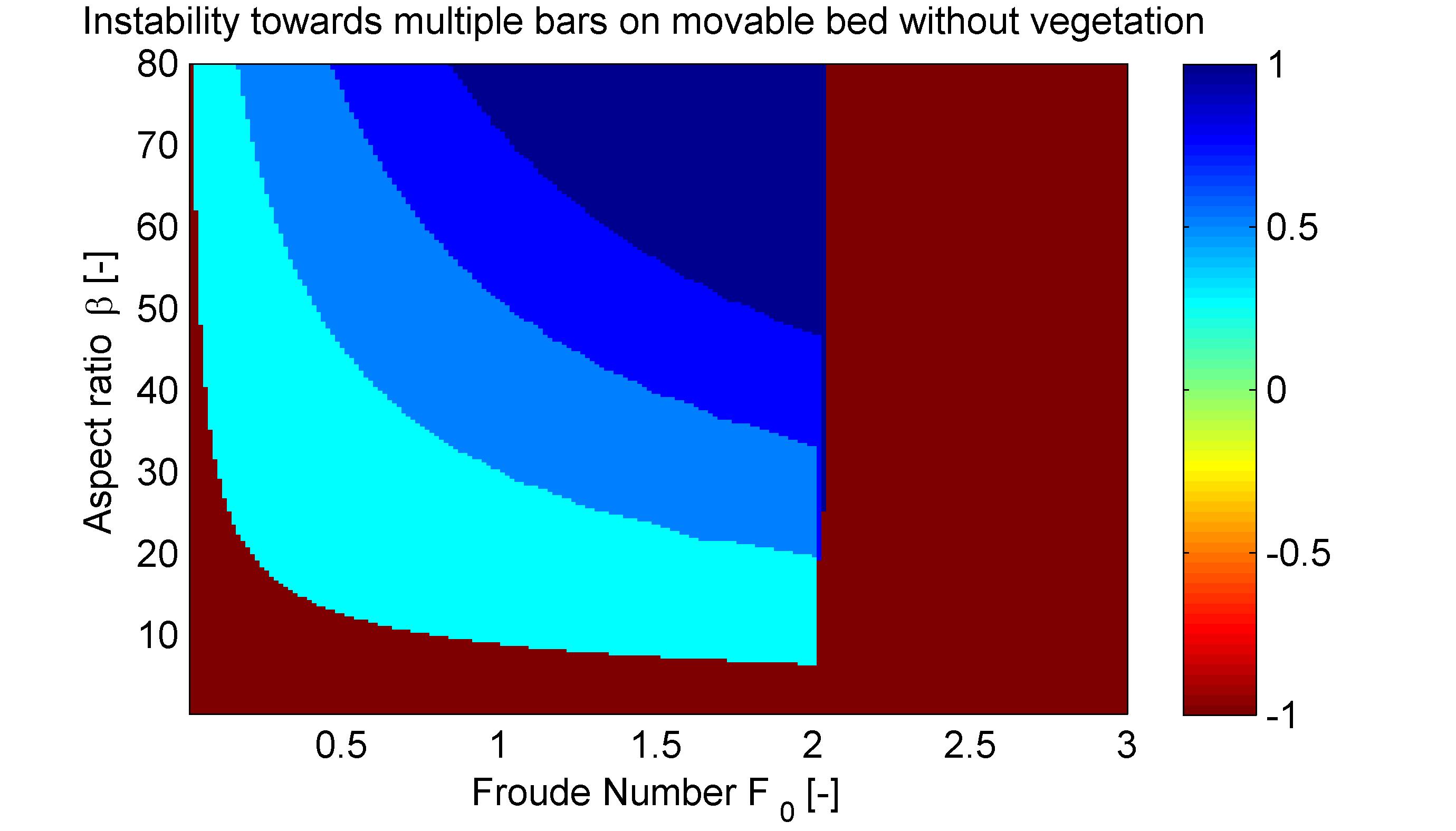} \\
    \caption{Multiple bar formation as a function of Froude number $F_0$ and aspect ratio $\beta$: no instability (red), alternate bars (light blue), multiple bars (darker blues for increasing bar order $m=2,3,4$); for parameter values see Table \ref{table:2D_sve}}
    \label{fig:2D_sve3}
\end{center}
\end{figure}
It was also observed in nature and found using linear stability theory (\cite{EnSk}) that the higher the aspect ratio of a river the higher the bar order (number of bars in the transverse direction) is. Figure \ref{fig:2D_sve3} depicts parameter domains where alternate bar and multiple bar regimes respectively dominate (based on the highestl growth rate). The aspect ratio $\beta$ seems to be the decisive parameter in a reasonable range of Froude numbers between 1 and 2. However, the Froude number for flooding in the Marshall River is between 0.3 and 0.4 according to \cite{To} and thus $F$ is also important to determine the bar regime of a river. And, once the Froude number exceeds a certain maximum value, instability towards bar formation does no longer exist. Overall, we can say that results of earlier stability analyses of bar instability could be reproduced in this work. Bar instability triggered by 2-dimensional sediment dynamics is sensitive to a river's aspect ratio (width to depth ratio) and also to Froude number for low values of $F$.
Finally, it is worth noting that the term accounting for gravitational effects of a weak lateral slope (second term of (\ref{eq:weak_slope})) is crucial in order to reproduce this well-known result. \cite{En81} was the first to propose this relation which was later confirmed experimentally by \cite{Ta} with both suggesting the parameter $r$ to be between 0.5 and 0.6.

\subsubsection{SVV Equations (fixed bed with vegetation)}\label{sec:2D_sta_SVV}
\begin{figure}
\begin{center}
    \includegraphics[width=130mm]{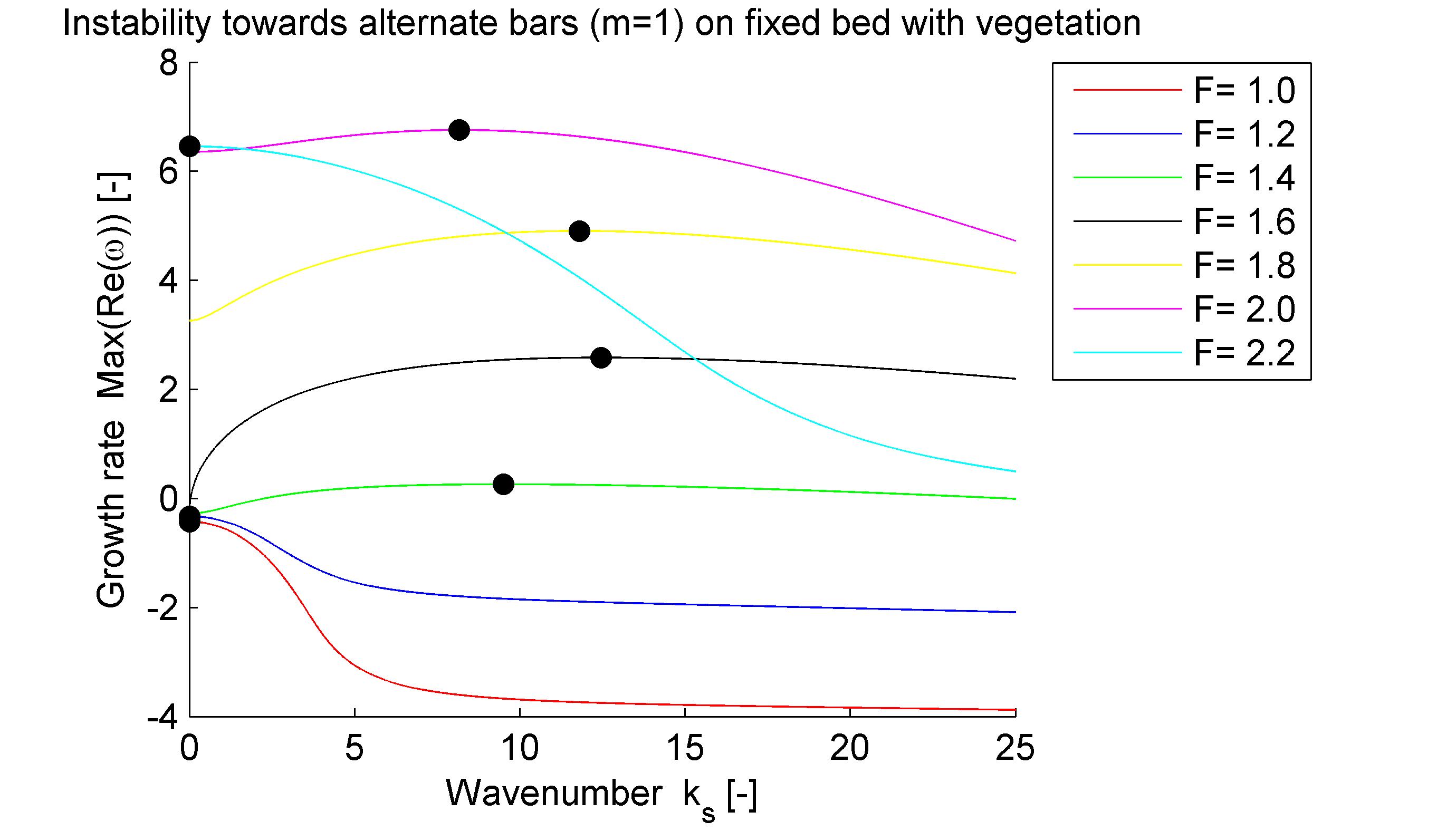} \\
    \caption{Maximum temporal growth rate of alternate bar formation (m=1) as a function of longitudinal wavenumber $k_s$ for different Froude numbers, the black dots mark the maximum of each curve; parameters used: $m=1$, $\beta=20$, $\tilde{\phi}_m=50\,\mathrm{m^{-2}}$, $\alpha_g=1\,\mathrm{m^2s^{-1}}$, $\alpha_d=1\,\mathrm{m^{-3}s}$, $D_s=0\,\mathrm{m^2s^{-1}}$, $D_n=0\,\mathrm{m^2s^{-1}}$ and Table \ref{table_param}\\}
    \label{fig:2D_svv_F}
\end{center}
\end{figure}
As it was done in the analysis of the 1-dimensional equations, we want to analyze separately the effect of vegetation dynamics in order to better understand its potential contribution to pattern formation. Thus, a fixed bed is presumed and the eigenvalues of equation (\ref{eq:2D_matrix}) removing line 4 and column 4 are analyzed in the following.\\
First, we want to look at instability towards alternate bars (m=1) and we can see in Figure \ref{fig:2D_svv_F} that, similarly to the 1D analysis, we can find the maximum growth rate to be at finite longitudinal wavenumber $k_s$ for a certain range of Froude numbers, which means that instability towards finite patterns exists. Yet, as explained before, in the 2D-case we are also interested in patterns with $k_s=0$ due to their similarity with channels in the Marshall River. Such longitudinally homogeneous patterns seem to occur at slightly higher Froude numbers than patterns with finite $k_s$, but not too high in order to still allow interaction between vegetation growth and mortality. Note that for the moment longitudinal as well as lateral vegetation diffusion (seeding and resprouting) are put to zero.\\
\begin{figure}
\begin{center}
    \includegraphics[width=130mm]{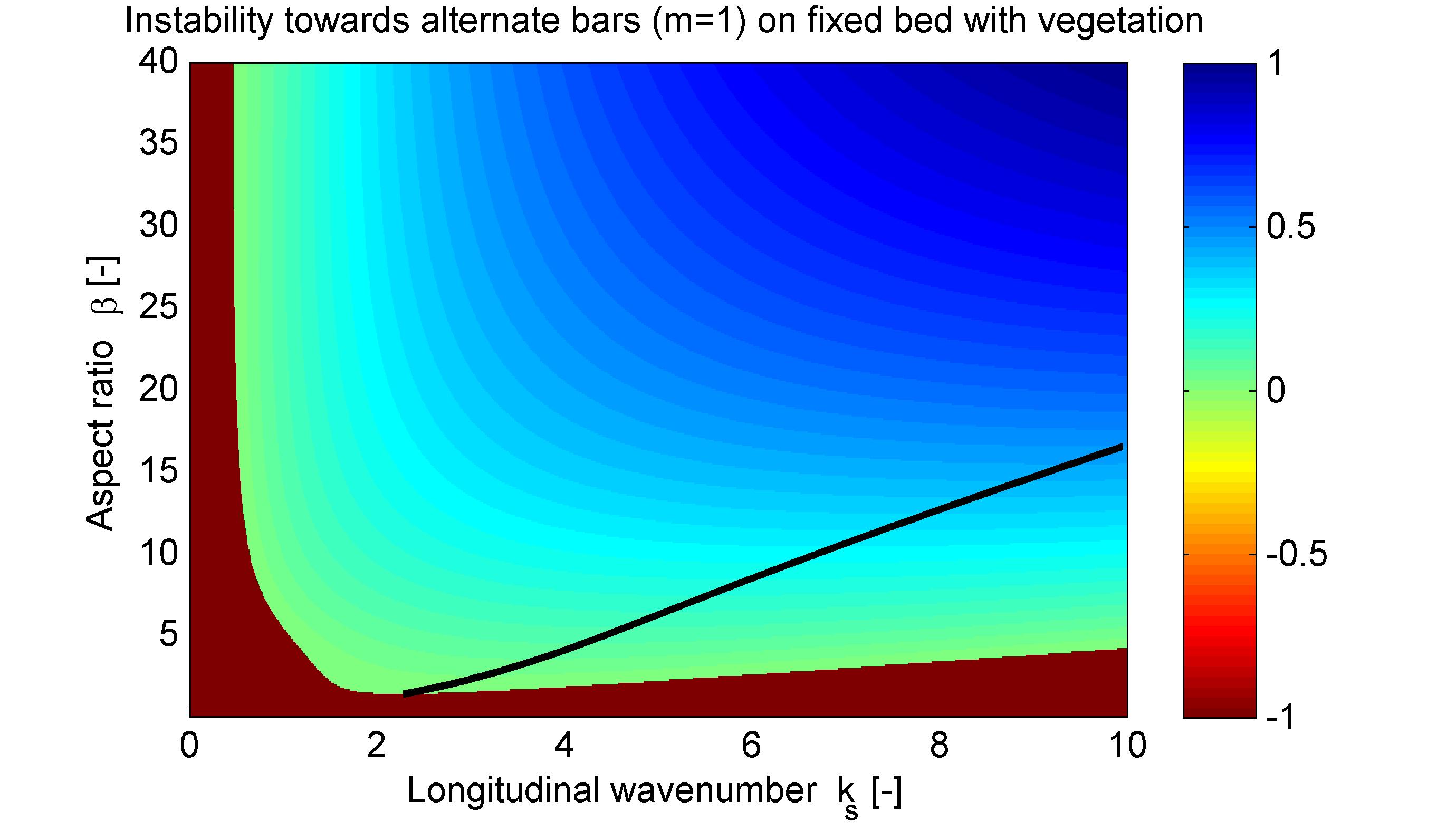} \\
    \caption{Relative growth rate (normalized with respect to the maximum growth rate occurring on the Figure) of alternate bar formation (m=1) as a function of longitudinal wavenumber $k_s$ and aspect ratio $\beta$: the color code indicates relative growth rate and a value of $-1$ (red) means that no patterns exist; the black line shows the maximum growth rate for given $\beta$ and thus indicates the value of the dominating longitudinal wavenumber; parameter values are $m=1$, $F_0=1.5$, $\tilde{\phi}_m=50\,\mathrm{m^{-2}}$, $\alpha_g=1\,\mathrm{m^2s^[{-1}}$, $\alpha_d=1\,\mathrm{m^{-3}s}$, $D_s=0\,\mathrm{m^2s^{-1}}$, $D_n=0\,\mathrm{m^2s^{-1}}$ and values indicated in Table \ref{table_param}}
    \label{fig:2D_svv1}
\end{center}
\end{figure}
\begin{figure}
\begin{center}
    \includegraphics[width=130mm]{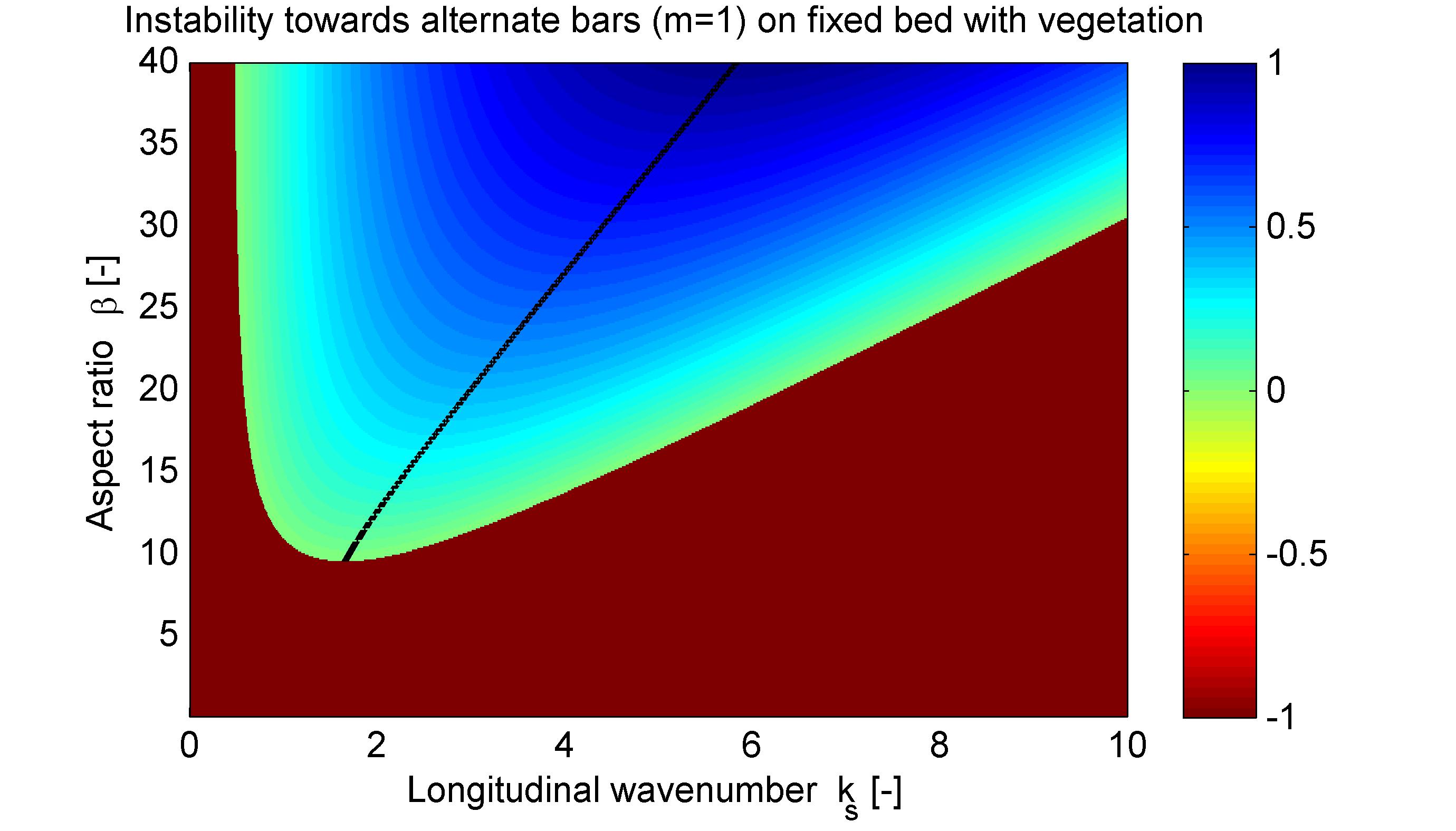} \\
    \caption{Relative growth rate (normalized with respect to the maximum growth rate occurring on the Figure) of alternate bar formation (m=1) as a function of longitudinal wavenumber $k_s$ and aspect ratio $\beta$: the color code indicates relative growth rate and a value of $-1$ (red) means that no patterns exist; the black line shows the maximum growth rate for given $\beta$ and thus indicates the value of the dominating longitudinal wavenumber; parameter values are $m=1$, $F_0=1.5$, $\tilde{\phi}_m=50\,\mathrm{m^{-2}}$, $\alpha_g=1\,\mathrm{m^2s^{-1}}$, $\alpha_d=1\,\mathrm{m^{-3}s}$, $D_s=100\,\mathrm{m^2s^{-1}}$, $D_n=100\,\mathrm{m^2s^{-1}}$ and values indicated in Table \ref{table_param}}
    \label{fig:2D_svv1_D}
\end{center}
\end{figure}
So at first glance, vegetation dynamics behaves quite alike in a 2D model than in the 1D one. To compare the instability created by vegetation dynamics to the one induced by sediment dynamics, we can have a look at Figure \ref{fig:2D_svv1}. Surprisingly, the pattern domain features some characteristics that are quite similar to the one depicted in Figure \ref{fig:2D_sve1}: a minimum value is required for the aspect ratio $\beta$ and the dominating longitudinal wavenumber increases with increasing aspect ratio. However, as visible in both Figures \ref{fig:2D_svv_F} and \ref{fig:2D_svv1} the maximum growth rate tends to occur at considerably higher $k_s$ for vegetated rivers which could be thought to be physically unrealistic (too short pattern wavelength could undermine the hypothesis of shallow water equations). One has to bear in mind though that in the 2D model length scales are normalized with respect to half-river-width $\tilde{B}$ and not normal water depth $\tilde{Y}_0$.  Thus, if we take $k_s=10$, $\beta=20$ and $\tilde{Y}_0=1\,\mathrm{m}$ for example we get
\begin{equation}
\tilde{\lambda}_s=\frac{2\pi}{\tilde{k}_s}=\frac{2\pi}{k_s/B}=12.6\, \mathrm{m}
\end{equation}
which is much smaller than river width $2\tilde{B}=2\beta \tilde{Y}_0=40\,\mathrm{m}$ but still in a reasonable order of magnitude. By tuning the parameters, we can quite easily get wavelengths that make more sense. For instance if we set the the vegetation diffusion rate $D_s=D_n=100$ we get dimensionless longitudinal wavenumbers on the order of 3 to 4 for $\beta=20$. This leads to a physical longitudinal wavelength between 30 and 40 meters which is close to the actual river width and thus much more realistic (compare Figure \ref{fig:2D_svv1_D} to Figure \ref{fig:2D_svv1} to see the influence of vegetation diffusion on the dominating wavenumber). We conclude that the inclusion of vegetation diffusion contributes to a more physically realistic result and we thus keep this value constant in the following analyses, keeping in mind though that it is not an absolutely indispensable part of the pattern producing mechanism.\\
\begin{figure}
\begin{center}
    \includegraphics[width=130mm]{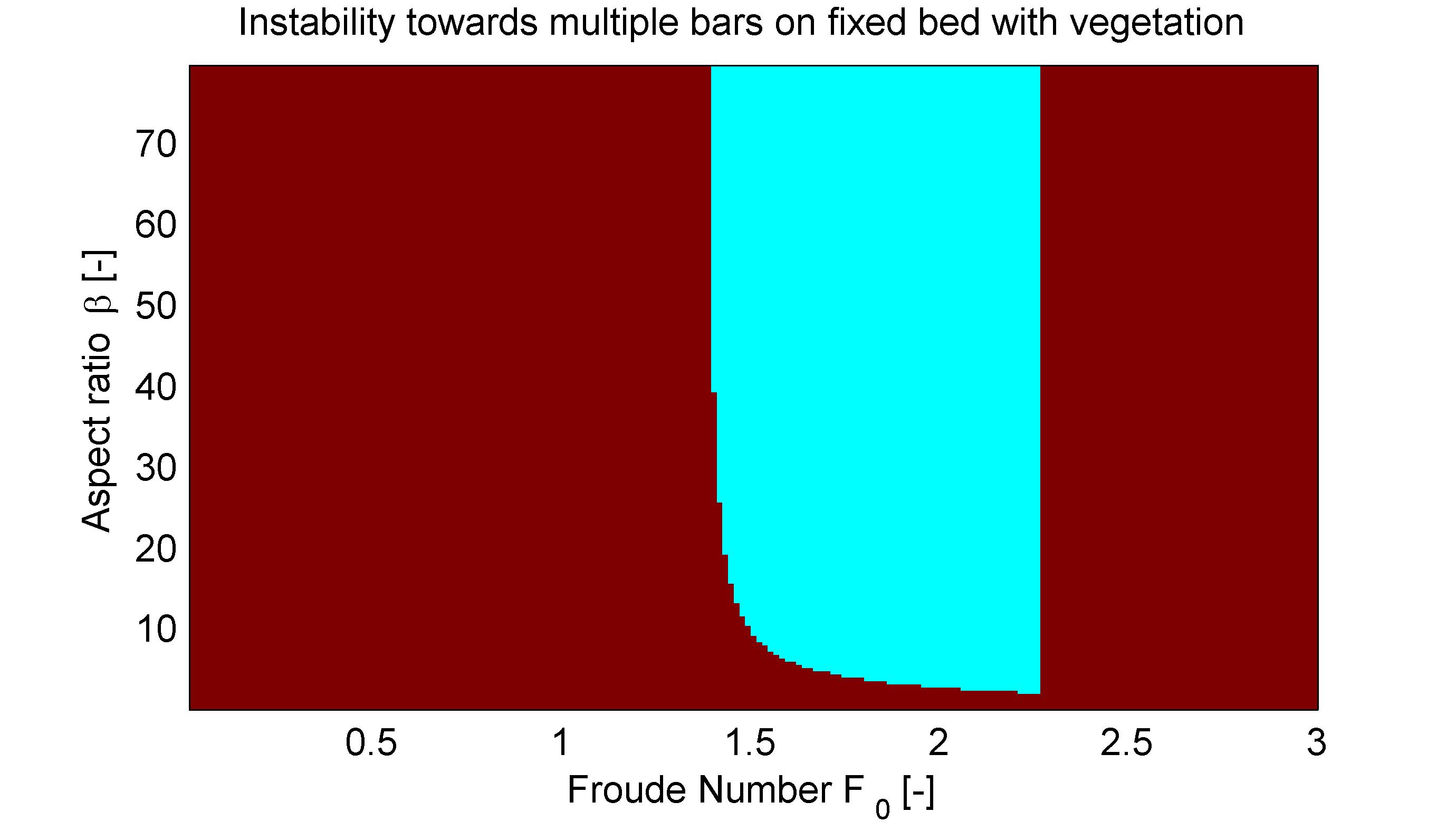} \\
    \caption{Multiple bar formation as a function of Froude number $F_0$ and aspect ratio $\beta$: no instability (red), alternate bars (light blue), multiple bars (darker blues) do not occur; parameter values are $\tilde{\phi}_m=50\,\mathrm{m^{-2}}$, $\alpha_g=1\,\mathrm{m^2s^{-1}}$, $\alpha_d=1\,\mathrm{sm^{-3}s}$, $D_s=100\,\mathrm{m^2s^{-1}}$, $D_n=100\,\mathrm{m^2s^{-1}}$ and values indicated in Table \ref{table_param}\\}
    \label{fig:2D_svv3}
\end{center}
\end{figure}
\begin{figure}
\begin{center}
    \includegraphics[width=130mm]{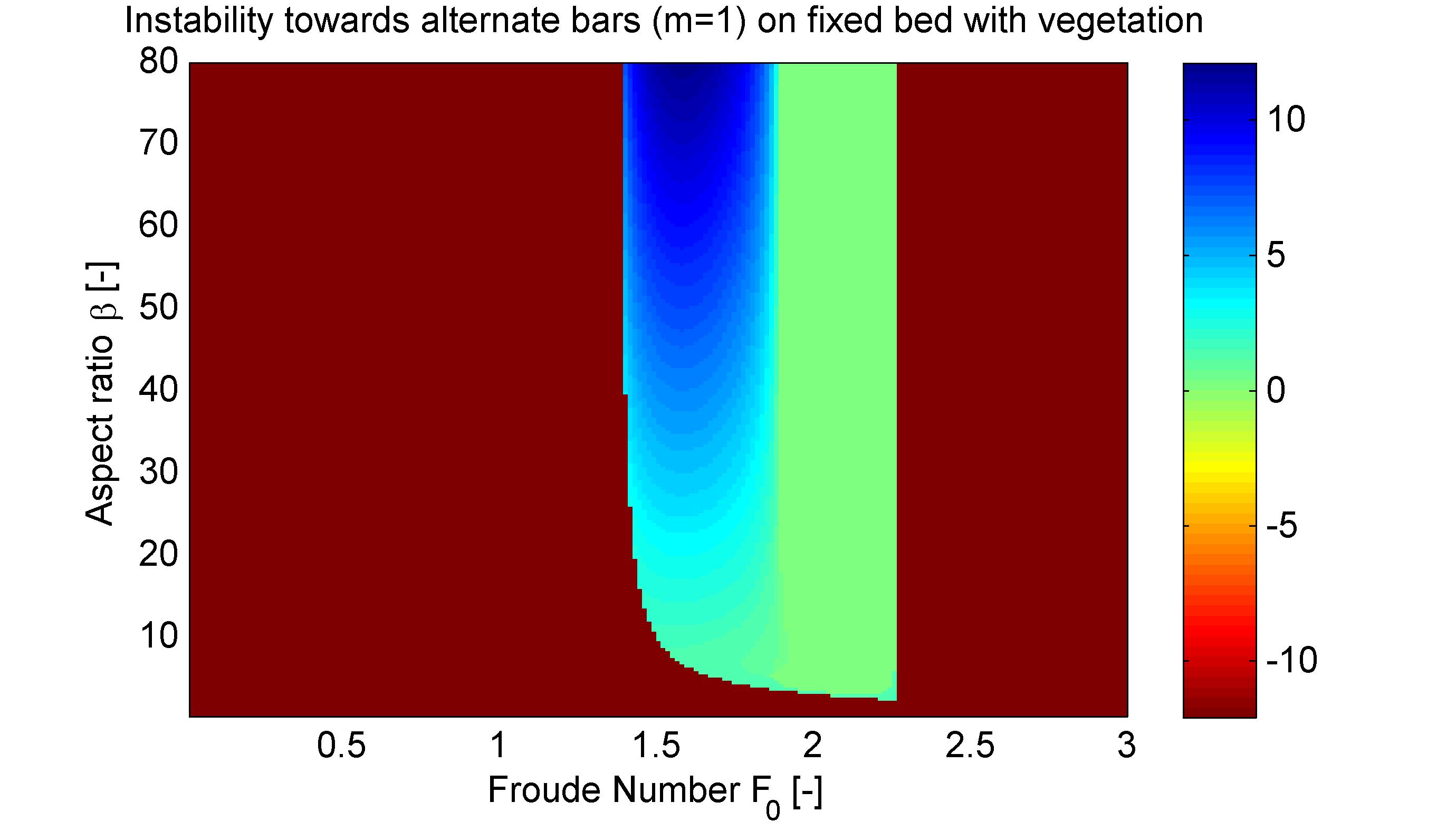} \\
    \caption{Alternate bar formation as a function of Froude number $F_0$ and aspect ratio $\beta$: the color code indicates the value of the most unstable longitudinal wavenumber $k_s$, negative numbers (red) mean no instability; parameter values are $m=1$, $\tilde{\phi}_m=50\,\mathrm{m^{-2}}$, $\alpha_g=1\,\mathrm{m^2s^{-1}}$, $\alpha_d=1\,\mathrm{m^{-3}s}$, $D_s=100\,\mathrm{m^2s^{-1}}$, $D_n=100\,\mathrm{m^2s^{-1}}$ and values indicated in Table \ref{table_param}\\}
    \label{fig:2D_svv3_1}
\end{center}
\end{figure}
\begin{figure}
\begin{center}
    \includegraphics[width=130mm]{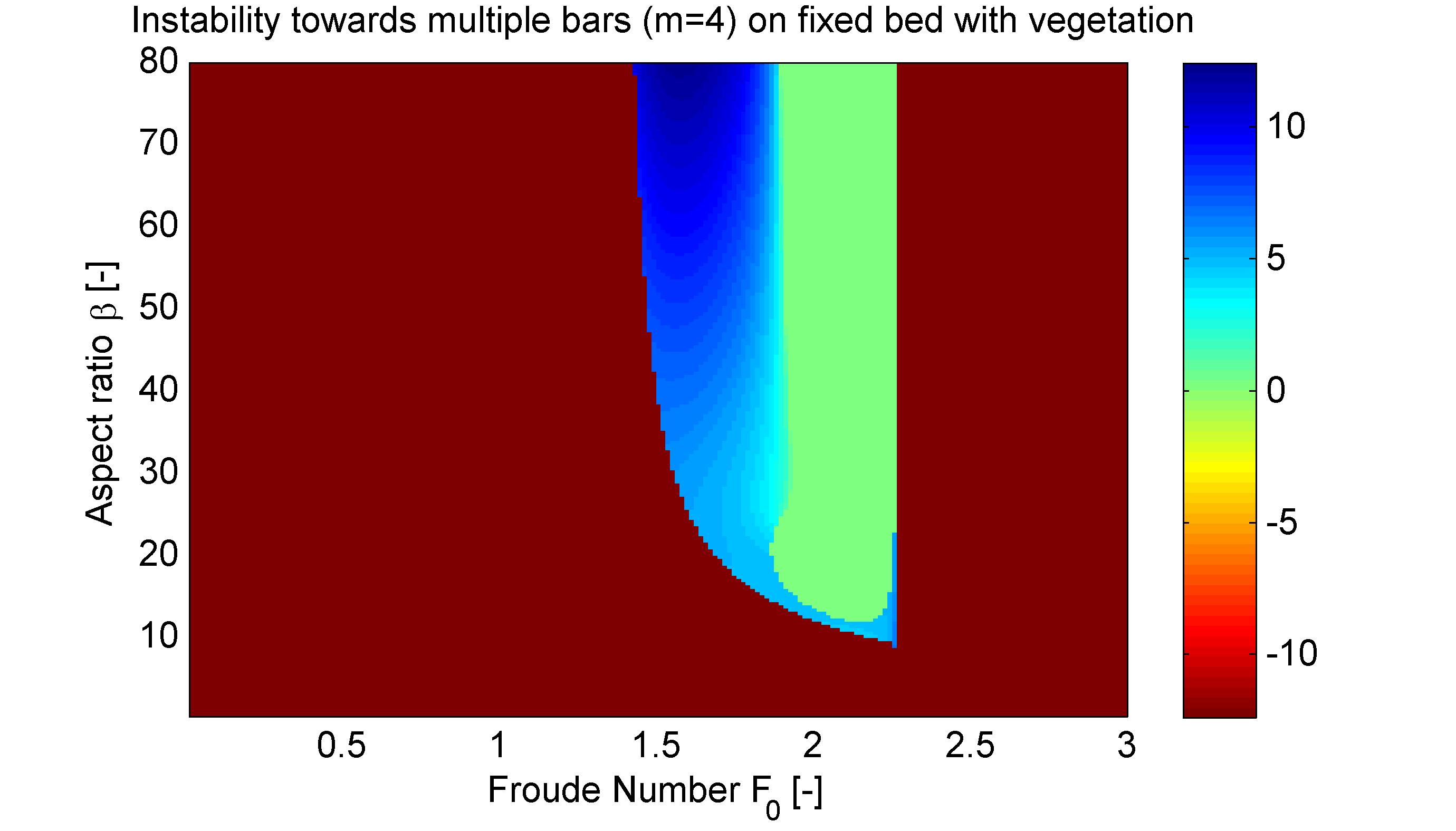} \\
    \caption{Multiple bar formation as a function of Froude number $F_0$ and aspect ratio $\beta$: the color code indicates the value of the most unstable longitudinal wavenumber $k_s$, negative numbers (red) mean no instability; parameter values are $m=4$, $\tilde{\phi}_m=50\,\mathrm{m^{-2}}$, $\alpha_g=1\,\mathrm{m^2s^{-1}}$, $\alpha_d=1\,\mathrm{m^{-3}s}$, $D_s=100\,\mathrm{m^2s^{-1}}$, $D_n=100\,\mathrm{m^2s^{-1}}$ and values indicated in Table \ref{table_param}\\}
    \label{fig:2D_svv3_4}
\end{center}
\end{figure}
Once we have seen the similarities of alternate bars inducing mechanisms of sediment and vegetation dynamics, we would like to know if that is still true for the formation of multiple bars. Thus, we repeat the multiple bar analysis of Figure \ref{fig:2D_sve3} including vegetation dynamics instead of sediment dynamics. The result can be seen in Figure \ref{fig:2D_svv3}: as expected, only a small range of Froude numbers allows pattern formation which is due to vegetation growth balance (as explained in section \ref{sec:1D_sta_SVV}). Surprisingly though, as opposed to the results of sediment dynamics, alternate bars grow always faster than multiple bars in the linear regime. This means that a fixed riverbed that is under the influence of vegetation dynamics only does not tolerate instability towards multiple bars. However, a number of vegetated patterns that occur in nature exhibit multiple bars (up to 10 for the Marshall River, even more in the case of rills on fluvial bars, see Figure \ref{fig:anabranch}). This apparent contradiction of the model and reality could still be due to the fact that sediment transport, whose influence will be analyzed in the next section, was not considered until here.\\
Figure \ref{fig:2D_svv3} only shows which kind of bars are occurring for a certain parameter configuration, but it does not give any information about the dominant longitudinal wavenumber. We could also wonder if in reality there is no instability towards multiple bars at all since we only can see the alternate bar domain in the former Figure. To answer these questions, we can have a look at Figures \ref{fig:2D_svv3_1} and \ref{fig:2D_svv3_4} which show the instability domain and most unstable longitudinal wavenumber for the case of alternate bars (m=1) and multiple bars (m=4) respectively. We can see that instability towards multiple bars does indeed exist but its growth rate being always smaller than the growth rate of alternate bars we can not perceive it in Figure \ref{fig:2D_svv3}. An interesting feature that is visible on both Figures is that about half the domain seems to have the dominating longitudinal wavenumber equal to 0 which means that riverbed vegetation (and also flow and depth) are longitudinally homogeneous in this domain.\\
\begin{figure}
  \centering
  \def\svgwidth{400pt}
  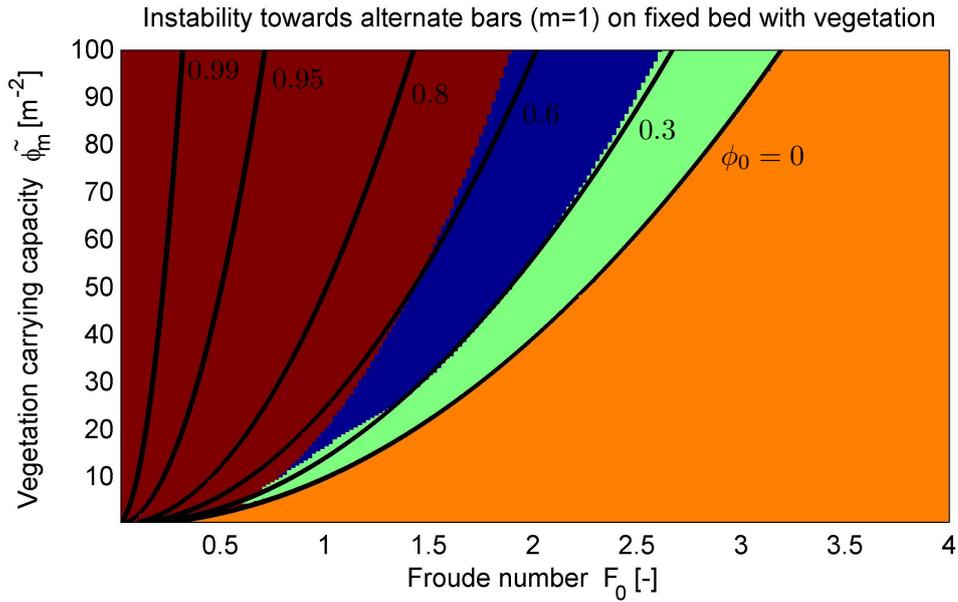
  \caption{Alternate bar formation as a function of Froude number $F_0$ and vegetation carrying capacity $\tilde{\phi}_m$ and contour lines of dimensionless homogeneous vegetation density $\phi_0$ in black: the domain with finite patterns is depicted in blue, pattern with $k=0$ in green, no pattern in red and unphysical solutions ($\phi_0<0$) in orange; parameter values are $m=1$, $\beta=20$, $\alpha_g=1\,\mathrm{m^2s^{-1}}$, $\alpha_d=1\,\mathrm{m^{-3}s}$, $D_s=100\,\mathrm{m^2s^{-1}}$, $D_n=100\,\mathrm{m^2s^{-1}}$ and values indicated in Table \ref{table_param}}
  \label{fig:2D_svv_phi_old}
\end{figure}
We then want to have a closer look at the vegetation growth balance which at first seems to be quite similar than in the 1-dimensional model. Indeed, the $F_0$ versus $\tilde{\phi}_m$ plot in Figure \ref{fig:2D_svv_phi_old} strongly resembles what we have seen before for vegetation dynamics in a 1D river. To the right (in orange), there is a domain where no physically possible solutions can exist while in the middle there is a domain (blue) where vegetation growth and death through uprooting are balanced to allow formation of vegetation patterns at finite longitudinal wavenumber. The only major difference lies in the fact that the domain with dominating longitudinal wavenumber $k_s$ equal to zero (green) may be interpreted as a pattern forming domain due to the lateral wavenumber being finite (as explained before). This domain with instability towards longitudinally homogeneous patterns ($k_s=0$) is slightly wider than the corresponding one we saw in the 1D-analysis (i.e. small red band in Figure \ref{fig_pattern_old}) which is due to the diffusion coefficients $D_s$ and $D_n$ being put to a non-zero value in the 2D-analysis. The fact that in the 2D-analysis the pattern domain directly borders on the domain with non-physical solution means that this domain boundary can be given analytically using the condition $\nu_g=\nu_d$, which yields:
\begin{equation}\label{eq:Froude_boundary}
\tilde{\phi}_m=\frac{\alpha_d\,g\tilde{Y}^2_0}{\alpha_g}F_0^2
\end{equation}
\begin{figure}
\begin{center}
    \includegraphics[width=130mm]{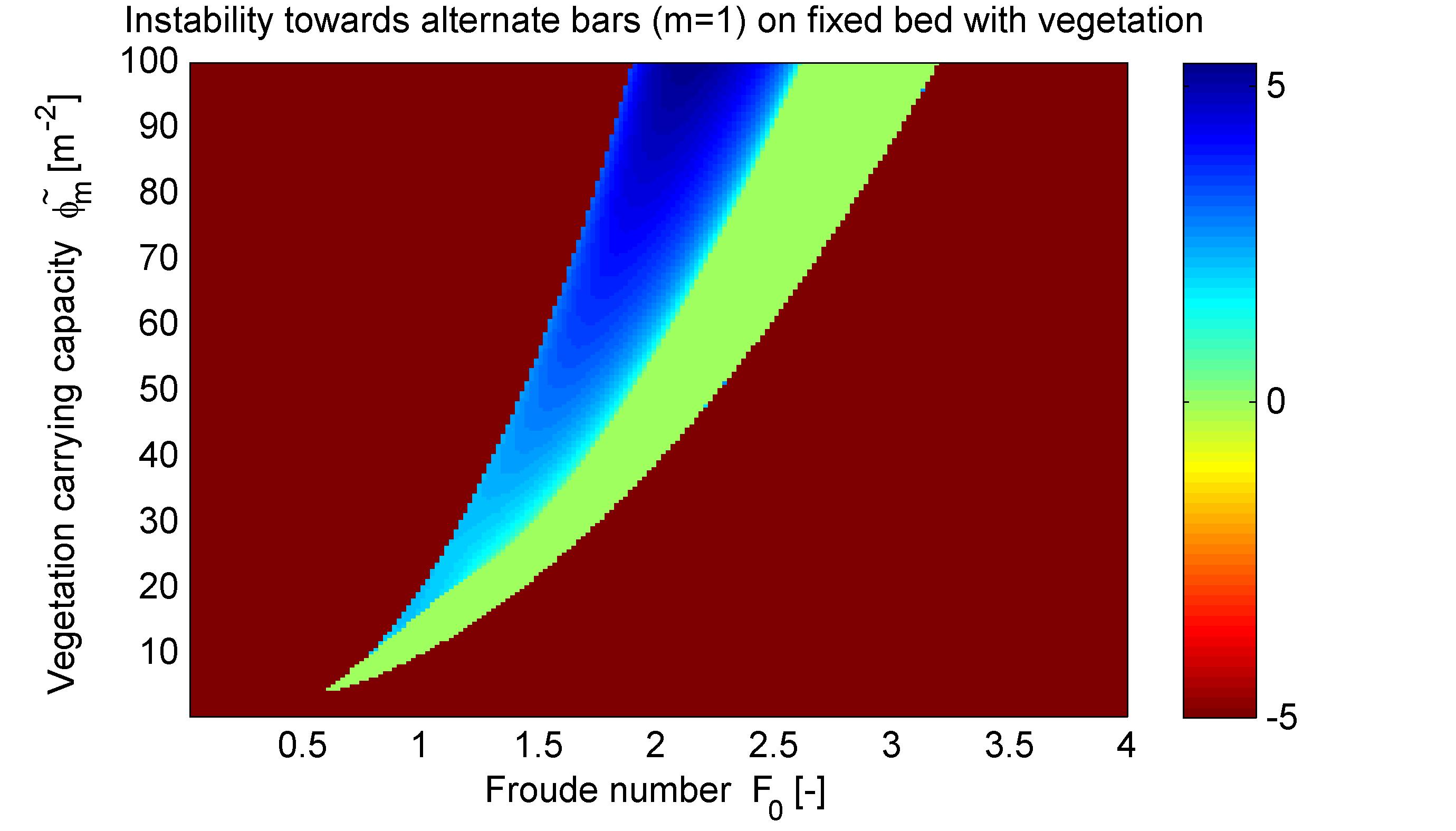} \\
    \caption{Alternate bar formation as a function of Froude number $F_0$ and vegetation carrying capacity $\tilde{\phi}_m$:  the color code indicates the value of the most unstable longitudinal wavenumber $k_s$, negative numbers (red) mean no instability; parameter values are $m=1$, $\beta=20$, $\alpha_g=1\,\mathrm{m^2s^{-1}}$, $\alpha_d=1\,\mathrm{m^{-3}s}$, $D_s=100\,\mathrm{m^2s^{-1}}$, $D_n=100\,\mathrm{m^2s^{-1}}$ and values indicated in Table \ref{table_param}}
    \label{fig:2D_svv_phi_c}
\end{center}
\end{figure}
We can also understand now why in the 1D analysis increasing the vegetation diffusion coefficient lead to a decreasing pattern domain (see Figure \ref{fig_pattern_D}). In fact, increasing the diffusion coefficient increases the part of the domain with the dominant longitudinal wavenumber $k_s$ equal to zero. But, in the 1D-context, a longitudinal wavenumber equal to zero means that no patterns exist since no lateral variability is possible. In contrast to Figure \ref{fig_pattern_D}, if we plotted $F_0$ versus the diffusion coefficient for the 2D-analysis, we would just get two straight boundaries at constant Froude number.\\
Equation (\ref{eq:Froude_boundary}) also allows us to calculate the boundary at the higher Froude number of Figures \ref{fig:2D_svv3_1} and \ref{fig:2D_svv3_4} (boundary is independent of bar order):
\begin{equation}
F_{0max}=\sqrt{\frac{\alpha_g\tilde{\phi}_m}{\alpha_dg\tilde{Y}^2_0}}=2.26
\end{equation}
Furthermore, Figure \ref{fig:2D_svv_phi_old} shows that longitudinally homogeneous patterns (vegetated, longitudinal channels) only occur at very low relative vegetation density. We write relative density because $\phi_0$ is normalized using $\tilde{\phi}_m$ which means that in case $\tilde{\phi}_m$ is large the real vegetation density does not necessarily have to be small. Figure \ref{fig:2D_svv_phi_c} shows the dominating longitudinal wavenumber in the $F_0-\tilde{\phi}_m$ space and we can see that the same conclusions regarding vegetation balance are true than for the 1D-analysis: the largest longitudinal wavenumber $k_s$ occurs if growth and mortality through uprooting are well balanced, larger $k_s$ occur for larger values of Froude number and vegetation carrying capacity. Concluding the analysis of the vegetation growth balance, we can say that, as in the 1D model, the pattern domain seems to be simply connected (one domain without holes) and it continues to open up as $\tilde{\phi}_m$ goes to infinity. Instability towards alternate bars could be detected, but not towards multiple bars.\\

\newpage
\subsubsection{SVEV Equations (movable bed with vegetation)}\label{sec:2D_sta_SVEV}
\begin{figure}
\begin{center}
    \includegraphics[width=130mm]{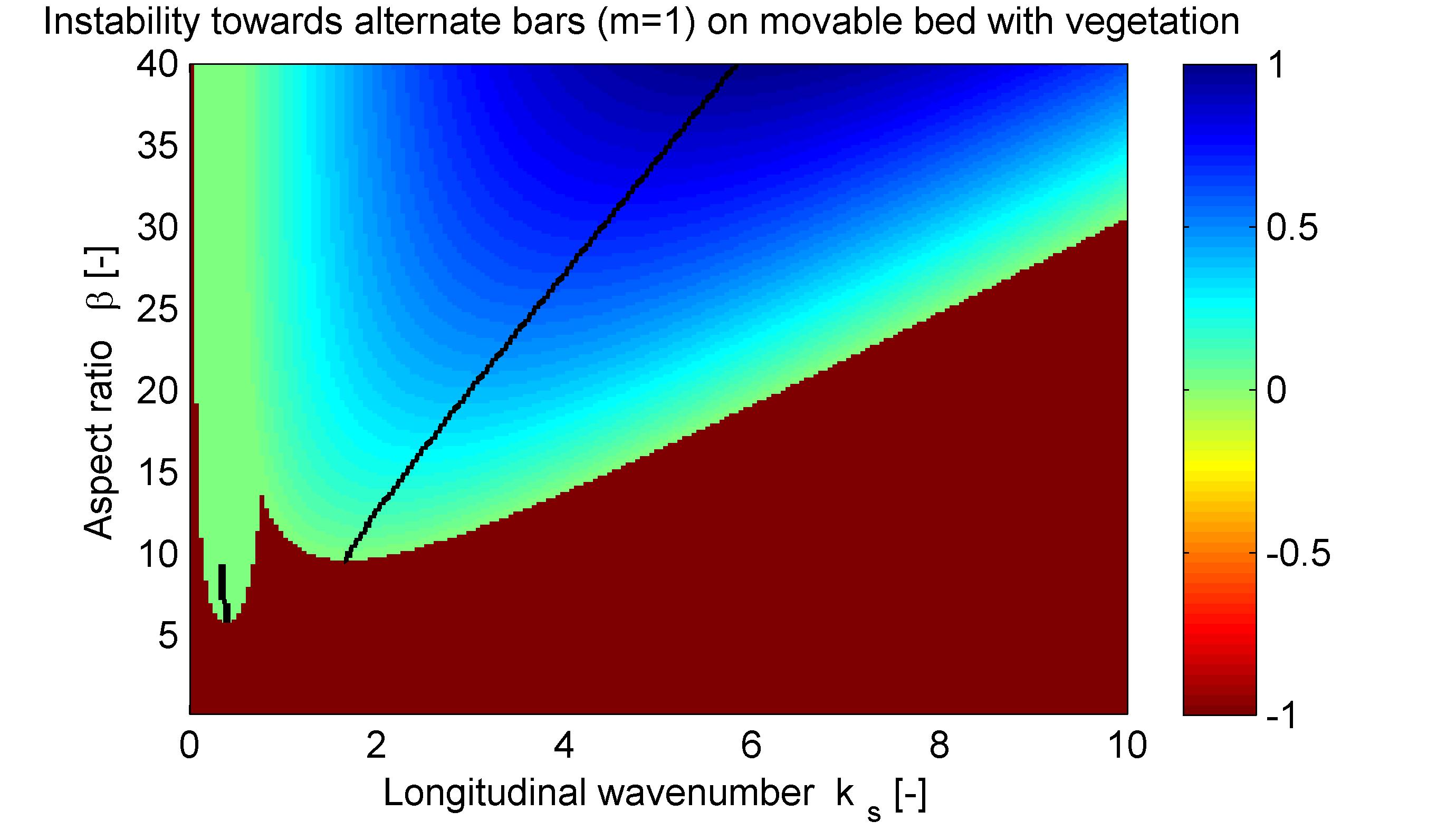} \\
    \caption{Relative growth rate (normalized with respect to the maximum growth rate occurring on the Figure) of alternate bar formation (m=1) as a function of longitudinal wavenumber $k_s$ and aspect ratio $\beta$: the color code indicates relative growth rate and a value of $-1$ (red) means that no patterns exist; the black line shows the maximum growth rate for given $\beta$ and thus indicates the value of the dominating longitudinal wavenumber; parameter values are $m=1$, $F_0=1.5$, $\gamma=10^{-3}$, $\tilde{\phi}_m=50\,\mathrm{m^{-2}}$, $\alpha_g=1\,\mathrm{m^2s^{-1}}$, $\alpha_d=1\,\mathrm{m^{-3}s}$, $D_s=100\,\mathrm{m^2s^{-1}}$, $D_n=100\,\mathrm{m^2s^{-1}}$ and values indicated in Tables \ref{table_param} and \ref{table:2D_sve}}
    \label{fig:2D_svev1}
\end{center}
\end{figure}
\begin{figure}
\begin{center}
    \includegraphics[width=130mm]{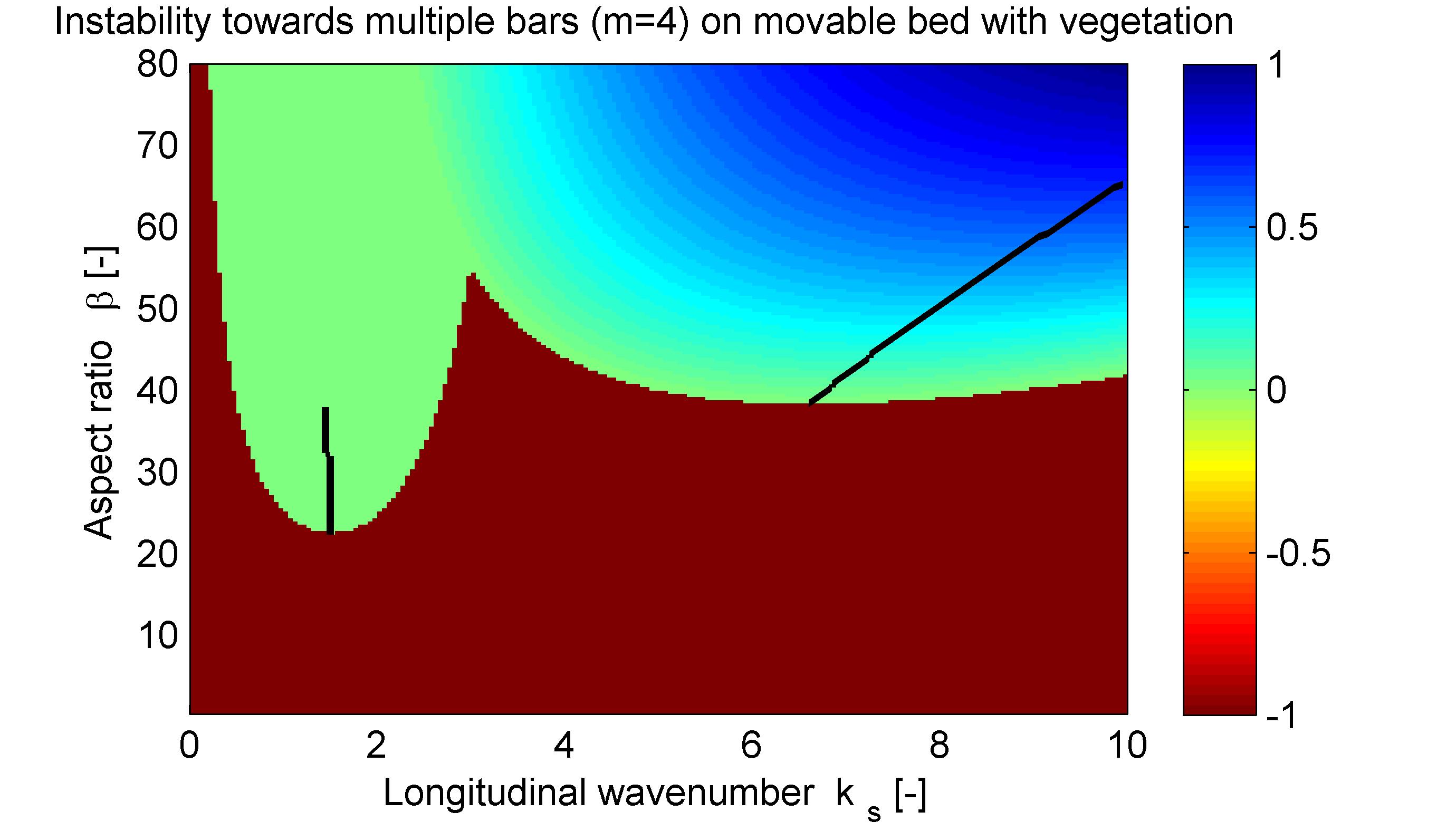} \\
    \caption{Relative growth rate (normalized with respect to the maximum growth rate occurring on the Figure) of multiple bar formation (m=4) as a function of longitudinal wavenumber $k_s$ and aspect ratio $\beta$: the color code indicates relative growth rate and a value of $-1$ (red) means that no patterns exist; the black line shows the maximum growth rate for given $\beta$ and thus indicates the value of the dominating longitudinal wavenumber; parameter values are $m=4$, $F_0=1.5$, $\gamma=10^{-3}$, $\tilde{\phi}_m=50\,\mathrm{m^{-2}}$, $\alpha_g=1\,\mathrm{m^2s^{-1}}$, $\alpha_d=1\,\mathrm{m^{-3}s}$, $D_s=100\,\mathrm{m^2s^{-1}}$, $D_n=100\,\mathrm{m^2s^{-1}}$ and values indicated in Tables \ref{table_param} and \ref{table:2D_sve}}
    \label{fig:2D_svev1_4}
\end{center}
\end{figure}
\begin{figure}
\begin{center}
    \includegraphics[width=130mm]{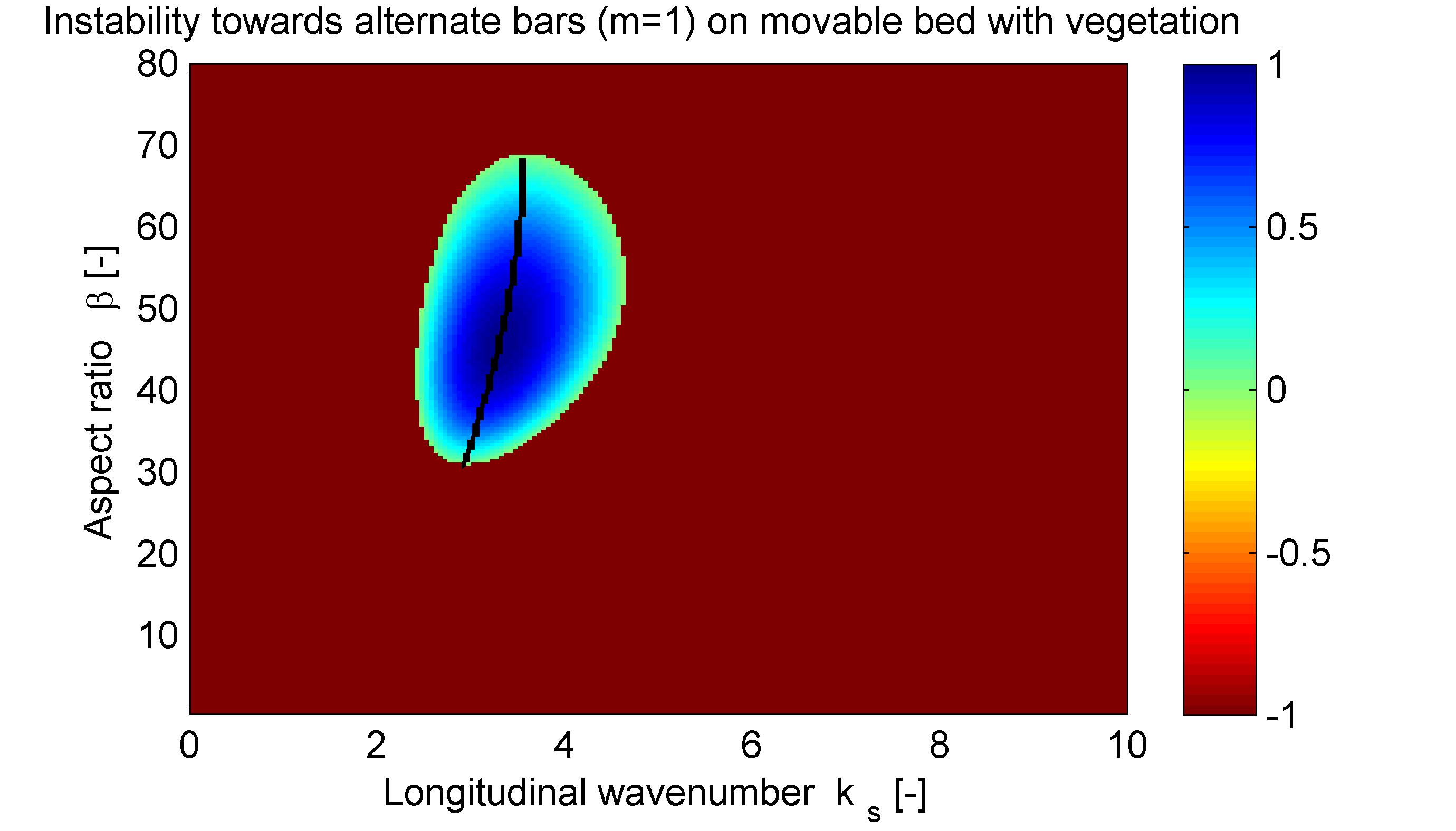} \\
    \caption{Relative growth rate (normalized with respect to the maximum growth rate occurring on the Figure) of alternate bar formation (m=1) as a function of longitudinal wavenumber $k_s$ and aspect ratio $\beta$: the color code indicates relative growth rate and a value of $-1$ (red) means that no patterns exist; the black line shows the maximum growth rate for given $\beta$ and thus indicates the value of the dominating longitudinal wavenumber; parameter values are $m=1$, $F_0=0.1$, $\gamma=10^{-3}$, $\tilde{\phi}_m=10\,\mathrm{m^{-2}}$, $\alpha_g=1\,\mathrm{m^2s^{-1}}$, $\alpha_d=1\,\mathrm{m^{-3}s}$, $D_s=100\,\mathrm{m^2s^{-1}}$, $D_n=100\,\mathrm{m^2s^{-1}}$ and values indicated in Tables \ref{table_param} and \ref{table:2D_sve}}
    \label{fig:2D_svev1_F01}
\end{center}
\end{figure}
\begin{figure}
\begin{center}
    \includegraphics[width=130mm]{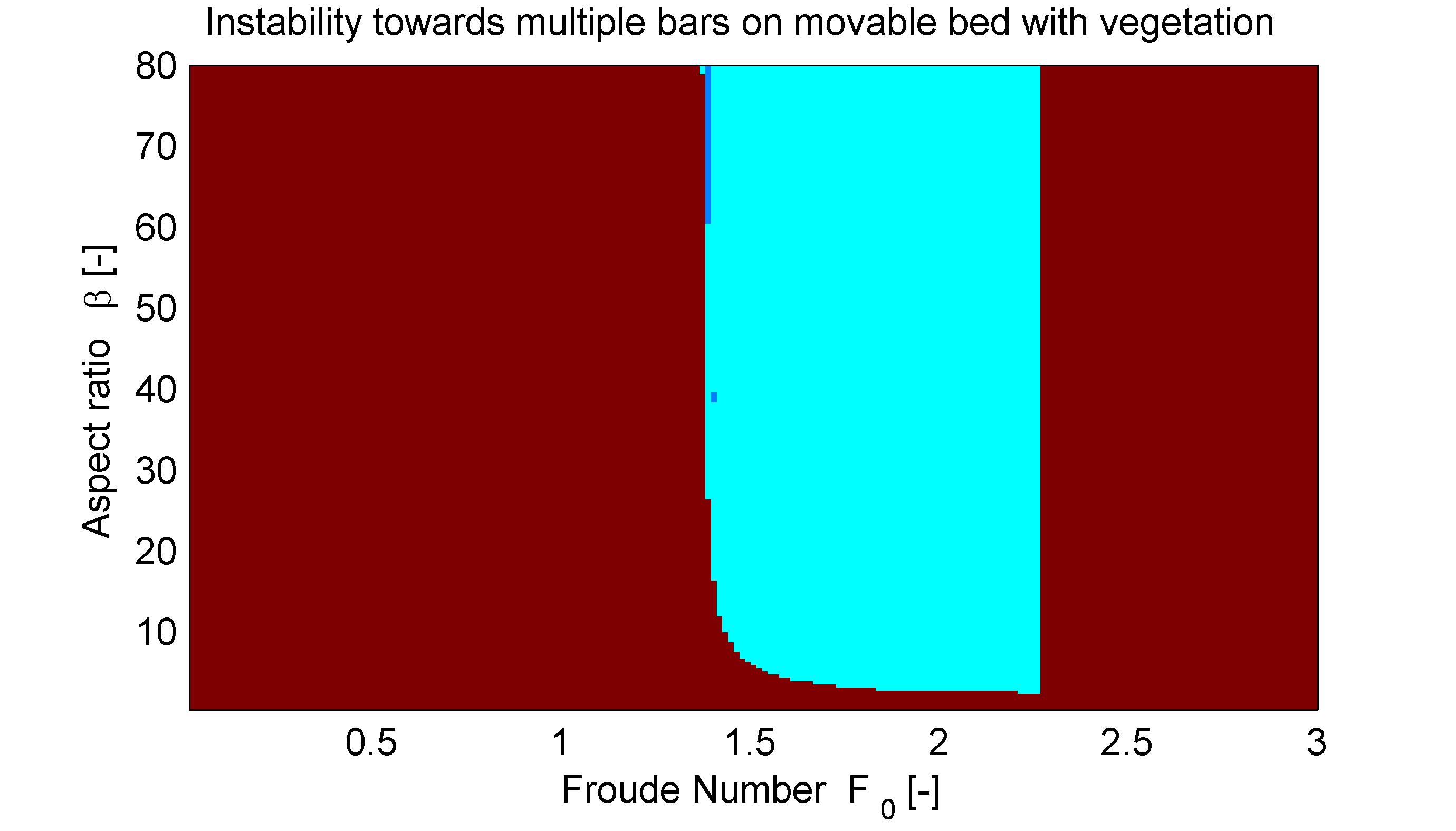} \\
    \caption{Multiple bar formation as a function of Froude number $F_0$ and aspect ratio $\beta$: no instability (red), alternate bars (light blue), multiple bars (darker blues for increasing bar order $m=2,3,4$); parameter values $\gamma=10^{-3}$, $\tilde{\phi}_m=50\,\mathrm{m^{-2}}$, $\alpha_g=1\,\mathrm{m^2s^{-1}}$, $\alpha_d=1\,\mathrm{m^{-3}s}$, $D_s=100\,\mathrm{m^2s^{-1}}$, $D_n=100\,\mathrm{m^2s^{-1}}$ and values indicated in Tables \ref{table_param} and \ref{table:2D_sve}\\}
    \label{fig:2D_svev3_mult}
\end{center}
\end{figure}
\begin{figure}
\begin{center}
    \includegraphics[width=130mm]{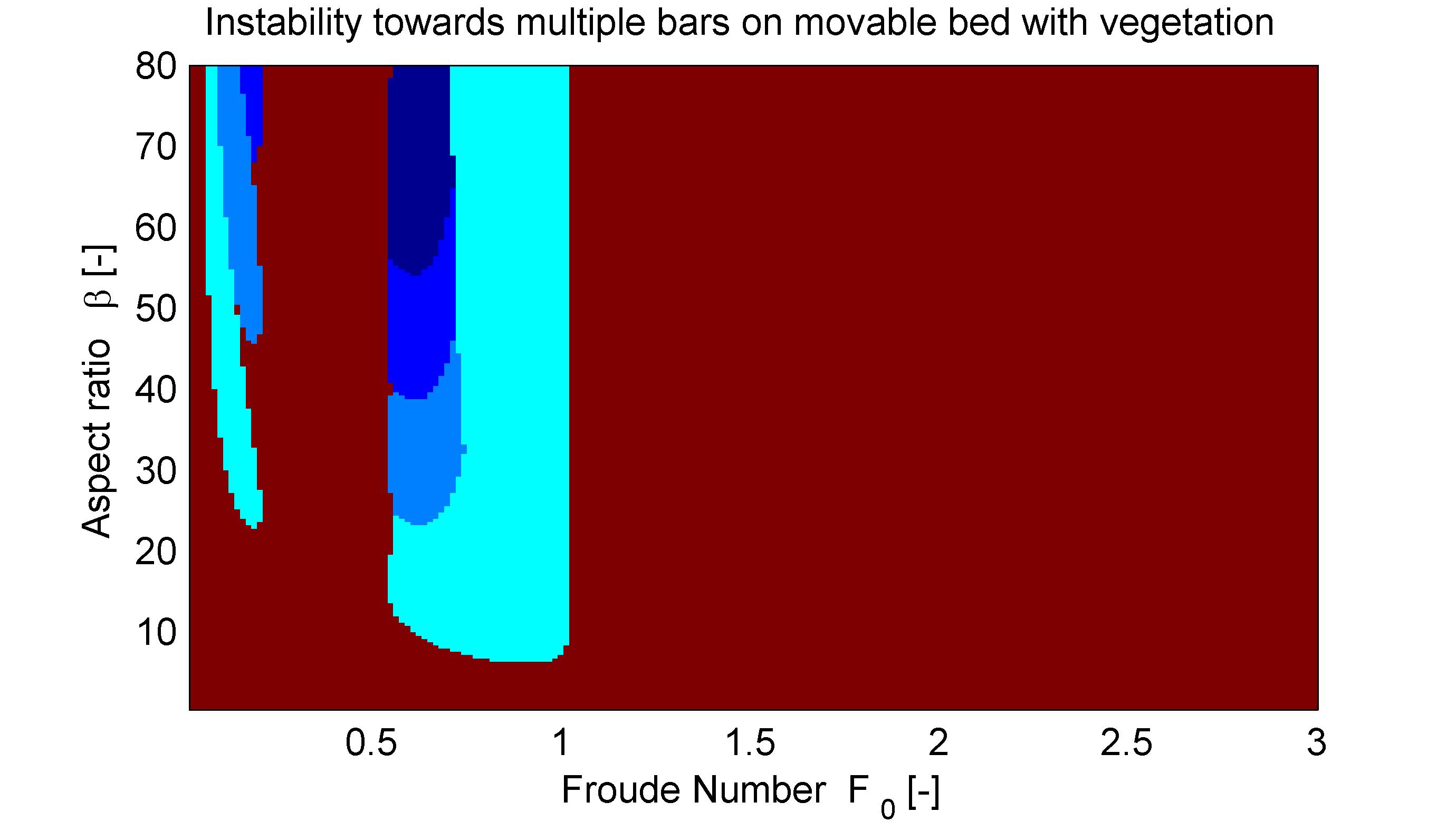} \\
    \caption{Multiple bar formation as a function of Froude number $F_0$ and aspect ratio $\beta$: no instability (red), alternate bars (light blue), multiple bars (darker blues for increasing bar order $m=2,3,4$); parameter values $\gamma=10^{-3}$, $\tilde{\phi}_m=10\,\mathrm{m^{-2}}$, $\alpha_g=1\,\mathrm{m^2s^{-1}}$, $\alpha_d=1\,\mathrm{m^{-3}s}$, $D_s=100\,\mathrm{m^2s^{-1}}$, $D_n=100\,\mathrm{m^2s^{-1}}$ and values indicated in Tables \ref{table_param} and \ref{table:2D_sve}\\}
    \label{fig:2D_svev3_mult_phi10}
\end{center}
\end{figure}
\begin{figure}
\begin{center}
    \includegraphics[width=130mm]{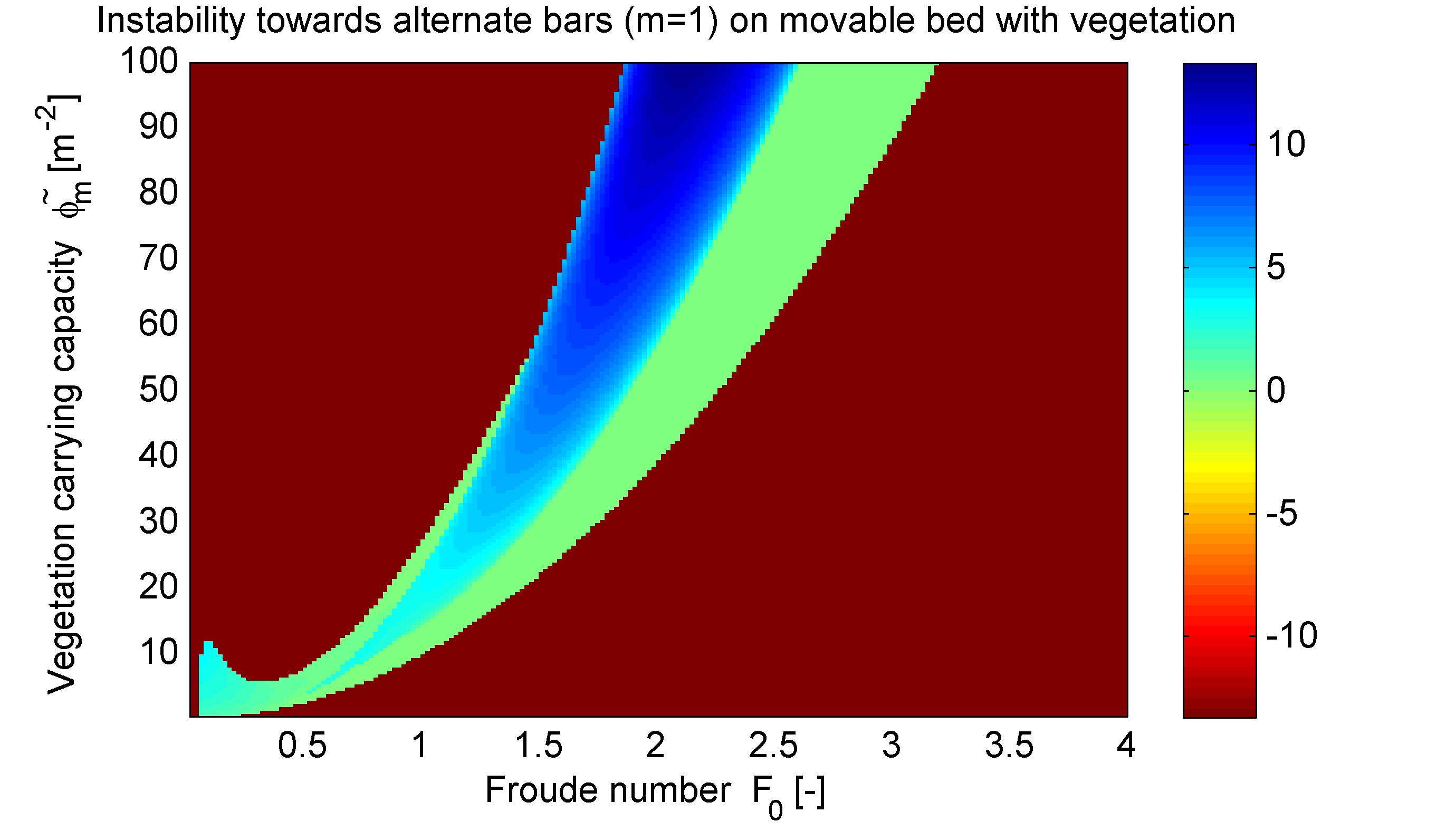} \\
    \caption{Alternate bar formation as a function of Froude number $F_0$ and vegetation carrying capacity $\tilde{\phi}_m$:  the color code indicates the value of the most unstable longitudinal wavenumber $k_s$, negative numbers (red) mean no instability; parameter values are $m=1$, $\beta=50$, $\gamma=10^{-3}$, $\alpha_g=1\,\mathrm{m^2s^{-1}}$, $\alpha_d=1\,\mathrm{m^{-3}s}$, $D_s=100\,\mathrm{m^2s^{-1}}$, $D_n=100\,\mathrm{m^2s^{-1}}$ and values indicated in Tables \ref{table_param} and \ref{table:2D_sve}}
    \label{fig:2D_svev_phi_c}
\end{center}
\end{figure}
\begin{figure}
  \centering
  \def\svgwidth{400pt}
  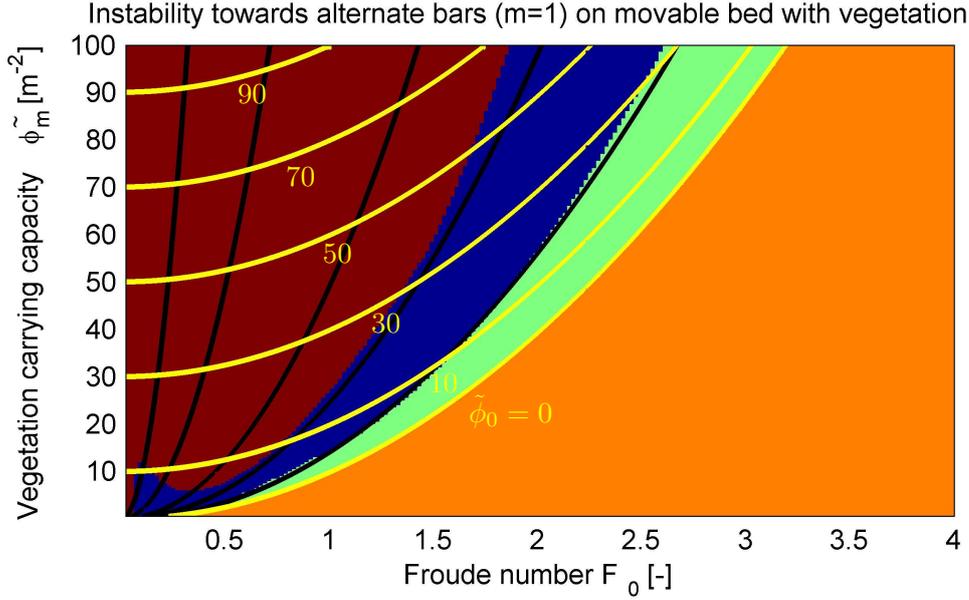
  \caption{Alternate bar formation as a function of Froude number $F_0$ and vegetation carrying capacity $\tilde{\phi}_m$ along with contour lines of dimensionless homogeneous vegetation density $\phi_0$ in black (line labels in Figure \ref{fig:2D_svv_phi_old}) and of physical homogeneous vegetation density $\tilde{\phi}_0$ in yellow: the domain with finite patterns is depicted in blue, pattern with $k=0$ in green, no pattern in red and unphysical solutions ($\phi_0<0$) in orange; parameter values are $m=1$, $\beta=50$, $\gamma=10^{-3}$, $\alpha_g=1\,\mathrm{m^2s^{-1}}$, $\alpha_d=1\,\mathrm{m^{-3}s}$, $D_s=100\,\mathrm{m^2s^{-1}}$, $D_n=100\,\mathrm{m^2s^{-1}}$ and values indicated in Tables \ref{table_param} and \ref{table:2D_sve}}
  \label{fig:2D_svev_phi_old}
\end{figure}

After having studied separately the effects on 2-dimensional river pattern formation of sediment dynamics and vegetation dynamics, we finally move on to the instability analysis of the complete 2D-model which includes sediment and vegetation dynamics and thus represents a river with movable bed and vegetation coverage. We saw before that sediment dynamics as well as vegetation dynamics are able to induce pattern formation in certain domains of the parameter space. We now want to know which of these effects remain or whether they even combine to form something not seen in the analysis of either sediment or vegetation dynamics alone.\\
We start with looking at Figure \ref{fig:2D_svev1} ($F_0=1.5$) which indeed shows positive growth rates for a range of aspect ratio $\beta$. Clearly, this pattern domain seems to be a superposition of Figures \ref{fig:2D_sve1} (note that this Figure is with $F_0=0.5$ instead of $F_0=1.5$) and \ref{fig:2D_svv1} with sediment influenced positive growth rates to the left and vegetation influenced ones to the right. Both parts of the pattern domain have a lower boundary (minimum aspect ratio $\beta$), but the domain allegedly created by sediment dynamics has positive growth rates at lower longitudinal wavenumbers (higher wavelenghts) than vegetation dynamics. Figure \ref{fig:2D_svev1_4} shows the same phenomena for multiple bars (m=4). In both Figures, we can see that for these parameters (see captions) sediment dynamics does determine the dominating longitudinal wavenumber $k_s$ (black line) for lower values of aspect ratio $\beta$ while vegetation dynamics is dominant at higher aspect ratios (around 10 for alternate bars and around 40 for multiple bars in Figure \ref{fig:2D_svev1_4}). 
\begin{figure}
\begin{center}
    \includegraphics[width=130mm]{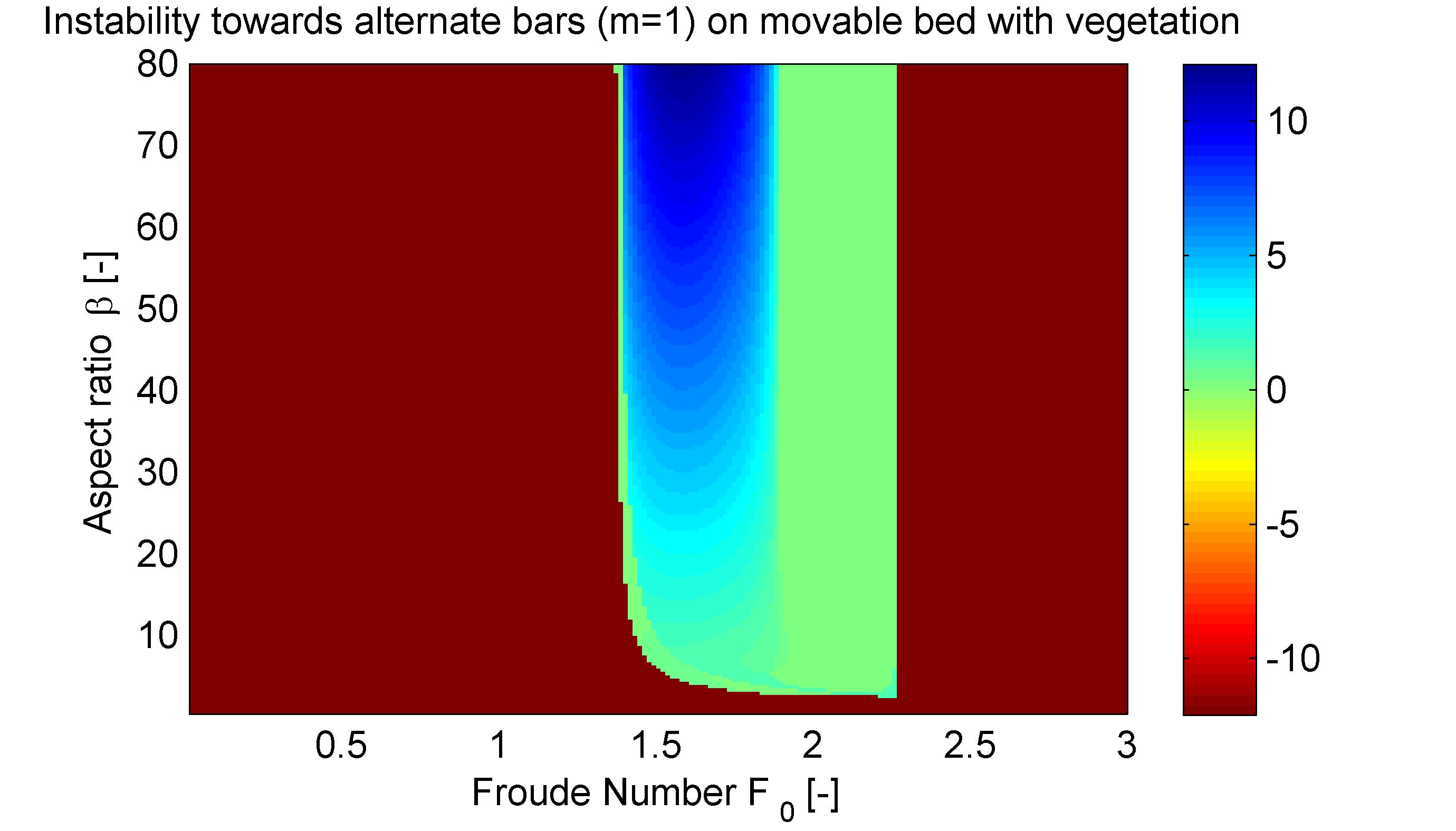} \\
    \caption{Alternate bar formation as a function of Froude number $F_0$ and aspect ratio $\beta$: the color code indicates the value of the most unstable longitudinal wavenumber $k_s$, negative numbers (red) mean no instability; parameter values are $m=1$, $\gamma=10^{-3}$, $\tilde{\phi}_m=50\,\mathrm{m^{-2}}$, $\alpha_g=1\,\mathrm{m^2s^{-1}}$, $\alpha_d=1\,\mathrm{m^{-3}s}$, $D_s=100\,\mathrm{m^2s^{-1}}$, $D_n=100\,\mathrm{m^2s^{-1}}$ and values indicated in Tables \ref{table_param} and \ref{table:2D_sve}\\}
    \label{fig:2D_svev3}
\end{center}
\end{figure}
\begin{figure}
\begin{center}
    \includegraphics[width=130mm]{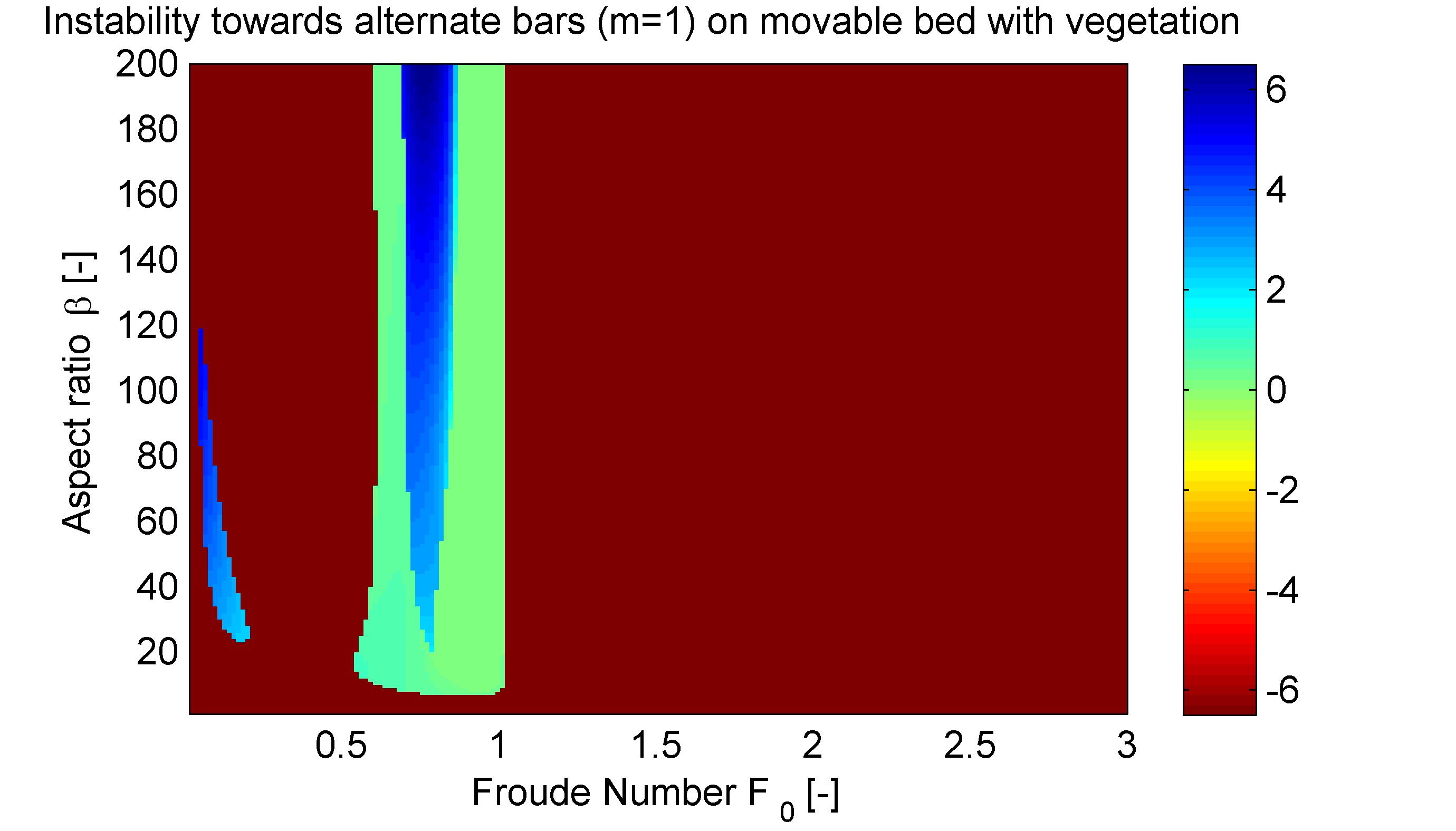} \\
    \caption{Alternate bar formation as a function of Froude number $F_0$ and aspect ratio $\beta$: the color code indicates the value of the most unstable longitudinal wavenumber $k_s$, negative numbers (red) mean no instability; parameter values are $m=1$, $\gamma=10^{-3}$, $\tilde{\phi}_m=10\,\mathrm{m^{-2}}$, $\alpha_g=1\,\mathrm{m^2s^{-1}}$, $\alpha_d=1\,\mathrm{m^{-3}s}$, $D_s=100\,\mathrm{m^2s^{-1}}$, $D_n=100\,\mathrm{m^2s^{-1}}$ and values indicated in Tables \ref{table_param} and \ref{table:2D_sve}\\}
    \label{fig:2D_svev3_phi}
\end{center}
\end{figure}
Then, if we take a much lower value for $F_0$ (along with a lower vegetation carrying capacity of $\tilde{\phi}_m=10$) we get a completely different picture which is shown in Figure \ref{fig:2D_svev1_F01}: all of a sudden, the pattern domain completely changes and we have only one domain that occurs with a lower and upper limit for the aspect ratio as well as a limited domain for longitudinal wavenumber that. Figure \ref{fig:2D_svev3} depicts the same situation from another angle, namely in the $F_0$-$\beta$ space. We can identify one single parameter domain leading to instability including a lower limit for a certain value for $\beta$. However, if we reduce vegetation carrying capacity (thus decreasing the influence of vegetation), we can see that two different instability domains appear in Figure \ref{fig:2D_svev3_phi} which was not the case in the 1D analysis. Not surprisingly though, the larger domain to the right resembles strongly the vegetation induced domain already seen in section \ref{sec:2D_sta_SVV} which also means that the domain to the left should probably be due to sediment dynamics. Comparing Figures \ref{fig:2D_svev3} and \ref{fig:2D_svev3_phi}, we can also see that the domain at larger Froude numbers decreases and is slightly more to the left in the latter Figure. This is another hint that the right domain comes from vegetation dynamics: when decreasing vegetation carrying capacity, the Froude number has to decrease as well otherwise the river's uprooting capacity would overwhelm vegetation growth.\\
\begin{figure}
\begin{center}
    \includegraphics[width=130mm]{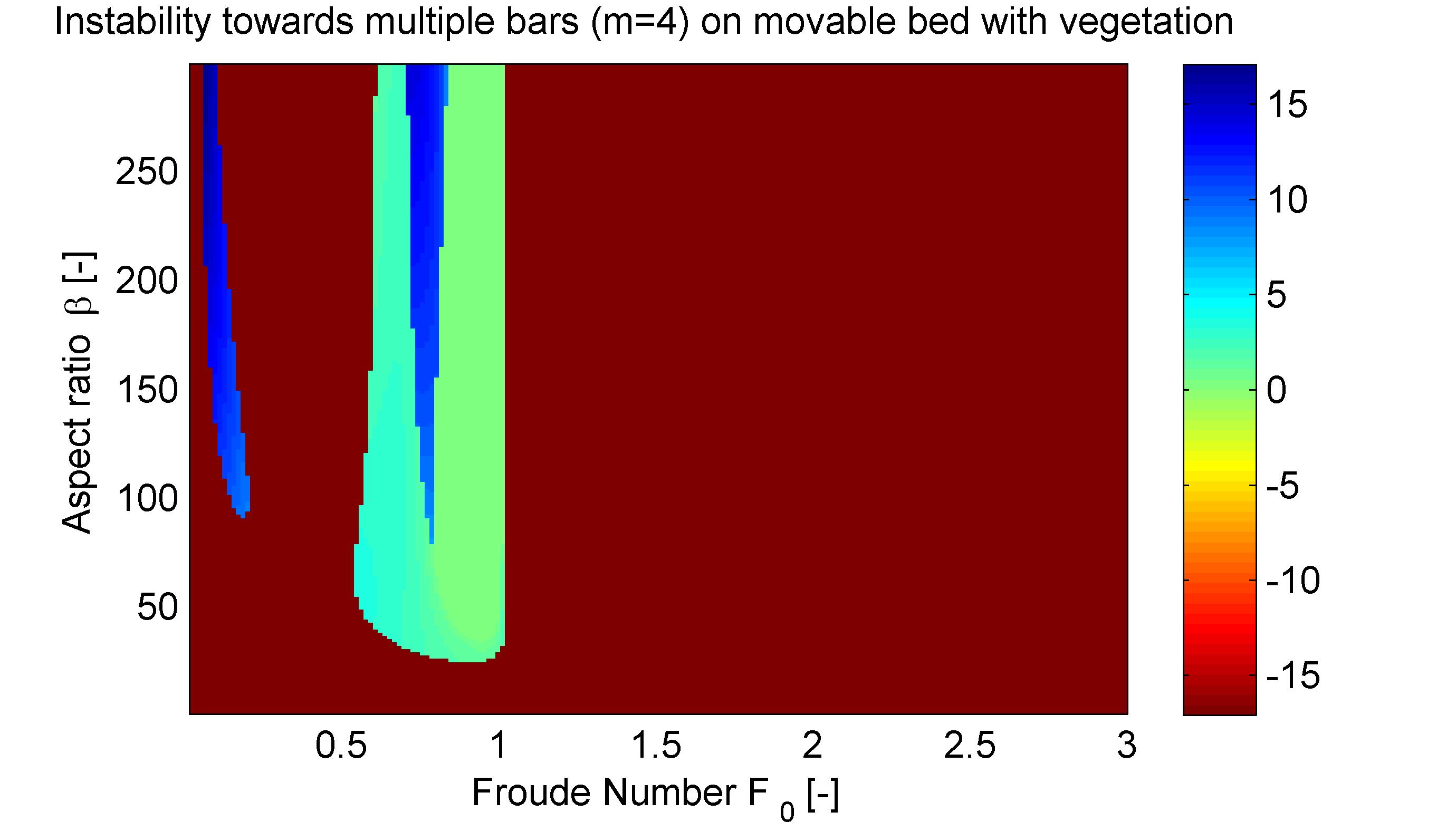} \\
    \caption{Multiple bar formation as a function of Froude number $F_0$ and aspect ratio $\beta$: the color code indicates the value of the most unstable longitudinal wavenumber $k_s$, negative numbers (red) mean no instability; parameter values are $m=4$, $\gamma=10^{-3}$, $\tilde{\phi}_m=50\,\mathrm{m^{-2}}$, $\alpha_g=1\,\mathrm{m^2s^{-1}}$, $\alpha_d=1\,\mathrm{m^{-3}s}$, $D_s=100\,\mathrm{m^2s^{-1}}$, $D_n=100\,\mathrm{m^2s^{-1}}$ and values indicated in Tables \ref{table_param} and \ref{table:2D_sve}\\}
    \label{fig:2D_svev3_phi_4}
\end{center}
\end{figure}
Next, we would like to know what happens to instability towards multiple bars. In the previous sections, we could show that instability towards multiple bars exists for flow dynamics coupled with sediment dynamics but not for flow dynamics coupled with vegetation dynamics. Figure \ref{fig:2D_svev3_phi_4} shows the domains and dominating longitudinal wavenumber for multiple bars (m=4). We can see that both domains are slightly shifted upwards (towards positive $\beta$) which was already seen before in the analysis with sediment dynamics. Moreover, higher order multiple bars develop higher longitudinal wavenumbers and thus shorter longitudinal wavelengths than alternate bars.\\
One important question remains though: which bar order will dominate in parameter domains where alternate bars and several orders of multiple bars can potentially exist? Figure \ref{fig:2D_svev3_mult} answers this question partially by showing that in the case of a movable bed with vegetation the domain to the right is always unstable towards alternate bars. Sediment dynamics induced patterns are not visible due to vegetation processes completely dominating river bed dynamics. Figure \ref{fig:2D_svev3_mult_phi10} shows what happens in not very highly vegetated riverbeds ($\tilde{\phi}_m=10$). As in the case of alternate bars (Figure \ref{fig:2D_svev3_phi}), there is sediment induced instability to the left, but the domain to the right seems to contain both sediment incued instability (towards higher order of multiple bars with increasing $\beta$) and vegetation induced instability to the very right. This means that instability towards multiple bars in vegetated rivers is indeed possible, but only at rather low Froude numbers (either at about F=0.2-0.3 or at F=0.6-0.7 in this case). Yet, as it can be seen in Figures \ref{fig:2D_svev3} to \ref{fig:2D_svev3_phi_4}, instability towards patterns with very low longitudinal wavenumbers occurs only in the domain at higher Froude number. At very low Froude numbers, only rather high wavenumbers are reached in the asymptotic limit and thus, longitudinal channels (which are characterized by $k_s$ close to zero) do not occur. Since the Froude number for flooding events in the Marshall River is rather low (0.3-0.4, \cite{To}), our results do not predict channels for this river which is contrary to reality. However, for not very highly vegetated riverbeds ($\tilde{\phi}_m=10\,\mathrm{m^{-2}}$ in Figure \ref{fig:2D_svev3_mult_phi10}) multiple bars may occur at Froude numbers up to about 0.7. In such a parameter configuration, we can thus have instability towards multiple bars for the Marshall River, but at rather high longitudinal wavenumbers, thus resembling more a braiding pattern than a channeled riverbed.\\
Interestingly, Figure \ref{fig:2D_svev3_phi} indicates that with decreasing vegetation carrying capacity $\tilde{\phi}_m$ the two parts of the domain move closer together which implies that they eventually could merge once $\tilde{\phi}_m$ falls below a certain threshold. We want to have a closer look at this by plotting $\tilde{\phi}_m$ against Froude number in Figure \ref{fig:2D_svev_phi_c}. Although it is difficult to separate sediment induced instability from vegetation induced, this can be done when the current Figure is compared to Figure \ref{fig:2D_svv_phi_c}. The sediment domain consists of the thickening of the domain to the very left as well as the thin greenish domain that joins the vegetation induced domain at the left. Then, it can be seen that the domains actually merge when $\tilde{\phi}_m$ is low enough. This means that although the pattern domain as a whole is not simply connected anymore (there are holes in the domain) it is still is connected.\\
To conclude, we again plot vegetation carrying capacity $\tilde{\phi}_0$ against $F_0$ in Figure \ref{fig:2D_svev_phi_old} and include contour lines for dimensionless homogeneous vegetation density $\phi_0$ (in black) as well as physical homogeneous vegetation density $\tilde{\phi}_0$ (plants per $\mathrm{m^2}$, in yellow). As explained earlier, we need well-developed vegetation (meaning $\phi_0$ well above zero) in order to not have negative vegetation density because of the sinusoidal oscillations. This means that the vegetation density wave amplitude always has to be smaller than the actual vegetation density. We don't have a way to know the wave amplitude, but still at least we have to wonder whether the model is valid for the part of the pattern domain close to the line $\phi_0=0$. This would mean that the part of the domain where the dominant $k_s=0$ could be actually non-physical due to the assumptions not met. Anyway, such patterns with dominant longitudinal wavenumber equal to zero would be alternate bars (remember that alternate bars always grow faster than multiple bars in this region of pattern domain). This would result in an asymmetric channel where either the left or the right side would be filled with sediment while water is flowing on the other side. Thus, what makes actually physically sense in Figure \ref{fig:2D_svev_phi_c} is instability towards alternate bars with finite $k_s$ at higher Froude numbers and instability towards multiple bars with rather high $k_s$ at lower Froude numbers.\\
We now quickly want to give an estimate for a longitudinal wavenumber of multiple bars. Looking at Figure \ref{fig:2D_svev3_phi_4} for example, we can see that a typical longitudinal wavenumber is $k_s=15$ for m=4 in the region where these multiple bars dominate. This yields for $\tilde{B}=200\,\mathrm{m}$ 
\begin{equation}
\tilde{\lambda}_s=\frac{2\pi}{k_s/B}=83.8\, \mathrm{m}
\end{equation}
which is not completely unreasonable as an order of magnitude, but too low in comparison of a river width of $2\tilde{B}=400\,\mathrm{m}$. Thus, the parameter values of the model would need to be checked and investigated order to get a more reasonable result.\\
Finally, we want to find out whether there is evidence that the pattern domain at low Froude numbers might be linked to actual vegetation density in the river. We can see on Figure \ref{fig:2D_svev_phi_old} that the domain to the left can best be characterized by a very high $\phi_0$ (black lines, see Figure \ref{fig:2D_svv_phi_old} for values). This is astonishing to some extent since this domain is thought to be governed by sediment dynamics. But then again, a very high relative (dimensionless) vegetation density does not necessarily mean a lot of variation. So, one could describe this domain to be representative for rivers with stable vegetation density close to carrying capacity. The vegetation would not be influenced much by flow due to low uprooting capacity (low Froude number) and sediment dynamics would thus be governing the river's instability mechanisms in this region of the parameter space as we suspected in the beginning.
\subsection{Synthesis and Conclusion of the 2D analysis} \label{sec:2D_concl}
We reproduced known results of river instability towards alternate and multiple bars using a 2-dimensional de Saint-Venant-Exner framework. As expected, the higher a river's width-to-depth ratio (aspect ratio $\beta$) the more bars a river tends to develop laterally, leading to the formation of multiple bars. The Froude number does has a crucial role at low F and no patterns exist above roughly $F_0=2$. When analyzing the 2D-de Saint-Venant equation combined with vegetation dynamics, we find, similarly to the 1-dimensional model, that there exists a domain where vegetation growth and mortality by means of uprooting compete and thus instability towards finite patterns prevails. However, a minimum value for the aspect ratio is required to induce such instability and only instability towards alternate bars occurs (the exponential growth rate of alternate bars in the linear regime always exceeds the growth rate of multiple bars of any order). The Froude number, which is directly proportional to stream velocity, is very important to balance vegetation dynamics since a higher velocity increases the river's uprooting capacity (see equation (\ref{eq:veg3})).\\
When analyzing the full model including sediment as well as vegetation dynamics, a pattern domain of essentially two parts is detected. One part occurs at low Froude numbers and high dimensionless vegetation density (independently of actual vegetation carrying capacity) and mainly possesses features of sediment transport induced instability: independence of Froude number, multiple bar order increases with increasing aspect ratio and higher dimensionless wavenumber for higher aspect ratios. The other part stems from vegetation growth balance and inherits equally its attributes: domain highly dependent on Froude number, instability towards alternate bars with rather low longitudinal wavenumbers at low vegetation density, but no instability towards multiple bars. The two parts of the pattern domain are separated for higher values of vegetation carrying capacity $\tilde{\phi}_m$ and linked when $\tilde{\phi}_m$ falls below a certain threshold (which depends on the other vegetation parameters $\{\alpha_g, \alpha_d, D_s, D_n\}$).

\newpage
\phantom{}
\newpage
\section{Conclusion and Perspectives} \label{sec:conclusion}
The goal of this work was to shed light on the influence of riparian vegetation on formation of morphological river patterns. We thus performed a linear stability analysis on the 1D and 2D ecomorphodynamic equations which include analytical models for flow, sediment as well as vegetation dynamics. The vegetation model was kept very simple in order to be suitable for a stability analysis and it included terms for vegetation growth, diffusion through seeding/resprouting and mortality by means of uprooting caused by flow shear. At first, this equation was developed for rivers with constant flow, but was shown to also apply (under certain conditions) to variable flow and even ephemeral rivers.\\
Our analysis of the 1D model showed that vegetated rivers indeed exhibit instability towards longitudinal sediment waves due to the competitive interaction between vegetation growth and mortality. Then, instability towards river patterns with lateral structure (bars) was assessed using the 2D ecomorphodynamic equations and it was discovered that two different kinds of instabilities occur. Instability at lower Froude numbers is mainly driven by sediment dynamics and leads to formation of alternate and multiple bars with the bar order increasing with increasing river width-to-depth ratio (aspect ratio). At higher Froude numbers, only instability towards alternate bars was detected, independently of the aspect ratio.\\
We also looked at the value of the most unstable longitudinal pattern wavenumber. Generally, it was found that the values identified were reasonable (corresponding wavelength has the same order of magnitude than river width) although sometimes at the higher limit of what is allowed by the model assumptions. However, we were not able to identify instability towards longitudinally infinite multiple channels as it occurs in some reaches of the Marshall River (Figure \ref{fig:anabranch} B). Actually, the longitudinal wavenumbers found for multiple bar instability at low Froude numbers were too high to form such channels. Other parameter configurations or more detailed modeling would probably be required to match reality in this case.\\ In general, experimental verification of the equation adopted and research on the parameter values would be needed to adjust our purely theoretical model to reality. In addition, flume experiments could also help to quantify other aspects of vegetation and sediment transport interaction that we did not take into account. For instance, we only model the effect of vegetation on sediment transport indirectly by increasing the river's bed roughness. Yet, other processes like scouring that increases sediment ablation around plants (see \cite{MeSu} for a scouring model around bridge piers) and riverbed stabilization by plant's root systems are probably important as well but too difficult to model analytically at this stage. Further improvement of the vegetation model could include finding expressions for vegetation uprooting by gradual exposure of plants root system (Type II mechanism of \cite{Ed}) and flow diversion produced mainly by rigid vegetation.\\
Finally, since we were dealing with a non-normal operator in our stability analysis, transient growth of the system can occur which was not considered in the present work. Therefore, as was done by \cite{Ca} for the morphodynamic equations, a non-modal analysis of our linear ecomorphodynamic operator is conceivable in the future to evaluate the importance of such transient growths. The reason is that sometimes this transient behavior can actually be more relevant in reality than the asymptotic fate depending on the timescale of interest. Moreover, in addition to performing a stability analysis which only takes into account growth or decay at the linear level, we could extend our research by adding a non-linear numerical simulation of the initial perturbations. In fact, as the perturbations amplify non-linearities of the system may become important and eventually dominate, thus determining the asymptotic fate of the system. In any case, all these possibilities of improvement of current modeling of interaction of riparian vegetation and river morphology show that we are still barely scratching the surface of this complex subject.

\newpage
\bibliographystyle{plainnat}

\end{document}